\documentclass[twocolumn,tighten]{aastex63_rt42}

\usepackage{graphicx}
\usepackage{amsmath}
\usepackage{multirow}
\usepackage{color}
\usepackage{icomma}
\usepackage{etipopsbl}

\def\Watt{\textrm{W}}

\def\Hz{\textrm{Hz}}
\def\GHz{\textrm{GHz}}

\def\Msun{M_{\odot}}

\def\Gyr{\textrm{Gyr}}

\def\ga{\gtrsim}

\def\endash{\text{--}}

\shorttitle{ETI Broadcast Populations I. Formalism}
\shortauthors{Lacki}

\begin{document}

\title{Artificial Broadcasts as Galactic Populations: I. A Point Process Formalism for Extraterrestrial Intelligences and Their Broadcasts}

\newcommand{\UCB}{Department of Astronomy,  University of California Berkeley, Berkeley CA 94720}

\correspondingauthor{Brian C. Lacki}
\email{astrobrianlacki@gmail.com}
\author[0000-0003-1515-4857]{Brian C. Lacki}
\affiliation{Breakthrough Listen, \UCB}

\begin{abstract}
Artificial broadcasts from extraterrestrial intelligences (ETIs) are a hypothetical class of celestial phenomena.  Unlike known astrophysical objects, the societies that generate them may be able to replicate on galactic scales through interstellar travel. Different galaxies could thus have drastically different populations, with abundance variations of many orders of magnitude. I present a probabilistic formalism to treat this shared history, in which societies and their broadcasts are described by distributions over basic properties like lifespan and energy released. The framework contains a hierarchy of objects related by a tree structure. Discrete societies, the sources of broadcasts, are organized into potentially interstellar ``metasocieties.'' The population of each type of object is represented by a random point process in an abstract parameter hyperspace, a ``haystack.'' When a selection like an observation draws a sample, the point process is thinned. Given assumptions of interchangeability and independence, observables are modeled with compound Poisson random variables. I present an example of how selection bias can favor sampling longer-lived objects. I rederive the Drake Equation for societies in the limit of no expansion. When interstellar replication is present, however, the mean number of detected broadcasts can depend quadratically on stellar mass, suggesting a search strategy favoring large galaxies.
\end{abstract}


\section{Introduction}
\label{sec:Intro}

The Search for Extraterrestrial Intelligence (SETI; \citealt{Tarter01,Worden17}) rests on the premise that the technosignatures of extraterrestrial intelligences (ETIs) are astrophysical phenomena, arising {in} and shaped by a cosmic environment and detectable over cosmic distances. ETIs are not ruptures in the laws of physics but consequences of them, forming a cosmic population, however rare they actually are. The properties of this population are constrained by physical limitations on technology and unknown internal factors. 

{Negative SETI results, although leaving plenty of room for ETIs, are at least strong enough to {eliminate many possible classes of broadcasts as the main contributor to the Milky Way's luminosity. N}arrowband radio beacons with isotropic luminosities around $10^{12} \endash 10^{{26}}\ \Watt$ cannot possibly make up a significant fraction of {its current} radio emission (\citealt{Enriquez17,Price20,Tremblay20,WlodarczykSroka20}; {\citealt{Gajjar21};} see Paper II). Optical SETI constraints also are sensitive to energy outputs far smaller than the {Galaxy's stellar luminosity} \citep{Horowitz93,Tellis17,Maire19}.} {Very little work has been done in SETI at higher energies like X-rays \citep{Corbet97,Hippke17-XRay}, or alternate messengers like neutrinos \citep{Subotowicz79,Learned94,Hippke18-Messenger}. But o}ur ability to interpret the cosmos as natural implicitly suggests that {artificial broadcasts {are} not the main emission process in} other {wave bands} {(admittedly, not a certainty as per \citealt{Cirkovic18-Catch})}. 

If we were {looking for conventional sources around stars}, we would conclude that {the} bright{est} examples of the source class are rare in all galaxies as they should {more or less} {just} trace stellar mass{, with additional biases from things like metallicity and age.} Only exotic environments like globular clusters and galactic nuclei might have much higher abundances than seen in the field, as for close binary star systems, stellar mergers, and their products (\citealt{Bailyn95,Pooley03,Muno05}; for ETIs, {see} \citealt{DiStefano16}).

Unlike all known astrophysical phenomena, however, ETIs potentially \emph{replicate}.\footnote{Life might also replicate on interstellar scales via panspermia \citep[e.g.,][]{Napier04}.} Interstellar travel can allow {intelligent beings who evolved on a single world} to spread to a large fraction of a galaxy {in well less than a billion years} if such propagation is practical \citep{Hart75,Tipler80,Jones81,Zackrisson15,CarrollNellenback19}. Widespread replication presents great opportunities for SETI, as it vastly increases the technosignature profile of a galaxy \citep{Kuiper77,Wright14-Paradox}. We have no evidence that this has happened in our Galaxy, a statement known as the Fermi Paradox ({\citealt{Brin83};} \citealt{Wright14-Paradox}; {\citealt{Webb15}}; \citealt{Cirkovic18-Book}; {\citealt{Forgan19,Lingam21}}). 

{However,} this does not rule out {it happening} in other {galaxies}. If ETIs with interstellar travel are rare but not impossible, we could end up with a situation where most galaxies ({perhaps} including our own) have {none}, but {a few} have millions {of inhabited worlds.} Galaxies that {look alike} to us could have divergent technosignature {populations} due to unobservable historical factors (perhaps {set by} the first ETIs as in \citealt{Scheffer94} {and \citealt{Hair11}}). {The menagerie of societies in galaxies populated by starfarers did not all arise independently, but have a shared origin {in} perhaps a single world, even if the resulting population is extremely diverse. This shared history can shape the entire population of descendant ETIs, and so the group might be treated as a whole -- neither a simple collection of noninteracting societies nor a generic galaxy solely described by astrophysical properties. I call these intermediate-level groupings \emph{metasocieties}.}\footnote{Compare analogous notions of \emph{metapopulations} and \emph{metacommunities} in ecology, both describing situations in which organisms are clustered in many small-scale habitats that influence each other through dispersal -- members of a single species for metapopulations \citep{Hanski98}, and sets of multiple interacting species for metacommunities \citep{Leibold04}. Metapopulation ecology bears {a} direct analogy with several {of} the dynamics of interstellar migration and metasocieties -- for example, the survival of a metapopulation {depends} on quality and number of habitat patches \citep[see][]{CarrollNellenback19}. {The analogy between interstellar migration and metapopulations is explicitly noted in \citet{Lingam21}, who argue it implies that not all worlds will be inhabited at any given time.}} {{Therefore,} t}he underlying parameters of the {technosignature} distribution in each galaxy are themselves the result of a stochastic process{, relating to the properties of the underlying metasociety}. As a result, constraints on ETIs in one galaxy could mean little for the presence or absence of ETIs in another.\footnote{Neglecting the possibility of intergalactic travel, which could lead to cosmological ``bubbles'' within which ETIs may synchronize \citep{Kardashev85,Armstrong13,Olson16}. Even then, different ``domains'' of the Universe could have varied technosignatures.} {SETI has set the first limits on technosignatures in other galaxies \citep[e.g.,][]{Horowitz93,Shostak96,Annis99,Griffith15,Gray17,Garrett23}, but these limits still lag far behind those in the Milky Way.}

Since ETIs could be biased toward specific galaxies where interstellar travel has flourished, this suggests a treatment of galactic ETI populations and their technosignatures is in order. We can use probability theory to calculate the expected emission from all the ETIs covered by an observation. Developing this population approach is the focus of this paper and its sequels. 

\subsection{The problem of notation}
\begin{figure*}
\centerline{\includegraphics[width=16cm]{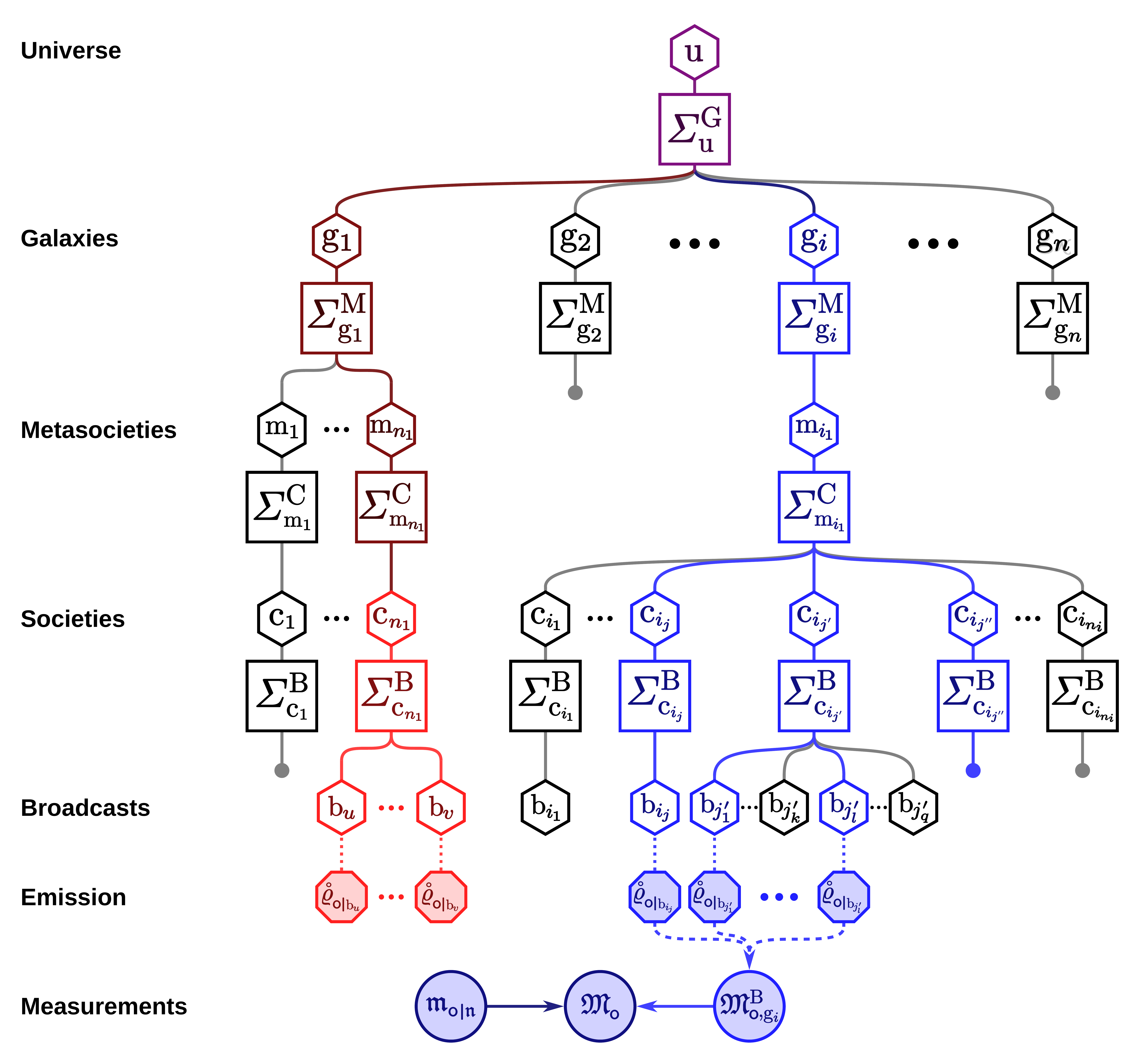}}
\figcaption{Overview sketch illustrating the tree structure of objects in this series. A selection draws samples of {galaxies}, metasocieties, societies, and broadcasts. The galaxy $\GalONEIndex$ is {in the} ``classic'' SETI {scenario} with isolated worlds, each developing its own metasociety, while $\GalIIndex$ has a single galaxy-spanning metasociety, and $\GalTWOIndex$ and {$\GalIndex_n$} are uninhabited (additional objects are represented by the ellipses). Here, an observation (${\ObsLabel}$) of a {galaxy} ($\GalIIndex$) {defines} a more restrictive selection $\Selection{\ObsLabel}{\GalIIndex}$ that picks a subsample of objects, highlighted in bright blue. The windowed emission $\mMeasureObsGalI$ from the broadcast sample $\bSampleObsGalI$ is mixed with background ($\kMeasureObs$) into an observed quantity, $\qMeasureObs$. {The selection also picks the ancestors of $\GalIIndex$ (dark blue).} Some societies and broadcasts in $\GalIIndex$ are excluded because they lie outside the ${\ObsLabel}$ window (in a different field or at a different time, for example). The objects and emission sampled by another selection, $\Selection{\AllLabel}{{\SocIndex_{n_1}}}$, {are} highlighted in red {(dark red for ancestors of {${\SocIndex_{n_1}}$})}. \label{fig:Tree}}
\end{figure*}

{The tables} in this paper list the important variables. There are many {{kinds} of objects with their own populations} and derived constraints (Figure~\ref{fig:Tree}). Each of these {levels} has its own associated parameters, which leads to a great proliferation of variables. Additionally, many of them already have {often conflicting} symbols in the literature. This can easily lead to a highly overloaded namespace: for example, many different types of quantity may lay claim to $N$ -- number of stars, number of societies, number of broadcasts, number of polarizations emitted, number of pointings, number of photons counted, and so on. Subscripts and superscripts help, but risk confusion when two completely different types of variables share the same symbol (e.g., $L$ for luminosity and longevity in the Drake equation), and lead to clutter when combined with the selection notation developed later. I use different fonts to {help} distinguish between different types of variables, as listed in Table~\ref{table:FontSummary}. {The chosen notation does not always match with their usage in other fields.}

\begin{deluxetable*}{cccp{10cm}c}
\tabletypesize{\footnotesize}
\tablecolumns{5}
\tablewidth{0pt}
\tablecaption{Summary of fonts used\label{table:FontSummary}}
\tablehead{\colhead{Font} & \colhead{Case} & \colhead{Examples} & \colhead{Use} & \colhead{Frame}}
\startdata
{Blackboard bold} & Upper   & $\fsReal$, $\MedianCore$, $\VarCore$
  & \vphantom{C}{Certain sets and operations}
	& \nodata\\
Hebrew                  & \nodata & {$\AncestorTypesOf{\KMark}$, $\TypeOf{\BcMark}$} & \vphantom{O}{Object type operations} & \nodata\\
Roman   	    & Any     & $\fWLambert$, $\fchi$, $\fDirac$
	& Mathematical functions and probability distributions
	& \nodata\\ 
	            & Lower   & $\JIndex$, $\MetaIndex$, $\BcIndex$
							          & \vphantom{I}{Indices for realized objects}
	& \nodata\\
         	    & Upper   & $\JMark$, $\GalMark$, $\MWIndex$ 
												& \vphantom{L}{Labels for types of objects, random objects, and certain named objects}
	& \nodata\\
San-serif               & Lower   & {$\ModeLabel$, $\ObsLabel$, $\GenLabel$, $\TimeLabel$}
	& \vphantom{L}{Labels for windows}
	& \nodata\\
	                      & Upper   & {$\oBandwidthObs$, $\oPolSetGen$, $\oGenOfAlt$}
	& \vphantom{V}{Variables defining a selection window}
	& Source\\
Italics       & Any     & $X$, $\TimeVar$, $f(x)$, $k_B$, $\ucHubbleZero$         
	& Generic variables and functions; {p}hysical and cosmological constants
	& \nodata\\
              & Lower   & $\jjSingleGen$, {$\uTupleUniv$, $\zTStartMeta$,} $\bEisoBc$
	& Quantities {describing} {intrinsic properties of individual objects}
	& Source\\
					    & Upper   & {$\jjDistK$}, {$\hMAggGal$, $\zSampleGen$, $\bjjAggLnuiso$}
	& {Q}uantities {describing} samples {of {objects}}
	& Source\\
              & \vphantom{L}{Lower}	& {$\yDistanceBc$, $\yRedshift$}
	& Distances{,} redshifts{, dilutions, and transmittances}
	& {\nodata}\\
\vphantom{F}{Fraktur} & Lower & $\lMeasureGenBc$, $\lFluenceEObs$
  & \vphantom{F}{Fluence, flux, or observables} for a single {object} or background
	& Observer\\
	            & Upper   & $\qMeasureGen$, $\kAmplitudeMode$
	& \vphantom{F}{Fluence, flux, or observables} from a sample, or in total
	& Observer\\		
\vphantom{C}{Calligraphic}            & Upper   & $\iAeff$, $\iNAntenna$
  & Quantities describing instrument
	& Observer
\enddata
\tablecomments{{Some characters, mainly the Greek letters, have no counterparts in the fonts used.} For these, the standard italic {or upright} character is used. A $^{\EarthFrameDecor}$ superscript attached to a source-frame variable changes it to an observer-frame variable.}
\end{deluxetable*}

\subsection{Outline of paper}
I begin the paper with a review of {point process theory} (Section~\ref{sec:ProbabilityBack}). {The construction of selections and distributions} is detailed in Section~\ref{sec:SelectionOverview}. {The nature of randomness in ETIs and broadcasts is discussed in Section~\ref{sec:PopAssumptions}.} A discussion of {the modeling of the universe, galaxies, and their stellar populations} (Section~\ref{sec:HighLevels}) is followed by the treatment of {the interplay of ETI metasocieties and societies (Section~\ref{sec:ETILevels})}. {Next is consideration of the properties of individual broadcasts} (Section~\ref{sec:BroadcastConsiderations}), {and} the results are assembled into calculations of broadcast populations and their total emission (Section~\ref{sec:BroadcastPopulations}). Sections~\ref{sec:BoxModelIntro} and~\ref{sec:ChordModelIntro} present the box and chord models of broadcasts for calculating observables. {A brief} discussion of noisy measurements comes in section~\ref{sec:Measurements}. Section~\ref{sec:WhichHostGalaxies} {demonstrates the formalism by considering how {a galaxy's} stellar mass {affects} the number of broadcasts in the face of sporadic interstellar travel.}  After the conclusion (section~\ref{sec:Conclusion}), appendices present additional details and derivations.

\section{Background on point processes}
\label{sec:ProbabilityBack}

{Point processes have a rich theoretical background. The concepts throughout this section are informed by \citet{Kingman93,Daley03,Daley08,Baddeley07,Chiu13,Haenggi13,Last17}. The reader may consult {them} for further details.} {Certain technicalities are given in footnotes for interested readers.}

{
\subsection{Random variables}
\label{sec:RVBasics}

Probability is defined over a sample space of all outcomes, each of which is a distinguishable configuration of all variables. Because outcomes are so particular, it is practical to group them into events defined by a shared quality. An event's probability $\fpP(\fpEvent)$ is summed (or integrated) over all outcomes in the event {$\fpEvent$} \citep{Wasserman04}. Outcomes can be summarized by a random variable, which assumes one value per outcome called a variate. Formally speaking, a random variable is defined as a function mapping each outcome onto a {set}, conventionally a real number \citep{Klenke20}. {Let's say we are interested in the positions of stars in the Milky Way. Every single possible configuration of stars is a distinct outcome. Those outcomes with two stars within one parsec of the Earth form an event; the number of stars within one parsec of the Earth is a random variable, which depends on what outcome is realized.}

Distributions describe the relative probability of different variates, among them the cumulative distribution function (CDF) $\CDF{X}(x) \equiv \fpP(X \le x)$ and probability density function (PDF) $\PDF{X}(x) \equiv d\CDF{X}(x)/dx$ for real-valued variables \citep{Wasserman04}\footnote{{The PDF is conventionally written as $f_X$, I use $\PDFCore$ to avoid confusion with other uses like the factors in the Drake equation.}}} (refer to Table~\ref{table:StatNotationSummary} for {the} statistical notation used in this series). 

{Random variables are very general, and can have completely arbitrary values. Among the simplest is the indicator variable: $\IndicatorOf{\fpEvent}$ is $1$ if event $\fpEvent$ happens and $0$ if it does not. Random variables can have a deterministic value regardless of outcome, like always being $0$, with a degenerate distribution. Thus, random variables \emph{per se} do not imply a lack of rhyme or reason. }

{Random variables are independent if we learn nothing new about the value of one if we know the value of the other. Often, however, random variables depend on an outside factor, without influencing each other. Conditional independence of $X$ and $Y$ means that, if we know the value of a third random variable $Z$, knowing what $X$ is tells us nothing more about the value of $Y$. It neither requires nor implies independence.\footnote{\vphantom{s}{Specifically, it means $\CDF{X, Y | Z}(x,y|z) = \CDF{X|Z}(x|z) \cdot \CDF{Y|Z}(y|z)$ \citep{Wasserman04,Klenke20}.}}} {Conditionalization is very important in this series because it can isolate the shared influences between random variables. The conditional mean $\Mean{X|Y}$ and variance $\Var{X|Y}$ come up frequently.} The law of total expectation states
\begin{equation}
\label{eqn:TotalExpectation}
\Mean{X} = \Mean{\Mean{X | Y}},
\end{equation}
{and} the law of total variance is
\begin{equation}
\label{eqn:TotalVariance}
\Var{X} = \Var{\Mean{X | Y}} + \Mean{\Var{X | Y}} 
\end{equation}
\citep{Brillinger69,Wasserman04,Bas19}. 

{
\subsection{Point processes}
\label{sec:PointProcess}
{An a}strophysical object may be described by a tuple $\fpPoint$ of quantitative parameters. Each possible tuple is a point in the set of all tuples, the state space $\fpSpace$. Even objects that are extended in an observable space are represented as points, their {internal structure} following from the point's location -- a core method in stochastic geometry. As the common metaphor goes, our observations trawl through that space to find the ``needles in the haystack'' \citep{Zwicky57,Harwit81,Tarter84,Djorgovski13,Wright18}.\footnote{In this series, these state spaces are those describing physical properties of objects, not the ``observable parameter spaces'' of \citealt{Harwit81} and \citealt{Djorgovski13} describing observational capabilities.}

{A population} is represented as a point process. Basically, a \emph{point process} is some random set $\fpRandomSet$ of points on a state space $\fpSpace$.\footnote{\vphantom{E}{Each bounded region has only a finite number of points (almost surely), which is no issue for realistic populations.} A point may be sampled multiple times, requiring some way to encode multiplicity. However, I assume the point processes are {``simple,''} each point included only once with probability $1$ {\citep{Baddeley07,Daley03}}.} {A random variable $\fpN(\fpSubset) = |\fpRandomSet \cap \fpSubset|$ counts the number of points appearing in each region $\fpSubset \subset \fpSpace$.}\footnote{\vphantom{F}{Formally, each region must be a Borel set \citep{Daley03}.} Any {rigorously defined} subset of $\fsRealN$ prone to be encountered in astronomy is Borel{:} {all continuous curves and surfaces, including fractals, whether unbounded or closed; their interiors and exteriors; Cantor dusts; all open sets; all closed sets; and countable unions of these, among others.}}

{A core property of each point process is the mean number of points\footnote{\vphantom{B}{Both $\fpN(\fpSubset)$ and $\Mean{\fpN(\fpSubset)}$ are measures. One consequence is that they are additive when taking the (countable) union of disjoint subsets.}} in each region $\fpSubset$. The means $\Mean{\fpN(\fpSubset)}$ are described by a} density distribution $\fpIntensity$ known as the intensity:
\begin{equation}
\Mean{\fpN(\fpSubset)} = \int_{\fpSubset} \fpIntensity(\fpPoint) d\fpPoint .
\end{equation}
{The intensity does not have to be a smooth function; it can include Dirac distributions to indicate exact locations where some points are found. Points can also fall along lines or surfaces in the space -- just like how the properties of galaxies and stars can {be idealized to} fall along lines or surfaces in graphs describing their properties, like the main sequence or the Tully-Fisher relation.}

Often, we wish to calculate the sum of random variables associated with each point $\fpPoint$ in the point process $\fpPointProcess$. Campbell's formula is a very general result that relates it to {the distribution}:
\begin{equation}
\label{eqn:CampbellFormula}
\Mean{\sum_{\fpPoint \in \fpRandomSet \cap \fpSubset} f(\fpPoint)} = \int_{\fpSubset} f(\fpPoint) \fpIntensity(\fpPoint) d\fpPoint .
\end{equation}

There are many ways to manipulate point processes. {The \emph{superposition} of several given point processes includes all the points from each of them in a new point process, with the $\fpIntensity(\fpPoint)$ summing together.\footnote{As long as the collection of superposed point processes is countable, there can even be infinitely many of them.}} \emph{Thinning} a point process {keeps some of the} points based on their locations in the space. {Independent thinnings keep or remove a point solely based on its own location and not on any other {point}'s.} {We can also use a point's position to \emph{map} it to another location in the space.}

Finally, we can \emph{mark} the points in a point process, attaching a random variable with additional information. If the marks are drawn from set $\fpMarkSpace$, a marked point process is equivalent to a point process on $\fpSpace \times \fpMarkSpace$. Furthermore, if the mark $\jSingle$ of a single point is drawn from the distribution $\PDF{\fpMarkPoint | \fpPoint}$, independent of the number of other points and their locations, then the intensity on this {new} space is $\fpIntensity{(\fpPoint)} \times \PDF{\fpMarkPoint|\fpPoint}$.  {That is} why the ``haystack'' concept works -- although objects are located in physical spacetime, their parameters serve as {``marks,''} which can then be interpreted as additional dimensions in an abstract space.\footnote{The reverse operation -- treating one or more of the dimensions as marks for a point process on a {lower-dimensional space} -- is only allowed if the resulting projected process does not have {infinite $\Mean{\fpN(\fpSubset)}$ for any bounded $\fpSubset$} {\citep{Baddeley07}}, but happens whenever we simplify distributions by marginalizing over a parameter.}
}

\subsection{Poisson point processes}
\label{sec:PoissonPointProcess}

{Poisson point processes are special point processes with many useful properties \citep{Kingman93}.\footnote{It is common to use ``Poisson process'' to refer to the specific case counting the number of hits along a {nonnegative} real line. A ``compound Poisson process'' {then} measures the accumulat{ed sum} of jumps occurring at each hit \citep[e.g.,][]{Ross96,Wasserman04,Embrechts13}.} They have seen use in modeling populations of ETIs \citep{Glade12,Kipping21}, {f}ast {r}adio {b}ursts \citep{Lawrence17}, and galaxies \citep{Neyman52,Martinez02}. In a Poisson point process, each of the $\fpN(\fpSubset)$ has a Poisson distribution:
\begin{equation}
\fpP(\fpN(\fpSubset) = n) = e^{-\Mean{\fpN(\fpSubset)}} \frac{\Mean{\fpN(\fpSubset)}^n}{n!} .
\end{equation}
Additionally, {the number of points in {nonoverlapping} regions are independent random variables. A Poisson point process is completely specified by its intensity.\footnote{\vphantom{A}{Additionally, a Poisson point process has $\Mean{\fpN(\{\fpPoint\})} = 0$ for any single point $\fpPoint$ (there are no ``atoms'' in the intensity{;} \citealt{Kingman93}). It is fine for lower-dimensional curves and surfaces to have $\Mean{\fpN(\fpSubset)} > 0$, however \citep{Daley03}.}} The Poissonian character of the point processes is unaffected by superposition, independent thinning, mapping, and displacing points by a random offset. The resulting intensities are what one naively expects -- added together for superposition, reduced by a position-dependent fraction in independent thinning, preserved in the image of a subset for mapping, and blurred for displacement -- subject to certain technical conditions \citep{Kingman93}. A marked Poisson process with points on $\fpSpace$ and marks from $\fpMarkSpace$ is equivalent to a new Poisson point process on $\fpSpace \times \fpMarkSpace$, as long as the marks are mutually independent. Other point processes converge to Poissonian when repeated randomizing operations are applied to them, most notably displacing their points independently and randomly \citep{Daley08}.}}

{These results formalize common sense notions in a very general way. Often in {astronomy,} we model a field of objects with a Poisson process, like counting stars in a patch of sky or transients observed by a radio telescope. These results show that we can add information, like the colors and magnitudes of stars, and still use the Poisson distribution. Poisson statistics also apply when we change the parameterization (e.g., frequency to wavelength), combine or filter out subclasses of objects (like {early-} and {late-type} stars), or have scatter (as from time delays or errors). There only has to be a determined intensity (Section~\ref{sec:OtherPointProcesses}).}

\subsection{Compound Poisson variables and point processes}
\label{sec:CompoundPoissonProcess}

{A compound Poisson random variable has the form}
\begin{equation}
\jAgg = \sum_{\JMark = 1}^{\fpN} {\jjSingle},
\end{equation}
where the ${\jjSingle}$ are {independent and identically distributed (i.i.d.), and also independent of $\fpN \sim \Poisson(\Mean{\fpN})$} (e.g., {\citealt{Adelson66,Barbour01}}; \citealt{Karlis05,GeringerSameth15}). If $\fpN = 0$, then $\jAgg = 0$. The distribution of $\jAgg$ {has a compound form as} the weighted sum {of the distribution for each fixed $\fpN$}:
\begin{equation}
\label{eqn:CompoundPoissonPDF}
\fpP(\jAgg \le x) = \sum_{n = 0}^{\infty} \fpP(\fpN = n | \Mean{\fpN}) \fpP({\jAgg \le x | \fpN = n}) .
\end{equation}
{Although this distribution} often has no closed form {expression, its Fourier transform, the} characteristic function, is
\begin{equation}
\label{eqn:CompoundPoissonCharacteristic}
\CF{\jAgg}(\tilde{x}) = {\Mean{e^{i \jAgg \tilde{x}}} =} \exp[\Mean{\fpN} (\CF{\jjSingle}(\tilde{x}) - 1)],
\end{equation}
where $\CF{\jjSingle}$ is the characteristic function of $\jjSingle$ \citep{Kemp67,Daley03}. An inverse Fourier transform then gives a numerical probability distribution \citep{GeringerSameth15}. Equation~\ref{eqn:CompoundPoissonCharacteristic} appears in $P(D)$ analyses, which implicitly use the compound Poisson distribution \citep{Scheuer57}. Both the mean and variance have simple {expressions}:
\begin{eqnarray}
\nonumber \Mean{\jAgg} & = \Mean{\fpN} \Mean{\jjSingle} \\
           \Var{\jAgg} & = \Mean{\fpN} \Mean{\jjSingle^2} 
\end{eqnarray}
\citep{Adelson66,Bas19}. 

{Whenever we have a Poisson point process $\fpRandomSet$, with each point marked by i.i.d. {nonnegative} real numbers, {the compound Poisson point process describes the sum of the marks for different subsets $\jAgg(\fpSubset)$} \citep{Daley03,Last17}. Each $\jAgg(\fpSubset)$ is a compound Poisson random variable. In fact, even} the ${\jjSingle}$ themselves can be compound Poisson random variables, representing hierarchical clusters of events, as long as they are i.i.d. and independent of $\fpN(\fpSubset)$ on each level.

{What if the marks are independent but their distribution varies with $\fpPoint$? Although the sum is no longer compound Poisson, Campbell's formula for the mean (equation~\ref{eqn:CampbellFormula}) still applies, and a second Campbell's theorem gives us
\begin{equation}
\label{eqn:CampbellVar}
\Var{\jAgg(\fpSubset)} = \int_{\fpSubset} \Mean{\fpMarkPoint(\fpPoint)^2} \fpIntensity(\fpPoint) d\fpPoint
\end{equation}
if $\Mean{\jAgg(\fpSubset)}$ exists {\citep[e.g.,][]{Kingman93}}. This equation is valid only for Poisson point processes, failing when the points are dependent or $\fpN(\fpSubset)$ is fixed.}

\begin{deluxetable}{cp{6.5cm}}
\tabletypesize{\footnotesize}
\tablecolumns{2}
\tablewidth{0pt}
\tablecaption{Summary of statistical notation used\label{table:StatNotationSummary}}
\tablehead{\colhead{Notation} & \colhead{Explanation}}
\startdata 
\cutinhead{Variables}
$X$                                    & Generic random variable\\
$\fpN$                                 & Generic {nonnegative} integer random variable\\
{$\IndicatorOf{\fpEvent}$}     & \vphantom{I}{Indicator random variable{;} $1$ if event $\fpEvent$ happens and $0$ otherwise}\\
$X | Y$                                & Random variable $X$, with outcomes limited by value of random variable $Y$\\
{$\jjSingleGen$}               & \vphantom{R}{Arbitrary singleton random variable describing object ${\JMark}$; quantity filtered by window ${\GenLabel}$}\\
{$\jjkAggGenAlt$}              & \vphantom{S}{Arbitrary aggregate random variable, summing $\jjSingleGen$ from $\JMark$-type objects selected by $\Selection{\AltLabel}{\KMark}$}\\
$|S|$                                  & Cardinality of set $S${, when $S$ is {nonnumeric}}\\
\cutinhead{Probability distributions and operations}
$X \sim \DISTRIBUTION(\alpha)$          & $X$ has $\DISTRIBUTION$ distribution with parameter $\alpha$\\
{$P (\fpEvent)$}                & Probability of some event {$\fpEvent$}\\
{$\DOperation{X}$}              & \vphantom{A}{Any operation on the distribution of the random variable $X$: {can} stand for event probability, cumulative distribution function (CDF), probability density function (PDF), characteristic function, mean, variance, {standard deviation}, and order statistics {including maxima and minima}}\\
{$\jkDOperationGen{X_{\JMark}}$} & \vphantom{E}{Ensemble selection-relative distribution operation on $X_{\JMark}$ {for a random} object ${\JMark}$ drawn by $\Selection{\GenLabel}{{\KMark}}$ (Section~\ref{sec:SelectionRel})}\\
{$\jkDOperationZAlt{\jjSingleGen}$} & \vphantom{M}{Multiwindow distribution operation on $\jjSingleGen$ {that lets} $\jkDOperationAlt{\jjSingleGenI}$ range over all the ${\GenILabel}$ windows in ${\ZLabel}$ (Section~\ref{sec:Multiwindow}).}\\
{${\jkPGen} (\fpEvent_{\JMark})$}  & \vphantom{E}{Ensemble probability of an event related to random object ${\JMark}$ selected by $\Selection{\GenLabel}{{\KMark}}$}\\
{$\CDF{X} (x)$}                 & \vphantom{C}{CDF of $X$, $P(X \le x)$}\\
{$\CDF{X | Y} (x | y)$}         & \vphantom{C}{Conditional CDF of $X$ conditionalized on $Y$, $P(X \le x | Y \le y)$}\\
{$\CDF{X, Y} (x, y)$}           & \vphantom{J}{Joint CDF of $X$ and $Y$, $P(X \le x\ \cap\ Y \le y)$. Can be extended to arbitrarily many variables.}\\
{$\PDF{X} (x)$}                 & PDF of $X$, {$d\CDF{X}(x)/dx$. Conditional and joint PDFs substitute $\PDFCore$ for $\CDFCore$ in the CDF notation.}\\
$\CF{X}(\tilde{x})$                     & Characteristic function of $X$\\
{$\jjSingleGenREGjkAlt$}        & \vphantom{R}{Regularization (trimming) of $\jjSingleGen$ to exclude values unlikely to occur in {$\jjkSampleAlt$}}\\
\cutinhead{Means and variances}
$\Mean{X}$                             & Mean (expectation value) of random variable $X${; simple mean for variables describing an object}\\
$\Var{X}$                              & Variance of random variable $X$ ($\Mean{X^2} - \Mean{X}^2$){; simple variance for variables describing an object}\\
\cutinhead{Order statistics}
$\Median{X}$                           & Median of random variable $X$\\
$\XOrderJ$                             & Order statistic: {$j$th} smallest in set of random variables $X_k$; distinct from unsorted $X_j$\\
$\XOrderMin$                           & Minimum of {the} set of random variables $X_k$\\
$\XOrderMaxNRV$                        & Maximum of {the} set of $N$ random variables $X_k$, where $N$ itself is a random variable
\enddata
\end{deluxetable}

{
\subsection{Other point processes}
\label{sec:OtherPointProcesses}
Actual counts may not be Poissonian. The binomial point process is what we get when the total number of points in a Poisson point process is fixed. The positions of the points are mutually independent, although the numbers of points in each $\fpSubset \in \fpSpace$ are dependent. 

While the Poisson point process is a mixture of binomial point processes with random $\fpN$, the Cox point process is a mixture of Poisson point processes where the intensity distribution $\fpIntensity$ itself is random. {O}nly one intensity field is ever realized for the point process. Looking at it, we would only ``see'' that realized intensity, and it would ``look like'' a Poisson point process, no matter how much we sampled it. Cox point processes are thus {\emph{nonergodic}}. Only if we had a family of identical Cox point processes could we see the true range of variability in them. Cox point processes have found use in modeling ecological populations when the configuration of offspring {is} hypothesized to cluster around parents \citep[e.g.,][]{Wiegand13}. The populations of societies and broadcasts in galaxies are regarded as Cox point process later in this paper as well.

Even more general are point processes where the random points interact. At the deterministic limit are lattices, with points ``repelling'' one another. Another example, the Gibbs point process, has a global energy function penalizing some configurations of points (e.g., forbidding those with points too close together), but practical results are difficult to come by. They remain outside of this work's scope.}

{In this work, objects of a single type usually appear randomly and do not interact (with some exceptions described in Section~\ref{sec:MetaScenarios}). Poisson point processes are natural when we have objects appearing independently according to a well-defined background rate. Several kinds of object populations behave this way (see discussion in Section~\ref{sec:PopAssumptions}), if their immediate host has specified properties, including broadcasts in a society. But sometimes we wish to consider the population hosted by a randomized collection of hosts. The intensity then is itself random, and the point process is Cox. The locations of stars in a specific galaxy are a Poisson point process \citep[e.g.,][]{Tonry88}, with the brightness profile serving as intensity. {However,} the star counts of ``a galaxy'' in an of itself is not defined, because the brightness profile itself varies with galactic mass, type, and so on -- it is more accurately described as a function that returns a Poisson point process given the galactic parameters. The star counts of a galaxy cluster are a Cox point process, since the gross properties of the cluster do not necessarily tell us about the characteristics of the member galaxies, only their \emph{distribution}. The star counts are a superposition of Poisson processes for all possible configurations of galaxies in the cluster. Broadcasts in galaxies are one example of a Cox point process in this paper because they are clustered into random societies.}

\section{Defining selections and samples}
\label{sec:SelectionOverview}

\begin{figure*}
\centerline{\includegraphics[width=14cm]{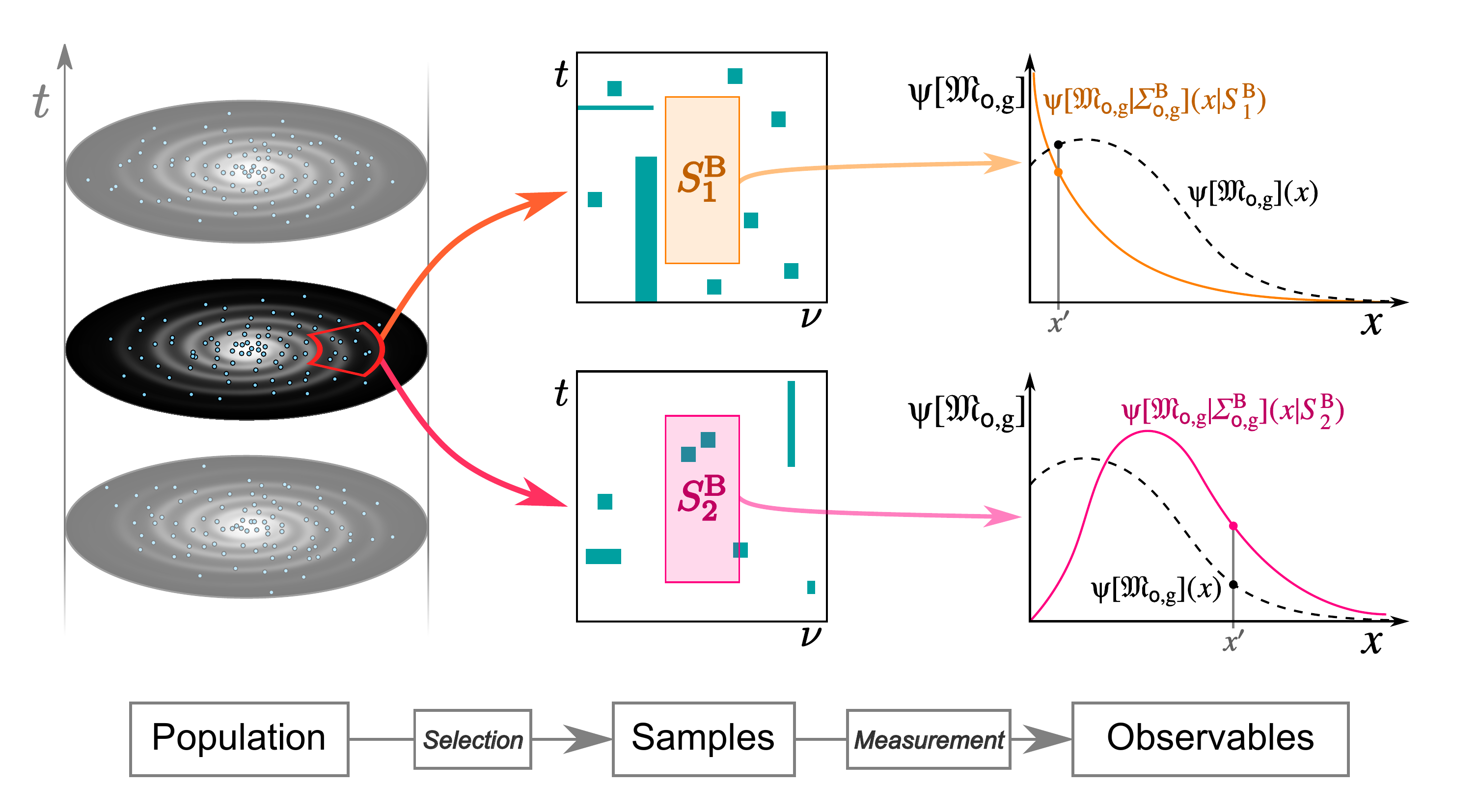}}
\figcaption{ETI broadcasts of a {galaxy} over its history form a population described by distributions (left). Surveys, and observations within the surveys, select subsets of the positions, times, and frequencies (and polarizations, not shown) spanned by this population, drawing samples treated as random {sets} (middle). In turn, {measurement} {yields} {observable} quantities, which are themselves random variables with probability distributions dependent on the sample drawn by the observation (right).  Both the sampling itself and the measurement conditionalized on the observed sample contribute variance to the final {quantity}. \label{fig:Sampling}}
\end{figure*}

We do not observe the properties of ETI broadcast populations directly{, only measure} limited samples of the entire population. {Measurements are treated as random variables that depend on, but are not determined by, the properties of the sampled broadcasts.} This introduces two forms of variance: the sampling variance due to {the} randomness {of} which objects are included, and noise variance introduced by microscopic fluctuations in observables for that sample (Figure~\ref{fig:Sampling}). {Describing how the sampling is modeled, and how it affects random variables, is the focus of this section.}

{
\subsection{Basic structure of populations}
\subsubsection{The tree and the haystacks}
}
A theory of the population of ETI technosignatures encompasses the habitat that hosts the ETIs, the ETIs themselves, and their technosignatures. {In this formalism, there {are two kinds of structures relating them.}

First is a {multilevel} \emph{tree}, assigning each object to a host. The ancestors of an object are its host, the hosts of the host, and so on, with the parent being the immediate node one level closer to the root. Likewise, each object can host a subpopulation of descendants, with the children being those that are one level deeper.}

{Five} ``levels'' {are} considered in this series{, each for a different type of object. The root is the \emph{universe} ({type} $\UnivMark$), containing all other objects.} {On the next level down are \emph{galaxies}} ({type} $\GalMark$). These contain \emph{metasocieties} ({type} $\MetaMark$), a collection of societies with a common origin or influence, which may be localized or extended. Within them, \emph{(communicative) societies} of ETIs ({type} $\SocMark$) may send out \emph{broadcasts} ({type} $\BcMark$). The objects in each level form a population of objects (Figure~\ref{fig:Tree}). 

{The levels are related by type operations. The ancestor-type $\AncestorTypesOf{\JMark}$ returns the set of object types that are ancestors of ${\JMark}$, with the parent-type operation $\ParentTypeOf{\JMark}$ returning the type one level up. The descendant-type $\DescendantTypesOf{\JMark}$ gives the set of object types that are lower on the tree, regardless of whether any such descendants exist, and the type-of operation $\TypeOf{{\JMark}}$ {(or $\TypeOf{\JIndex}$)} {which} returns an object's type. Thus, for the metasocietal type $\MetaMark$, $\AncestorTypesOf{\MetaMark} = \{\UnivMark, \GalMark\}$, $\ParentTypeOf{\MetaMark} = \GalMark$, $\TypeOf{\MetaMark} = \MetaMark$, and $\DescendantTypesOf{\MetaMark} = \{\SocMark, \BcMark\}$.}

{The other ingredient is a series of state spaces for objects, a collection of \emph{haystacks} \citep[see][]{Wright18}. Every type of object $\JMark$ has a haystack $\jjHaystack$ (Figure~\ref{fig:PointProcessHaystacks}). A single object {${\JMark}$} is described fully by a corresponding tuple {$\jjTuple$} in its haystack. Each tuple is a Cartesian product of parameters describing the object.} These can include basic quantities like the lifespan of the object, its location, energetics, and so on. 

{Included among the intrinsic properties of the object are parameters describing the} statistical properties of its internal {child} population. So, for every object ${\KIndex}$ of type $\KMark$ with $\JMark \in \DescendantTypesOf{\KMark}$, $\jkTupleK$ specifies the {\emph{distribution}} of the {descendant}s on the corresponding haystack $\jjHaystack$, but generally not which $\jjTuple$ are included in the actual subpopulation.\footnote{{This framework echoes others developed in the past literature across science, without directly following them. In the Neyman-Scott process, clusters of galaxies are represented by random points in real space, and associated with each is a collection of galaxies represented by points randomly distributed by some distribution. The clusters themselves can be clustered this way, to arbitrarily high order \citep{Neyman52,Neyman58,Martinez02,Haenggi13}. Also related are multiplicative population chains, modeling the growth of population between generations \citep{Moyal62-Mult,Moyal62-Gen,Daley03}. Unlike those {formalisms}, however, the underlying type of state space changes between ``generations'' because each level is a new type of object.}}

\subsubsection{Models}
\label{sec:ModelNature}
{A model describes the populations within some object of interest. It picks some {node ${\KMark}$} in the tree, whether the universe or a single {host} that we are studying, {to serve as the root of a subtree. The model} specifies enough parameters that the {descendant} populations of that node are statistically characterized. {The model can be identified with the tuple $\jkTuple$ describing this subtree root object ${\KMark}$. Thus,} the model is essentially a generalized random variable. The actual population of objects that exists is a realization of the model.}

{Only one set of model parameters actually describes the model's root. We may know some of these parameters -- the stellar mass of a model galaxy, or the cosmological parameters of the universe -- but others, particularly those describing ETI subpopulations, are unknown. However, the random nature of the model root is different than the randomness of the {descendant} populations. The distributions of objects on lower levels have a frequentist character. The subpopulation of {$\JMark$ objects} within a host ${\KMark}$ is a single trial; if we had an infinite number of trials, with an ensemble of hosts all with the same $\jkTuple$, the mean density of {$\JMark$ objects} in the haystack really would converge to $\jjkDist$. We use $\jkTuple$ to predict the properties of the descendant subpopulation.}

{In contrast, there is only one root object, as is clear when it is the universe as a whole, and cannot repeat ``trials.'' Yet we are uncertain about the true values of $\jkTuple$, so we have to consider many possible models and compare the kinds of populations that result with reality. To the extent there is any ``distribution'' for the root object, the probabilities involved are Bayesian, describing our level of belief in each hypothetical model; the goal is inference, not prediction. The results of SETI surveys are phrased as} statement{s} about likelihood, the consistency of each model with the data. The classic rate-versus-luminosity plots of SETI \citep[e.g.,][]{Enriquez17} {are an example, describing the consistency with an empirical null result.}

\subsubsection{Random and realized objects}
\label{sec:RandomVsRealizedObjects}
Random variables describe {the observables of} objects, but sometimes we need a general result -- the object itself is a ``random variable'' of sorts. {These are the random objects, in opposition to realized objects. With a random object, we mean ``for anything of this {type,}'' but for a realized object, we mean ``\emph{this} (possibly hypothetical) {object.}'' Samples draw random objects from the haystack, where the object's tuple itself is the random variable.}

Each dimension of a haystack is a {coordinate} variable; realized objects have specific values for {each coordinate, while random objects leave them as free parameters.\footnote{\vphantom{T}{The interpretation of the basic parameters as coordinates and not random variables is implicit in their direct use in the distributions (e.g., $d\Mean{\jjkNGen}/d\jjDuration$). The same is true of deterministic combinations of the basic parameters (like broadcast isotropic luminosity from total energy release and lifespan), which can be regarded as using a different coordinate system. {However,} these intrinsic quantities also could be interpreted as random variables with degenerate distributions, as in section~\ref{sec:Singleton}. The notation for these basic parameters is the same as for the random quantities that mark each notation, however, reflecting this ambiguity. If we wished to be absolutely rigorous, we would make a distinction between the basic parameter as coordinate variable and a random variable {``wrapper,''} but I ignore the distinction to avoid further clutter.}} Every quantity associated with a realized object that is not a deterministic combination of these coordinate parameters is a random variable, with a distribution that is fixed by their values. With random objects, these quantities} are not formally random variables, {as} {we cannot specify a single distribution without the object's properties being known}. Instead, {they actually are} \emph{functions} that yield random variable{s} when the tuple describing a realized object is fed into them as input. The {number of photons intercepted from} random broadcast ${\BcMark}$ {during window ${\GenLabel}$}, ${\lPhotonObs}$, is an example of one of these random functions, and should in all strictness be written as ${\lPhotonObs (\bTuple)}$. When $\bTuple$ is assigned a single value $\bTupleBc$, then ${\lPhotonObsBc \equiv \lPhotonObs(\bTupleBc)}$ is a \emph{bona fide} random variable with a single distribution. {In practice, the random functions may be written as if they are variables to avoid cluttering notation, but the dependence is still implicitly there.}

{The notation makes this distinction in the way it indexes random variables. An uppercase index designating a type of object also refers to a random object of that type. {When} the index is a lowercase letter, it denotes a realized object. Thus, for a random variable $X$ and a {$\JMark$}-type object ${\JIndex}$ described by a parameter tuple $\jjTupleJ$,
\begin{equation}
X_{\JIndex} = X_{\JMark} (\jjTupleJ) .
\end{equation}
This distinction is very general, applying even to distributions, point processes, and {the selection-relative operations} -- so $\jjkDist (\jjTuple | \jkTuple)$ is the distribution of $\JMark$-type objects in a random $\KMark$-type host, while $\jjDistK (\jjTuple)$ is the distribution in a realized host ${\KIndex}$.}

{Most of the results that follow apply to both random and realized objects, and can be interconverted between the two by substituting notation and using the appropriate dependence on the random object's tuple}{:
\begin{align}
\nonumber \jjkDist (\jjTuple | \jkTuple) & \rightarrow \jjDistK (\jjTuple) = \jjkDist (\jjTuple | \jkTuple = \jkTupleK)\\
\nonumber \jkPDFAlt{\jjSingleGen} (\jSingleCore) & \rightarrow \begin{aligned}[t]
								  & \PDFAltK{\jjSingleGen} (\jSingleCore) \\
								=~& \jkPDFAlt{\jjSingleGen} (\jSingleCore | \jjTuple; \jkTuple = \jkTupleK)\end{aligned}\\
\nonumber \jjSingleGen (\jjTuple)        & \rightarrow \jjSingleGenJ = \jjSingleGen(\jjTuple = \jjTupleJ)\\
          \jjkAggGen (\jkTuple)          & \rightarrow \jjAggGenK = \jjkAggGen(\jkTuple = \jkTupleK) .
\end{align}}

\subsubsection{Populations as point processes}
{The population of each kind $\JMark$ of objects is described by a point process on its haystack $\jjHaystack$ (Figure~\ref{fig:PointProcessHaystacks}). Furthermore, a point process exists for every possible host. The point processes are equivalently described by a random set on the haystack space, $\jjSample$, or by {its resultant collection of} $\jjN{(\fpSubset)}$ (section~\ref{sec:PointProcess}). Only one of the possible sets {is actually} drawn, the realized sample or sample variate $\jjSampleReal$. 

\begin{figure*}
\centerline{\includegraphics[width=14cm]{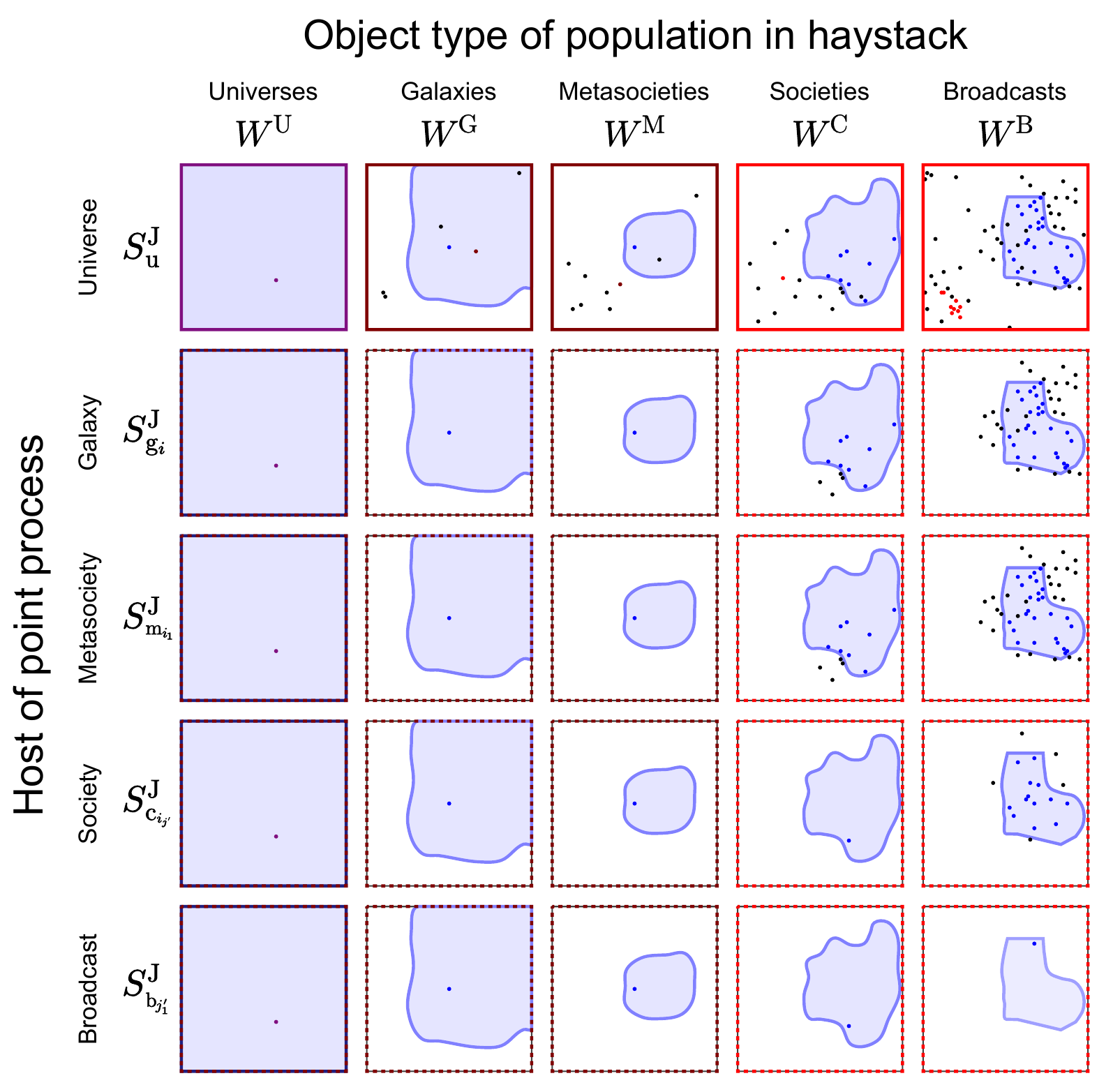}}
\figcaption{{Example i}llustration of the various haystacks for (left to right) the universe, galaxies, metasocieties, societies, and broadcasts, depicted here as having only two dimensions. Each host, chosen along one branch of Figure~\ref{fig:Tree}, has its own point process for each type of haystack. The realized samples illustrated here represent one possible population, each object a single point. Those objects that are selected by {$\Selection{\GenLabel}{\GalIIndex}$} (blue region) or {$\Selection{\AllLabel}{\SocIndex_{n_1}}$} (red outline) are colored blue and red, respectively. In some cases, an object falling in the ${\GenLabel}$ window is missed because its ancestors on the tree (its hosts) are not selected {(black dots in blue regions)}. \label{fig:PointProcessHaystacks}}
\end{figure*}

Let ${{\KMark}}$ be the parent of $\JMark$-type children. These children form a random set ${\jjkSample(\jkTuple)}$ on $\jjHaystack$, the $\JMark$-haystack.\footnote{\vphantom{O}{Or, to be pedantic about it, $\jjkSample(\jkTuple)$ is a random function that yields a random set (point process) based on $\jkTuple$; for the realized host ${\KIndex}$, $\jjSampleK \equiv \jjkSample(\jkTuple = \jkTupleK)$ is a random set (section~\ref{sec:RandomVsRealizedObjects}).}} So the broadcasts produced by a {realized} society ${\SocIndex}$ are characterized by a point process $\bSampleSoc$ on the broadcast haystack $\bHaystack${, for instance}. Of course, the children of an object are constrained by the properties of the parent. {Descendant} objects cannot be born before their ancestors, for example. As a result, the region of the haystack where children objects may be found is restricted.

Each point process has an intensity distribution}{
\begin{equation}
\label{eqn:JKIntensity}
{\jjkDist}(\jTupleCore | {\jkTuple}) = \frac{d\Mean{{\jjkN (\jkTuple)}}}{d\jjTuple} (\jTupleCore),
\end{equation}
where ${\jjkN}$ is the number of $\JMark$-type objects descended from {the random host ${\KMark}$}.\footnote{\vphantom{T}{The haystack is multidimensional, and the tuples are more or less vectors with many components. Even some of the parameters themselves are multidimensional (e.g., spatial position). The number of dimensions depends on the type of object and model, however; for an $n$-dimensional variable $\boldsymbol{x}$, $d/d\boldsymbol{x}$ should be understood to be the {$n$th} order derivative, {and $d\boldsymbol{x}$ as a hypervolume element.}}} This equation applies generally, but it serves as a definition for the distribution of child objects in a parent. Because the characteristics of the parent {determines} the statistical properties of the child subpopulation, the value of the intensity function at each point $\jjTuple$ in the $\JMark$-haystack is specified by the model and ${\jkTuple}$.} Note that the{se} mean numbers refer to the population of objects over the entire history of the parent, \emph{not} the typical number of objects that are active at any one time (contrast the $N$ in the Drake equation). Finally, the distribution is an average over an ensemble {of} all possible realizations of the subpopulations in ${{\KMark}}$.

{Of course, we are not just interested in the subpopulation in the parent, but also {in} the higher-order ancestors. We often wish to know the total number of broadcasts in an entire galaxy, not just one society, for instance. These populations are just the superposition for the point processes of {descendant} hosts -- for an $\LMark$-type {random} host with $\LMark \in \AncestorTypesOf{\KMark}$ and $\KMark \in \AncestorTypesOf{\JMark}$, 
\begin{align}
\nonumber {\jjlSample(\jlTuple)}        & = \bigcup_{{\jkTuple \in \jklSample(\jlTuple)}} \jjkSample(\jkTuple)\\
{\jjlN}(\fpSubset {| \jlTuple}) & = \sum_{{i} = 1}^{{\jklN}(\fpSubset {| \jlTuple})} {\jjkNKI(\fpSubset | \jkkTupleI})
\end{align}
for each {region} $\fpSubset \subset \jjHaystack$. The number of $\KMark$-type subancestors is itself a random variable, and ${\jklSample(\jlTuple)}$ is {random too.} {Now, the mean value of $\jjlN(\fpSubset|\jlTuple)$ is closely related to $\jjlDist(\jjTuple | \jlTuple) = d\Mean{\jjlN(\jjHaystack | \jlTuple)}/d\jjTuple$ through a derivative. The distribution is thus also a weighted sum over the possible $\KMark$ hosts.} The intensity for {the total $\JMark$-object} population is
\begin{equation}
\label{eqn:AncestorRelDist}
{\jjlDist(\jjTuple | \jlTuple)} = \int_{\jkHaystack} \jjkDist(\jjTuple | \jkTuple) {\jklDist(\jkTuple | \jlTuple)} d\jkTuple .
\end{equation}
This follows essentially from applying Campbell's formula (equation~\ref{eqn:CampbellFormula}), because $\jjkDist(\jjTuple | \jkTuple)$ is a random variable at each $\jkTuple$. {This way, we} can build up the distribution functions {up} from the parent all the way to the universe, given a model \citep[see][]{Moyal62-Mult}. The distribution of objects in their parent-type hosts is fundamental. Given these, we can derive all the other distributions for objects in ancestors, informing us about the populations we expect to be contained within each host.

What if the ``host'' object type is not ancestral, though?  The $\JMark$-type population in a $\JMark$-type object is the self-population, namely, itself: ${\jjSampleReal(\jjTuple) = \jjSample(\jjTuple) = \{\jjTuple\}}$. It is a Dirac point process, picking the object with certainty:
\begin{equation}
\jjjDist (\jTupleCore | \jjTuple) = \fDirac(\jTupleCore - \jjTuple) .
\end{equation}

What about the point process of ancestors ``contained'' within an object? It helps to distinguish between what ancestors an object \emph{could have} had, and what ancestors an object \emph{actually} has. Only the latter, the realized sample, is used, and it only contains one point for each level. When ${\KIndex}$ is a $\KMark$-type ancestor of ${\JIndex}$, $\jkSampleRealJ = \jkSampleRealK = \{\jkTupleK\}$.
}

\subsection{Selections}
Any realistic observation program cannot observe the entire realized population of objects throughout the universe, much less all possible objects {in the haystack}. An observation, survey, or other selection instead draws a sample of them based on position, time, frequency, and other constraints, as well as which ancestors they have (Figure~\ref{fig:Sampling}). A \emph{selection} makes these cuts. Each selection $\Selection{\GenLabel}{{\JMark}}$ is the combination of a \emph{window} ${\GenLabel}$ and a \emph{host} object ${{\JMark}}$. Notation for windows is listed in Table~\ref{table:SelectionNotation}.

\begin{deluxetable}{cp{7cm}}
\tabletypesize{\footnotesize}
\tablecolumns{2}
\tablewidth{0pt}
\tablecaption{Notation for selection windows \label{table:SelectionNotation}}
\tablehead{\colhead{Notation} & \colhead{Explanation}}
\startdata
$\Selection{\GenLabel}{{\JMark}}$         & \vphantom{T}{The selection picking objects and emission using window $\GenLabel$ from a population hosted by object ${{\JMark}}$}\\
\cutinhead{Useful windows}
${\GenLabel}$, ${\AltLabel}$, ${\ZLabel}$ & Denote generic windows; can be substituted with any other unless otherwise indicated.\\
${\ModeLabel}$                           & Picks objects and emission covered by in one electromagnetic field mode {(see Paper II)}\\
${\ObsLabel}$                            & Picks objects and emission covered by an observation\\
${\BeamLabel}$                           & Picks objects and emission covered by a single beam or resolution element with {a} fixed sky position in a series of observations\\
${\TuneLabel}$                           & Picks objects and emission covered by a {``tuning,''} a series of observations covering a fixed frequency range (e.g., in a single filter in optical)\\
${\PointLabel}$                          & Picks objects and emission covered by a pointing, a series of observations over a fixed sky field\\
${\SurvLabel}$                           & Picks objects and emission covered by a survey\\
${\FundLabel}$                           & \vphantom{F}{{Function} returning a window that picks} all objects and emission coincident with a fixed value of an arbitrary basic variable $\FundVar$\\
${\TimeLabel}$                           & Picks all objects and emission active at one instant as viewed from Earth (i.e., along our past light cone){; a function of $\TimeVar$} \\
${\HistLabel}$                           & Picks all objects and emission in past history, within our past light cone{; a function of $\TimeVar$}\\
${\FreqLabel}$                           & Picks all objects and emission at one source-frame frequency, regardless of time{; a function of $\FreqVar$}\\
${\TimeFreqLabel}$                       & Picks all objects active at one instant as viewed from Earth at one source-frame frequency{; a function of $\TimeVar$ and $\FreqVar$}\\
${\AllLabel}$                            & \vphantom{A}{ALL window: p}icks all objects and emission, regardless of parameters\\
\cutinhead{Quantities describing window definition}
$\oDurationGen$                          & Duration of window ${\GenLabel}$\\
$\oBandwidthGen$                         & Bandwidth covered by ${\GenLabel}$\\
$\oTStartGen$                            & Central time of ${\GenLabel}$\\
$\oNuMidGen$                             & Central frequency of ${\GenLabel}$\\
$\oDriftRateGen$                         & Drift rate of ${\GenLabel}$ window; used for describing dedrifted observations\\
$\oPolSetGen$					                   & Set of polarizations covered by ${\GenLabel}$\\
$\oNumPolGen$                            & Number of independent polarizations covered by ${\GenLabel}$\\
$\oSkyFieldGen$                          & Sky field covered by ${\GenLabel}$\\
$\oVolumeGen$                            & Spatial volume covered by ${\GenLabel}$\\
\cutinhead{Window manipulation}
$\oGenOfAlt$                             & \vphantom{S}{Set} of windows of type ${\GenLabel}$ that are fused into ${\AltLabel}$\\
$\oNGenAlt$                              & $|\oGenOfAlt|$, number of component windows of type ${\GenLabel}$ making up ${\AltLabel}$\\
{$\FuseSelection{\GenLabel}{\AltLabel}$}   & \vphantom{F}{Fused selection, inclusive of both ${\GenLabel}$ and ${\AltLabel}$}\\
$\JointSelection{\GenLabel}{\AltLabel}$  & Joint selection, intersection of ${\GenLabel}$ and ${\AltLabel}$\\
$\ProjSelection{\GenLabel}{\FundVar}$, $\ProjSelection{\JIndex}{\FundVar}$ & Projection of ${\GenLabel}$ or ${\JIndex}$ into basic variable $\FundVar$; picks all objects and emission that have a $\FundVar$ value in the range covered by ${\GenLabel}$ or ${\JIndex}$
\enddata
\end{deluxetable}

\subsubsection{Windows}
\label{sec:Windows}
{A window ${\GenLabel}$} selects things {that pass} cuts on {variables} like location{, lifespan, or frequencies covered. They include} the reach of a survey {or} observation, but the {cuts} can also be more abstract, picking all broadcasts that are active at {an arbitrary} frequency{, for example}. A window is {thus a kind of selection defined on the haystack of each object type.}

{Each window acts to thin a point process according to position in the haystack}. {A} window {${\GenLabel}$} {has two functions. First, it includes a filter on observable quantities, generally a bounded region over some basic variables so that any object that ``touches'' it is selected (see Section~\ref{sec:Singleton}). Second, a window restricts object selections: formally it includes a collection of probability} functions $\ojPGen (\jjTuple)$ for each type of object $\JMark$, giving the probability that the window selects an object with tuple $\jjTuple$. Basically, $\ojPGen$ gives the completeness of the window selection for an object with $\jjTuple$. All selected objects fall within a subset of the haystack,\footnote{Windows do not have to be contiguous, but that is assumed in the box and chord models.}
\begin{equation}
\jjHaystackGen = \{\jjTuple \in \jjHaystack : \ojPGen(\jjTuple) > 0\} . 
\end{equation}
Whether an object ${\JMark}$ is selected by the window depends solely on {$\ojPGen(\jjTuple)$}. It does not {otherwise} matter whether an object is detectable or distinguishable from a background. For example, the leakage from a cell phone call in a distant galaxy would be selected by a survey window if the survey looks at that galaxy at the right frequency when it happens, unless {the} window {also imposes} a luminosity {or fluence} cut.

\begin{figure}
\centerline{\includegraphics[width=8.5cm]{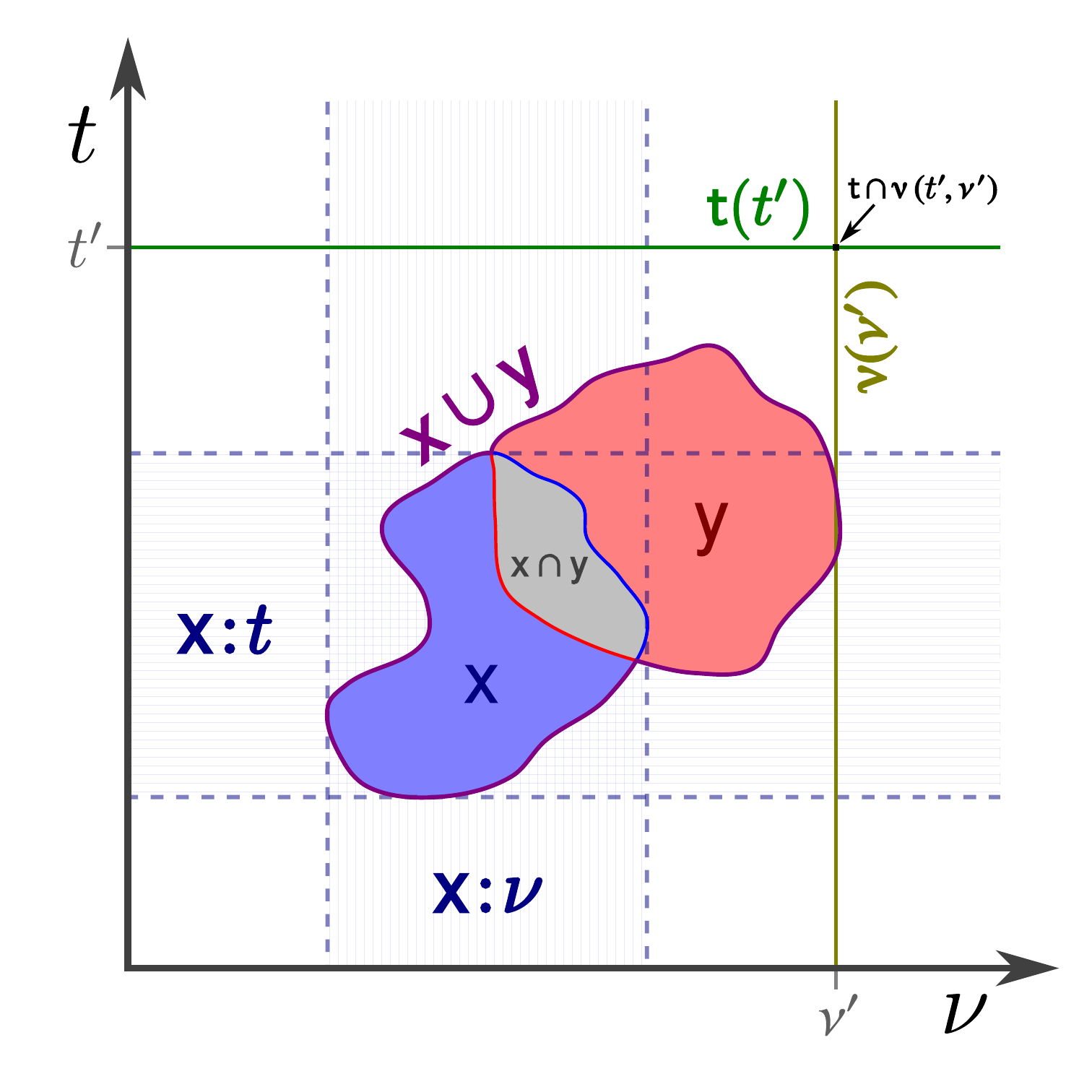}}
\figcaption{{{Illustration} of how different selection windows could manifest in a spectrogram. {window} ${\GenLabel}$ (shaded blue) can be projected onto time ($\TimeOFGenLabel$) and frequency ($\FreqOFGenLabel$). A second window ${\AltLabel}$ (shaded red) can be fused with ${\GenLabel}$ (violet outline), or a joint selection can be made ({gray} shading). Also shown are windows picking a specific time (green line), frequency (gold line), or both (black dot).} \label{fig:WindowOperations}}
\end{figure}

{A few fundamental variables -- position, time, and polarization -- define an arena in which things are situated (Table~\ref{table:BasicVariables}). {Windows and objects may cover} extended {regions} in this space (Figure~\ref{fig:WindowOperations}), {and} their arrangement {is} a problem of stochastic geometry. A basic kind of window is one that picks objects along a fixed value of one of these variables. For each variable $\FundVar$, a function} $\FundLabel {(\FundVar)}$ {returns} a window that intersects with a {given} fixed value {of it. The most commonly used one is the time window function, ${\TimeLabel} {(\TimeVar)}$,} select{ing} every object that exists during a particular moment {$\TimeVar$} in time as viewed from Earth.

\begin{deluxetable}{cl}
\tabletypesize{\footnotesize}
\tablecolumns{2}
\tablewidth{0pt}
\tablecaption{Notation for the basic general variables \label{table:BasicVariables}}
\tablehead{\colhead{Notation} & \colhead{Explanation}}
\startdata
$\TimeVar$                               & Time\\
$\FreqVar$     						               & Frequency\\
$\PolVar$ 				                       & Polarization\\
$\DistanceVar$                           & \vphantom{D}{Distance}\\
{$\PosVar$}											 & \vphantom{P}{Position in space}\\
$\SkyAngVar$                             & \vphantom{L}{Location on receiver sky}\\
$\FundVar$															 & Arbit{r}ary basic variable\\
\enddata
\end{deluxetable}

{In this work}, an observation ${\ObsLabel}$ is the lowest level {window} for which independent data are analyzed: each distinct combination of angular resolution element, channel, and time yields a separate measurement and can count as an observation. Other useful windows {are the survey ${\SurvLabel}$, a complete collection of observations; and ${\HistLabel}{(\TimeVar)}$, picking} all objects in past history to the present as viewed from Earth. Still more are listed in Table~\ref{table:SelectionNotation}.\footnote{Although not used in this work, windows can be random. These would be a point process on their own {``haystack.''} {In fact, \citet{Wright18} basically describe a survey window haystack \citep[see also][]{Djorgovski13}.}}

A collection $\oGenOfAlt$ of windows can be fused to form another, 
\begin{equation}
\jjHaystackAlt = \bigcup_{\GenILabel \in \oGenOfAlt} \jjHaystackGenI,
\end{equation}
{and if the{ir selections} are independent,}
\begin{equation}
\ojPAlt(\jjTuple) = 1 - \prod_{\GenILabel \in \oGenOfAlt} (1 - \ojPGenI(\jjTuple)) .
\end{equation}
{T}he number of these {subwindows} {is} $\oNGenAlt$. For example, a survey (${\SurvLabel}$) is divided into $\oNObsSurv$ observation windows ({generically labeled ${\ObsILabel}$}) from the set $\oObsOfSurv$. {The window resulting from fusing two subwindows ${\GenLabel}$ and ${\AltLabel}$ in this way can be written $\FuseSelection{\GenLabel}{\AltLabel}$.} A joint window $\JointSelection{\GenLabel}{\AltLabel}$ {instead} picks an event only if it is selected by both {subwindows}.

The last operation on windows is projection, working on a window or {host} and a fundamental variable $\FundVar$ {like time or frequency}. The projection $\ProjSelection{\GenLabel}{\FundVar}$ (or $\ProjSelection{{\KMark}}{\FundVar}$) makes a selection according to the range of $\FundVar$ covered by ${\GenLabel}$ or {${\KMark}$} (Figure~\ref{fig:WindowOperations}). {The projection of a window is given by
\begin{equation}
\jjHaystackFundOFGen = \bigcup_{\FundVar: \jjHaystackGen \cap \jjHaystackFund(\FundVar) \ne \varnothing} \jjHaystackFund(\FundVar)
\end{equation}
and the projection of an object is
\begin{equation}
{\jjkHaystackFundOF(\jkTuple) = \bigcup_{\FundVar: \jkTuple \in \jkHaystackFund(\FundVar)}} \jjHaystackFund(\FundVar)
\end{equation}
with $\ojPFundOFGen (\jjTuple; \FundVar) = {\max[}\ojPGen({\jTupleCore}) \ojPFund(\jjTuple; \FundVar) {| \jTupleCore,} \jjTuple \in \jjHaystackFund(\FundVar){]}$ and ${\ojkPFundOF} (\jjTuple; \FundVar {, \jkTuple}) = \IndicatorOf{\jjTuple \in {\jjkHaystackFundOF(\jkTuple)}}$.} Projection is a versatile operation, letting us make selections according to a single shared quantity. For example, $\bEisoFreqOFGenBc$ is the effective isotropic energy release of broadcast ${\BcIndex}$ at \emph{any} time or polarization within the bandwidth of ${\GenLabel}$; $\aNTimeOFSocGal$ is the number of communicative societies {in galaxy ${\GalIndex}$} that are ever active during the lifespan of the specific society ${\SocIndex}$. {Projection is particularly useful when describing broadcasts that can run in and out of a window, like in the chord model later on (Appendix~\ref{sec:ChordModelDetails}).}

By default, this series uses the ALL window ($\AllLabel$) selecting every event in a population when no other window is specified or implied.

\subsubsection{Hosts}

{Specifying a host ${{\KMark}}$ in a selection restricts the population to its subpopulation, as defined by its point process: it draws $\JMark$-type objects from ${\jjkSample(\jkTuple)}$. The host-based selection of objects} is determined solely by {their position on the tree -- the object must be the direct ancestor or {descendant} of the host, or be the host itself.} Objects are indexed, and the indices serve to denote their associated {host selection} as well. Most variables in this series use a {host}, but when none is otherwise specified, the cosmic {host} ${\UnivMark}$ is assumed.

{
\subsubsection{Samples and selection-relative intensities}
\label{sec:Samples}
A full selection $\Selection{\GenLabel}{{\LMark}}$, as a combination of a window and a host, selects {$\JMark$ objects} from the ${\GenLabel}$-thinning of the point process ${\jjlSample(\jlTuple)}$ over the set $\jjHaystackGen$. The resulting point process, ${\jjlSampleGen(\jlTuple)}$, is the random sample drawn by $\Selection{\GenLabel}{{\LMark}}$. An object ${{\JMark}}$ is selected by $\Selection{\GenLabel}{{\LMark}}$ (${\jjTuple \in \jjlSampleGen(\jlTuple)}$) if and only if:}
\begin{itemize}
\item {It is within the random sample $\jjSampleGen$; that is, its parameter tuple ${\jjTuple}$ is within $\jjHaystackGen$ and selected by the ${\GenLabel}$-thinning.}
\item {It is a member of the subpopulation of its host (${\jjTuple \in \jjlSample(\jlTuple)}$), picked by the host selection ${{\LMark}}$.}
\item {For each ancestor ${{\KMark}}$ of ${{\JMark}}$ up to and including ${\LMark}$, ${\jkTuple \in \jklSampleGen(\jlTuple)}$; so that each of these ancestors is \emph{also} selected by the window.}
\end{itemize}
{This final condition is a somewhat tricky one, but an object can be missed because its ancestor was not sampled. Suppose we defined a window ${\GenLabel}$ that picked all {galaxies} and metasocieties in a patch {of} the sky, but only societies that are older than a million years old, and all their broadcasts. A radio broadcast from a society that is only one thousand years old would not be selected by $\Selection{\GenLabel}{{\UnivMark}}$ even though ${\obPGen(\bTuple) = 1}$ for all ${\bTuple}$, because its host society is excluded.}

{If ${\LMark}$ is a} {host for objects of type $\KMark \in \AncestorTypesOf{\JMark}$, the intensity relative to a selection $\Selection{\GenLabel}{{\LMark}}$ is
\begin{equation}
{\jjlDistGen}(\jjTuple{|\jlTuple}) = \olPGen({\jlTuple}) \int_{\jkHaystack} {\jklDist}(\jkTuple {| \jlTuple}) \jjkDistGen(\jjTuple | \jkTuple) d\jkTuple,
\end{equation}
with
\begin{equation}
\jjjDistGen(\jTupleCore | \jjTuple) = \ojPGen(\jTupleCore) \jjjDist(\jTupleCore | \jjTuple) = \ojPGen(\jTupleCore) \fDirac(\jjTuple - \jTupleCore).
\end{equation}
Integrating gives us the expected number of $\JMark$-type objects sampled by the selection:
\begin{equation}
\label{eqn:MeanNGenericSelection}
{\Mean{\jjlNGen} \equiv} \Mean{{\jjlNGen (\jlTuple)}} = \int_{\jjHaystack} {\jjlDistGen} (\jjTuple {| \jlTuple}) d\jjTuple .
\end{equation}
} 

The process of selection often {introduces} biases {in object properties, so ${\jjlDistGen}$ is not usually proportional to ${\jjlDist}$}. Generally, objects that are ``larger'' are overrepresented in a sample {(see section~\ref{sec:LifespanBias})}.

{
\subsubsection{Simplifying distributions through marginalization, rates, and abundances}
\label{sec:Marginalization}

The full intensities often involve many parameters that are of no interest in a given problem. The distribution is simplified by marginalizing these irrelevant quantities, treating them as marks to be ignored (Section~\ref{sec:PointProcess}). If a parameter tuple $\jjTuple$ can be divided into components that we are interested in, $\jjParamsSimple$, and those that we are not, $\jjParamsOther$, with $\jjTuple = \jjParamsSimple \times \jjParamsOther$, then
\begin{equation}
\label{eqn:DistMarginalize}
{\jjkDistSimpleGen} (\jjParamsSimple {| \jkTuple}) = \int {\jjkDistGen} (\jjParamsSimple \times \jjParamsOther {| \jkTuple}) d\jjParamsOther.
\end{equation}
Of course, this equation applies to ${\jjkDist}$ itself, which uses the $\AllLabel$ window.}
 
{The most common of these marginalized intensities is the rate that objects form, a function over {time,} and possibly other fundamental variables like frequency. When $\jjTStart$ is the time an object is born, the rate is
\begin{equation}
{\jjkRateTotal} (\TimeVar, \FreqVar {| \jkTuple}) = \frac{d\Mean{{\jjkNFreq}(\FreqVar{, \jkTuple})}}{d\jjTStart} (\jjTStart = \TimeVar) ,
\end{equation}
where ${\jjkNFreq}(\FreqVar{, \jkTuple})$ is the number of objects hosted by {${\KMark}$} and selected by a window picking all objects that cover the frequency $\FreqVar$. Broadcasts have limited bandwidth, but other types of objects (including societies, metasocieties, and {galaxies}) are not confined by frequency and have ${\jjkNFreq}(\FreqVar{, \jkTuple}) = {\jjkN (\jkTuple)}$. If the number of selected objects is proportional to the number of stars (e.g., more stars means more inhabited worlds), a convenient variable is the stellar rate,
\begin{equation}
\label{eqn:jRate}
{\jjkRate} (\TimeVar, \FreqVar {| \jkTuple}) = {\jjkRateTotal (\TimeVar, \FreqVar | \jkTuple) / \Mean{\hkNTime(\TimeVar, \jkTuple)}},
\end{equation}
where ${\hkNTime}(\TimeVar{, \jkTuple})$ is the number of stars extant at time $\TimeVar$ within the host ${\KMark}$.

If ${\jjkDistGen}$ is marginalized over all parameters, then the resulting {zero-dimensional} distribution simply gives $\Mean{{\jjkNGen}}$ as per Campbell's formula (equation~\ref{eqn:CampbellFormula}). This itself can be regarded as a parameter, with $\Mean{{\jjkNGen}}$ defined for some well-characterized window ${\GenLabel}$ scaling any other distribution. For objects that trace stars, the most useful of these scaling parameters is the stellar abundance 
\begin{equation}
\label{eqn:jAbund}
{\jjkAbund} (\TimeVar, \FreqVar {| \jkTuple}) = \Mean{{\jjkNTimeFreq} (\TimeVar, \FreqVar{, \jkTuple})} / \Mean{{\hkNTime}(\TimeVar{, \jkTuple})} ,
\end{equation}
with $\TimeFreqLabel$ picking all objects active at both time $\TimeVar$ and, for broadcasts, frequency $\FreqVar$. It is the mean number of objects per star that ``cover'' a point on a spectrogram.}

\subsection{Random variables}
\label{sec:ObjectRVs}

\subsubsection{Singleton variables}

\label{sec:Singleton}
A \emph{singleton variable} $\jjSingle$ is a random variable that describes a single object ${\JMark}$.\footnote{\vphantom{T}{Technically, for a random object ${\JMark}$, $\jjSingle(\jjTuple)$ is a random function that outputs a random variable from the input $\jjTuple$ (Section~\ref{sec:RandomVsRealizedObjects}). The same caveat applies to aggregate ``variables'' with random hosts.}} {I assume that it is independent of all other objects, depending only on $\jjTuple$. Examples include position of an object, luminosity, flux received at Earth, and total number of photons collected in a telescope from that object in one pixel of a detector. The value of a singleton variable is not always determined by $\jjTuple$, only its distribution is. The number of photons counted even from a perfectly coherent laser beacon has detector and shot noise, for example.}

The basic notation for singleton variables is $\jjSingleGen$, where $\jSingleCore$ may be replaced with another symbol, usually a lowercase letter.  {In some cases, a random variable depends on how long we are observing, in what frequencies, at what polarizations, as is true for emission measured.} The \emph{quantity window} ${\GenLabel}$ sets the bounds of integration for $\jSingle$ {(Figure~\ref{fig:MeansComparison})}. {Generally,} an object with {no activity or} emission within this window is {not} selected by ${\GenLabel}$,
\begin{equation}
\label{eqn:SingletonNonOverlap}
\jjSingleGen{(\jjTuple)} = 0~{\Rightarrow}~\jjTuple \notin \jjHaystackGen {,}
\end{equation}
{although an object can be excluded for other reasons.} {When no quantity window is specified, the ALL window implicitly is used ($\jjSingle = \jjSingleAll$). The basic parameters for each object {can be interpreted as} special singleton variables that naturally use the ALL window; they are intrinsic properties, unaffected by observation.}

{A calculation with singleton variables over a population ${\jjkSampleAlt(\jkTuple)}$ is performed by marking each point with the value of $\jjSingleGen{(\jjTuple)}$. As noted in Section~\ref{sec:PointProcess}, this is equivalent to considering a point process in $\jjHaystack \times \fsReal$ for a real-valued random variable. As long as the marks are independent of any and all other points in the process, the intensity of this new process is
\begin{equation}
\frac{d^2 \Mean{{\jjkNAlt}}}{d\jjTuple d\jjSingleGen} (\fpPoint, \jSingleCore {| \jkTuple}) = {\jjkDistAlt}(\fpPoint {| \jkTuple}) {\cdot} \PDF{\jjSingleGen} (\jSingleCore | \fpPoint) .
\end{equation}
}

{
\subsubsection{Aggregate variables}
}
An \emph{aggregate variable} $\jAgg$ is the sum of the singleton variables for a sample of objects drawn from a {host} ${\KMark}$:
\begin{equation}
{\jjkAggGenAlt (\jkTuple)} \equiv {\jkSumJAltFULL} \jjSingleGen{(\jjTuple)} ,
\end{equation}
where $\jAgg$ may be replaced with another symbol, usually an uppercase letter. In addition to the quantity window ${\GenLabel}$, a potentially distinct \emph{object window} is combined with the {host} ${\KMark}$ to define {a selection that picks out which {of} the objects included in a sum. Each ${\jjkAggGenAlt}$ is treated using the marked point process for the corresponding $\jjSingleGen$.}

To simplify notation, the quantity window may be omitted to indicate that it is the same as the object window: {$\jjkAggGen \equiv \jjkAggGenGen$ (Table~\ref{table:Simplify})}. Applying equation~\ref{eqn:SingletonNonOverlap} {allows} us {to} ``expand'' or ``contract'' the object {window}:
\begin{equation}
\label{eqn:AggregateExpanded}
{{\jjkAggGen = \jjkAggGenAlt}~\text{if}~(\forall\jjTuple \in \jjHaystackGen)~\ojPGen(\jjTuple) = \ojPAlt(\jjTuple)} .
\end{equation}

{A}ggregate and singleton variables can be interconverted,
\begin{equation}
\label{eqn:SingleToAgg}
{\jkSingleAlt \leftrightarrow \jjkAggGenAlt},
\end{equation}
where the object window of a singleton variable is always identical to its quantity window. {These simply reflect different ways of viewing the quantity: the aggregate variable interprets it as a sum over a population, while the singleton variable interprets it as an intrinsic property of the host. A common example in extragalactic astronomy is the mass of a galaxy's stars -- whether it is viewed as the collective property of a stellar population or a property of a single galaxy, it is still the same quantity.} This lets us build high-level aggregates recursively, like the total emission from all broadcasts in all societies in a galaxy. 

Number variables {can be interpreted as a special case of} aggregate variables:
\begin{equation}
{\jjkNAlt (\jkTuple)} = {\jkSumJAltFULL} 1 = |{\jjkSampleAlt(\jkTuple)}| .
\end{equation}

\subsection{Selection-relative probabilities, means, and variances: The effects of selection bias}
\label{sec:SelectionRel}
{A distribution operation ${\DOperationCore}$ takes a variable and returns a function or number derived from its probability distribution -- it needs a random variable, and cannot stand on its own. These operations include the CDF, PDF, mean, and variance. In this section, I consider how these operations are affected by selection biases.}

\subsubsection{Selection-relative probabilities}
{Let us say we want to know the probability of an event $\fpEvent_{\JMark}$ describing some property of $\jjSingleGen$. Now, this probability is associated with a random object ${\JMark}$ and thus is actually a random function of $\jjTuple$. Our general assumption has been that the probability distribution of any random variable $\jjSingleGen$ is fully determined by $\jjTuple$, automatically implying independence between other such variables and the number of objects. This is the simple probability of the event, $\fpP(\jjEvent) \equiv \fpP(\jjEvent | \jjTuple)$.

In turn, the probability $\fpP(\fpEvent_{\JMark})$ is also a function of $\jjTuple$. {However}, any distribution for the $\JMark$ objects themselves allows us to generate a mixture distribution for $\jjSingleGen$, and then we can evaluate the probability of $\fpEvent_{\JMark}$ for this mixture.} {This includes distributions modified by selection windows.} When the value of a random variable $\jjSingleGen$ for a single object is independent of all other objects, {the selection-relative probability is thus defined}
\begin{equation}
\label{eqn:BiasedP}
{\jkPAlt}(\fpEvent_{\JMark}) \equiv \frac{1}{\Mean{{\jjkNAlt}}} \int_{\jjHaystackAlt} {\jjkDistAlt (\jjTuple | \jkTuple)} \fpP(\fpEvent_{\JMark} | \jjTuple) d\jjTuple .
\end{equation}
{This is the probability that $\fpEvent_{\JMark}$ occurs for a $\JMark$-type object fairly drawn from the random sample $\jjkSampleAlt(\jkTuple)$. It is an ensemble probability, averaging over all possible selections, not just for a realized sample.} {It} reflects the bias of a selection. 

{An archetypal event is $\jjSingleGen$ being within some specified range. The previous equation immediately yields a selection-relative CDF, with ${\jkCDFAlt{\jjSingleGen}(x) \equiv \jkPAlt(\jjSingleGen \le x)}$. Differentiating then gives the selection-relative PDF,}
\begin{multline}
\label{eqn:BiasedPDF}
{\jkPDFAlt{\jjSingleGen}} ({\jSingleCore}) \equiv  \\
\frac{1}{\Mean{{\jjkNAlt}}} \int_{\jjHaystackAlt} {\jjkDistAlt (\jjTuple | \jkTuple)} \PDF{\jjSingleGen} ({\jSingleCore} | \jjTuple) d\jjTuple ,
\end{multline}
{where $\PDF{\jjSingleGen}$ is the (unbiased) simple probability distribution for $\jjSingleGen$ determined by $\jjTuple$. An analogous equation can be written for aggregate variables. }

{Note that the objects sampled by a selection are random. Thus, selection-relative probabilities and operations are used on things describing random objects, although the selection itself may have a realized host.}\footnote{\vphantom{I}{If for some reason a selection-relative operation were applied to a variable describing a realized object, the selection would drop out, and it would become a simple operation described in the next subsection (e.g., $\jkPDFAlt{\jjSingleGenJ} = \PDF{\jjSingleGenJ}$).}}

{
\subsubsection{A notational convention for simple and selection-relative operations}
\label{sec:SelRelOperationRules}
The simple probabilities are selection-relative probabilities where the {self-point} process is being used. This rule extends to all distribution operations on the probability:
\begin{align}
\nonumber \DOperation{\jjSingleGen} & = \DOperation{\jjSingleGen | \jjTuple} = \jjDOperationGen{\jjSingleGen}\\
         \DOperation{\jjkAggGenAlt} & = \DOperation{\jjkAggGenAlt | \jkTuple} = \jkDOperationAlt{\jjkAggGenAlt}
\end{align}
including mean, variance, and order statistics.

A similar simplification is used in this series for windows -- a selection-relative distribution operation ``missing'' its window inherits it from the variable's object window (which is the quantity window too for singleton variables):
\begin{equation}
{\jlDOperation{\jjSingleGen}} \equiv {\jlDOperationGen{\jjSingleGen}}~\text{and}~{\jlDOperation{\jjkAggGenAlt}} \equiv {\jlDOperationAlt{\jjkAggGenAlt}} ,
\end{equation}
including mean and variance {(Table~\ref{table:Simplify})}. Contrast with the convention for variables, which stand on their own and have no other variable to draw a window from. This convention is motivated by practicality: in most windows we use, the vast majority of objects do not contribute to aggregate quantities (equation~\ref{eqn:SingletonNonOverlap}) and do not concern us. When we want to know the mean broadcast fluence over an observation, we do not want to include those that were only visible during the Jurassic, for instance.\footnote{If we \emph{do} want to include these objects, we apply the ALL window operator, ${\jkDOperationAll{X}}$ {(compare Figure~\ref{fig:MeansComparison})}.} Remember, if the variable itself does not specify a window, it uses the ALL ($\AllLabel$) window (Section~\ref{sec:Windows}), which the mean then inherits (${\jkDOperation{\jjSingle}} = {\jkDOperationAll{\jjSingleAll}}$). This applies to the intrinsic quantities describing objects like their lifespan and position.}

{
\subsubsection{Simple means for a single object}
\label{sec:SimpleMeans}
The most straightforward averaging operation is the simple mean. Say we have a singleton quantity $\jjSingleGen$ associated with a random object of type $\JMark$ with a parameter tuple $\jjTuple$. Its simple mean is:
\begin{equation}
\label{eqn:SimpleSingleMean}
\Mean{\jjSingleGen} \equiv \Mean{{\jjSingleGen(\jjTuple)} | \jjTuple} = \int {\jSingleCore} \cdot \PDF{\jjSingleGen} (\jSingleCore | \jjTuple) d\jSingleCore .
\end{equation}
This is a function of $\jjTuple$. The simple mean of an aggregate variable is, with the help of equation~\ref{eqn:SingleToAgg},
\begin{align}
\nonumber \Mean{\jjkAggGenAlt} & \equiv \Mean{\jjkAggGenAlt(\jkTuple) \middle| \jkTuple} \\
\label{eqn:SelfAggMean}
                               & = \int {\jAggCore} \cdot \PDF{\jjkAggGenAlt} (\jAggCore | \jkTuple) d\jAggCore .
\end{align}
Recall that $\jkTuple$ does \emph{not} determine the subpopulation of {${\KMark}$'s} {descendant}s, only its statistical properties. The mean is therefore an average over all realizations of this subpopulation, conditionalized on the properties of the host. This sampling variance within the host is regarded as {``noise.''} 

The average is found by applying Campbell's formula (equation~\ref{eqn:CampbellFormula}) to the intensity of the marked point process:
\begin{equation}
\label{eqn:BasicAggMean}
\Mean{{\jjkAggGenAlt}} = \int_{\jjHaystackAlt} \Mean{\jjSingleGen} {\jjkDistAlt (\jjTuple | \jkTuple)} d\jjTuple .
\end{equation}
The integral for $\Mean{\jjSingleGen}$ in equation~\ref{eqn:SimpleSingleMean} is just a special case of this, summing over $\jjjSample {(\jjTuple)}$.

\subsubsection{Selection-relative means}
\label{sec:SelectionRelMeans}

But what if we want to average, not just over the ``noise'' in a random variable, but over different possible ``draws'' of its associated object? This is what a selection-relative mean does (Figure~\ref{fig:MeansComparison}). For a selection $\Selection{\ZLabel}{{\LMark}}$, the selection-relative mean is
\begin{align}
\nonumber {\jlMeanZ{\jjSingleGen}}   & \equiv \Mean{\jjSingleGen {(\jjTuple)} | \jjTuple \in {\jjlSampleZ(\jlTuple)}}\\
          {\jlMeanZ{\jjkAggGenAlt}}  & \equiv \Mean{\jjkAggGenAlt {(\jkTuple)} \middle| \jkTuple \in {\jklSampleZ(\jlTuple)}} .
\end{align}
Means with respect to {realized} hosts have analogous definitions. Unlike the simple means, the selection-relative means averages over all the different possible parameter tuples for the object being described by the random variable, weighted by the probability of it being a part of the sample drawn by the selection. It is an \emph{ensemble} average, the mean that we would get if we repeated the selection again and again.

To give a concrete example, suppose we had a society ${\SocIndex}$ in a galaxy ${\GalIndex}$, and we observe it over a window ${\ObsLabel}$. The total energy received from all broadcasts in that society over ${\ObsLabel}$ is $\mEnergyObsSoc$. Then $\Mean{\mEnergyObsSoc}$ is the mean value of the energy we receive from the broadcasts in that society, or more precisely, the average for an ensemble of societies each with an identical parameter tuple $\aTupleSoc$. {However,} $\MeanGal{\maEnergyObs}$ is the mean value over all the possible societies in the galaxy, roughly the value of $\maEnergyObs$ expected from a typical society.

\begin{figure*}
\centerline{\includegraphics[width=18cm]{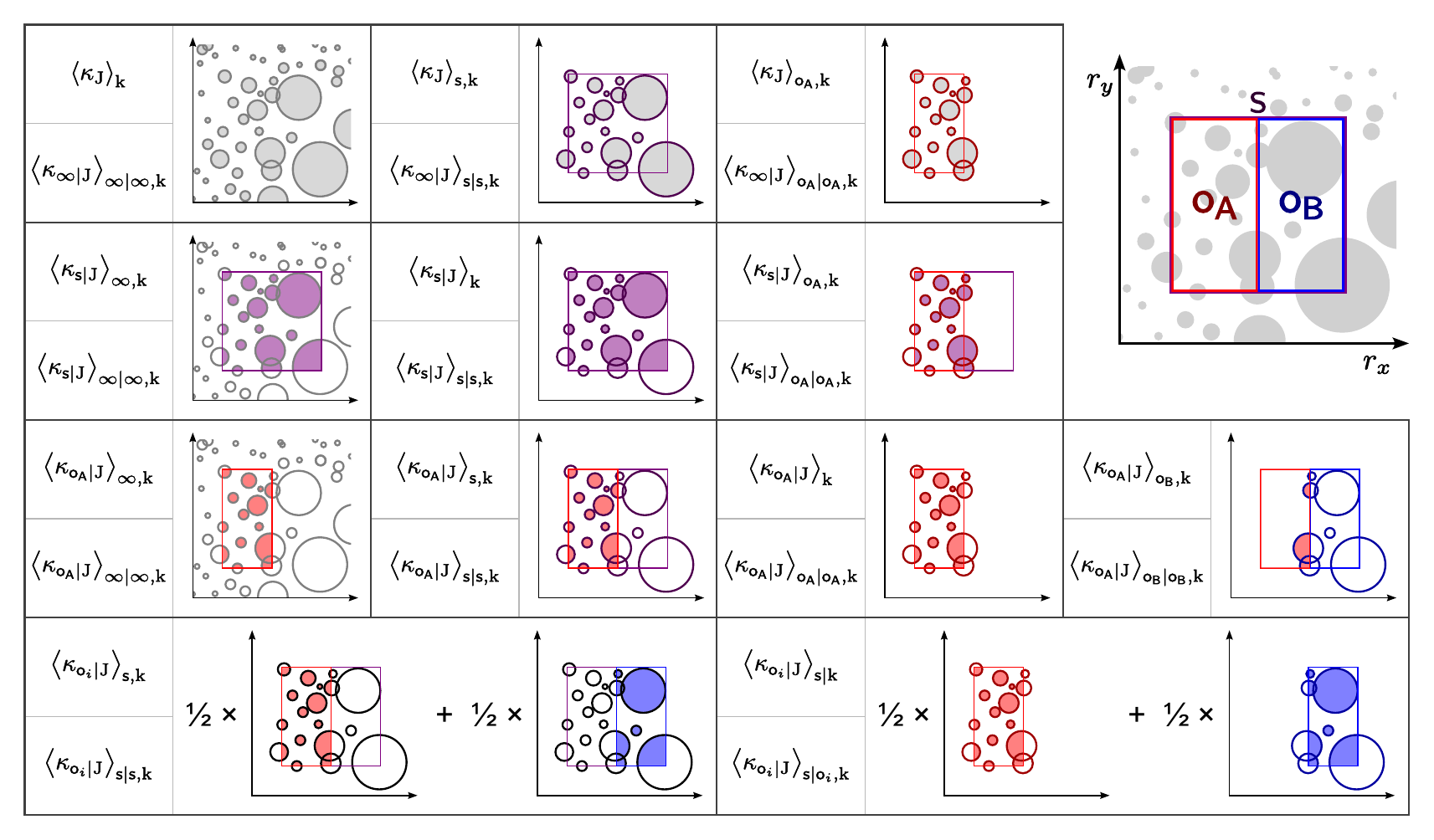}}
\figcaption{{{Comparison} of different quantity windows and selections applied to a variable and an averaging operation. Here we have a cluster of $\JMark$-type spherical objects in a host ${\KIndex}$, observed in {the} sky field as pictured at upper right. A survey ${\SurvLabel}$ covers this field with two observations, ${\ObsALabel}$ and ${\ObsBLabel}$, each an aperture. Any object {that} touches a window is selected by it. The subplots depict different multiwindow selection-relative means, with simplified (above) and full (below) notation. The underlying variable $\jSingle$ is a quantity {that} is integrated over its quantity window (e.g., starlight from a galaxy that falls within the window), with the amount falling within its bounds indicated by the shading. The means average this quantity over only some of the objects, the ones outlined by rings. The multiwindow means on {the} bottom average over several {observations}; ${\ObsILabel}$ is a placeholder for ${\ObsALabel}$ and ${\ObsBLabel}$. Note however that the means are actually ensemble averages, over all possible configurations of the objects, whereas only one realized configuration is shown. Other operations like the variance and PDF also have multiwindow selection-relative versions analogous to the means shown here.\label{fig:MeansComparison}}}
\end{figure*}

The selection-relative mean is found using the selection-relative (biased) PDF (equation~\ref{eqn:BiasedPDF}). Assuming all the variables are mutually independent, equation~\ref{eqn:SimpleSingleMean} gives us
\begin{equation}
\label{eqn:BiasMean}
{\jlMeanZ{\jjSingleGen}} = \frac{1}{\Mean{{\jjlNZ}}} \int_{\jjHaystackZ} \Mean{\jjSingleGen} {\jjlDistZ (\jjTuple | \jlTuple)} d\jjTuple ,
\end{equation}
with the selection-relative mean of an aggregate variable found with an analogous integral (equation~\ref{eqn:SingleToAgg}). A consequence of Campbell's formula is that
\begin{equation}
\Mean{{\jjkAggGenAlt}} = \Mean{{\jkSumJAltFULL} \jjSingleGen{(\jjTuple)}} = \Mean{{\jjkNAlt}} {\jkMeanAlt{\jjSingleGen}} .
\end{equation}
}

Once completed, the averages may themselves be averaged{, most naturally when the {inner average's} selection host is a random object}.\footnote{\vphantom{I}{If the operation's host is a realized object ${\KIndex}$, then the {selection-relative mean} (or variance) has a fixed value determined by $\jkTupleK$ (i.e., it is a degenerate random variable); whereas, if the operation's host is random object ${\KMark}$, then it is a function that yields a single value for each $\jkTuple$, or equivalently, the degenerate random variable that always attains that value.}} As a random variable, this variable inherits the object window and {host} from the averaging operator itself: $\jkMeanAlt{X_{\JMark}} \rightarrow {\jkSinglePRIMEAlt} {(\jkTuple)}$.

Note that, generally,
\begin{equation}
\label{eqn:RecursiveAverageInequality}
{\jlMeanZ{\jkMeanAlt{\jjSingleGen}} \ne \jlMeanZ{\jjSingleGen}},
\end{equation}
and similarly for {aggregate} variables. That is because the left-hand side averages over subselections before averaging over the outermost selection. Some subselections may have more $\JMark$-type objects than others, but these are all weighted equally on the left-handed side, despite their different population sizes. By evenly sampling the $\KMark$-type {host} objects, the left-hand side fails to evenly sample the $\JMark$-type objects {like} the right-hand side. This effect is related to Simpson's Paradox \citep{Haigh13}. {{However,} it \emph{is} true that
\begin{equation}
{\jlMeanZ{\Mean{\jjSingleGen}} = \jlMeanZ{\jjSingleGen}},
\end{equation}
because the inner mean's selection necessarily has a population size of $1$ -- this is just the law of total expectation {and} an expression of Campbell's formula for marked point processes}.

\subsubsection{Simple and selection-relative variances}
\label{sec:VarSelectionRel}
{The simple variance of a variable is found using the distribution of that variable that comes from its tuple. Selection-}relative variances {then average over a distribution of random objects. Their definition follows from that of} a general conditional variance, $\Var{X | Y} = \Mean{X^2 | Y} - \Mean{X | Y}^2$ \citep{Wasserman04,Bas19}. Thus ${\jkVarGen{X} = \jkMeanGen{X^2} - \jkMeanGen{X}^2}$.

{The law of total variance is an important tool in evaluating both types of variances.} {By first conditionalizing on $\jjkNAlt(\jkTuple)$, we derive the simple variance of $\jjkAggGenAlt(\jkTuple)$:}
\begin{equation}
\label{eqn:IndependentVarAggregate}
{\Var{\jjkAggGenAlt} = \jkMeanAlt{\jjSingleGen}^2 \Var{\jjkNAlt} + \Mean{\jjkNAlt} \jkVarAlt{\jjSingleGen},}
\end{equation}
{as long as all of the variables are independent. When ${\jjkNAlt (\jkTuple)}$ {for a fixed $\jkTuple$} is also Poisson, {the variance} reduces to $${\Mean{\jjkNAlt} \jkMeanAlt{\jjSingleGen^2} = \int_{\jjHaystack} \Mean{\jjSingleGen^2} \jjkDistAlt(\jjTuple | \jkTuple) d\jjTuple},$$ equivalent to Campbell's second formula (equation~\ref{eqn:CampbellVar}). The law also gives us, for selection-relative variances,}
\begin{equation}
\label{eqn:VarAggregate}
{\jlVarZ{{\jjkAggGenAlt}} = \jlVarZ{\Mean{{\jjkAggGenAlt}}} + \jlMeanZ{\Var{{\jjkAggGenAlt}}} .}
\end{equation}

\subsection{Multiwindow operations}
\label{sec:Multiwindow}
{
An observing program is usually a collection of many observations: a survey ${\SurvLabel}$ might {take} several pointings ${\PointILabel}$, each in turn involving thousands of distinct observations ${\ObsJLabel}$. The problem is that the statistical properties of a sample may depend on which subwindow we are considering. Differences between subwindows also introduce still more variance. When mapping a spiral galaxy, for instance, the stars are far more dense in the core than the outer disk, increasing the variance in star counts because the underlying density changes. We need to be able to ask questions like ``{What} is the expected number of stars per observation averaged over all observations in the survey?'' and ``{What's} the maximum broadcast fluence during an observation among all the observations we do?''

Just as the selection-relative operations have an object window that picks which objects are included in the operation, the multiwindow operations have a quantity window, which defines which subwindows are considered. The most general operation is written $\jkDOperationZAlt{X}$, where ${\ZLabel}$ is its object window, and $\Selection{\AltLabel}{\KMark}$ is its selection. 

For our purposes, there are a fixed number of subwindows with specified properties. I define the PDF as a mixture of the PDFs for the individual windows:
\begin{equation}
\jkPDFZAlt{\jjSingleGen} (\jSingleCore | \jkTuple) \equiv \frac{1}{\oNGenZ} \sum_{i = 1}^{\oNGenZ} \jkPDFAlt{\jjSingleGen} (\jSingleCore | \jkTuple).
\end{equation}
The general multiwindow mean is
\begin{equation}
\jkMeanZAlt{\jjSingleGen} \equiv \Mean{\jkMeanAlt{\jjSingleGenI} | \GenILabel \in \oGenOfZ}.
\end{equation}
We simply take the mean over the given selection for each object window, then take the average over the object windows (Figure~\ref{fig:MeansComparison}). The general multiwindow variance is, by the law of total variance,
\begin{multline}
\jkVarZAlt{\jjSingleGen} \equiv \jkMeanZAlt{\jjSingleGen^2} - \jkMeanZAlt{\jjSingleGen}^2\\
 = \Mean{\jkVarAlt{\jjSingleGenI} | \GenILabel \in \oGenOfZ} + \Var{\jkMeanAlt{\jjSingleGenI} | \GenILabel \in \oGenOfZ}.
\end{multline}
The second term in the final equality is {the} variance due to the changes in the population between ${\GenILabel}$ windows. Finally the general multiwindow maximum is
\begin{equation}
\jkMaxZAlt{\jjSingleGen} \equiv \Max{\jkMaxAlt{\jjSingleGenI} | \GenILabel \in \oGenOfZ},
\end{equation}
and likewise for the minimum. Thus, the mean energy fluence from a broadcast during a typical observation in a survey ${\SurvLabel}$ among the broadcasts {the survey} selects from a galaxy ${\GalIndex}$ is written $\MeanSurvSurvGal{\lFluenceEObsI}$, for instance.

The populations in different windows are not generally independent. Adjacent windows can cover the same objects: the sidelobes of one pointing can intersect the primary beam of another for instance. Sometimes the objects themselves are extended, crossing into different windows, as when a frequency-drifting narrowband broadcast cuts across many channels. The greatest effect of this cross-correlation is on the maxima and minima because they reduce the effective number of independent samples. If the subwindows are very broad, the variance can be reduced as well, because the same objects are covered in all pointings, averaging out each measurement.

When working with multiwindow operations, there is a distinction between general unspecified windows and specific windows, similar to that between random objects and realized objects. We can assign a unique label to a unique window, like ${\ObsONELabel}$ or ${\ObsTWOLabel}$, and apply it to a variable. This fixes a unique subwindow; the operation's object window becomes irrelevant because the subwindow is one of a kind.  In other cases, we really do want to consider the full gamut of subwindows in a collection, and to emphasize this, we might use a subwindow label like ${\ObsILabel}$, where $i$ is a free variable. Most often, I assume that the statistical properties of the population do not change appreciably from window to window. Then the multiwindow selection-relative means and variances simply collapse to the selection-relative means and variances, and the multiwindow maxima are a simple application of extreme value theory (Appendix~\ref{sec:ExtremeValue}). Only then can we elide the distinction between unspecified and specified windows: any observation is just an observation ${\ObsLabel}$.

The multiwindow operations come with their own notational simplifications. The quantity window can be inherited from the operation's selection but not the variable, while an operation labeled \emph{solely} by a window uses it as a quantity window:
\begin{align}
\nonumber {\jkDOperationAlt{\jjSingleGenI}} & \equiv {\jkDOperationAltAlt{\jjSingleGenI}}\\
          {\DOperationAltNULL{\jjSingleGenI}}   & \equiv {\jjDOperationAltGenI{\jjSingleGenI}} 
\end{align}
Table~\ref{table:Simplify} summarizes all of the notation rules, and Figure~\ref{fig:MeansComparison} provides an illustrated example.} {As before, the object window of an operation is inherited from the variable, even when a separate quantity window is specified:
\begin{equation}
\jkDOperationAltNULL{\jjSingleGenI} \equiv \jkDOperationAltGenI{\jjSingleGenI} .
\end{equation}
}

\begin{deluxetable}{ll|ll}
\tabletypesize{\footnotesize}
\tablecolumns{4}
\tablewidth{0pt}
\tablecaption{Simplifications in notation for variables and operations \label{table:Simplify}}
\tablehead{\multicolumn{2}{c}{Singleton variable} & \multicolumn{2}{c}{Aggregate variable} \\ \colhead{Short} & \colhead{Full} & \colhead{Short} & \colhead{Full}}
\startdata
\cutinhead{Variables}
$\jjSingle$ & $\jjSingleAll (\jjTuple)$       & $\jjkAgg$        & $\jjkAggAllAll (\jkTuple)$\\
            &                                 & $\jjkAggGen$     & $\jjkAggGenGen (\jkTuple)$\\
						& 																& $\jjkAggGenNULL$ & $\jjkAggGenAll(\jkTuple)$\\
						& 																& $\jjAggGen$      & $\jjuAggGenGen(\uTuple)$\\
						& 																& $\jjAggGenAlt$   & $\jjuAggGenAlt(\uTuple)$\\
\cutinhead{Operations on variables}
$\DOperation{\jjSingleGen}$          & $\jjDOperationGenGen{\jjSingleGen}$    & $\DOperation{\jjkAggGenAlt}$        & $\jkDOperationAltAlt{\jjkAggGenAlt}$\\
$\jkDOperation{\jjSingleGen}$        & $\jkDOperationGenGen{\jjSingleGen}$    & $\jlDOperation{\jjkAggGenAlt}$      & $\jlDOperationAltAlt{\jjkAggGenAlt}$\\
$\DOperationAltNULL{\jjSingleGen}$   & $\jjDOperationAltGen{\jjSingleGen}$    & $\DOperationZNULL{\jjkAggGenAlt}$   & $\jkDOperationZAlt{\jjkAggGenAlt}$\\
$\jkDOperationAlt{\jjSingleGen}$     & $\jkDOperationAltAlt{\jjSingleGen}$    & $\jlDOperationZ{\jjkAggGenAlt}$     & $\jlDOperationZZ{\jjkAggGenAlt}$\\
$\jkDOperationAltNULL{\jjSingleGen}$ & $\jkDOperationAltGen{\jjSingleGen}$    & $\jlDOperationZNULL{\jjkAggGenAlt}$ & $\jlDOperationZAlt{\jjkAggGenAlt}$\\
\enddata
\tablecomments{{The simplifications are applied from the inside out. Thus, $\jkMeanAlt{\jjSingle} \to \jkMeanAlt{\jjSingleAll} \to \jkMeanAltAlt{\jjSingleAll}$. When the host of a selection-relative operation (including a general multiwindow operation) is a random object, the result should be {understood} as a function of the host's properties.}}
\end{deluxetable}

\subsection{Regularization of variables with broad distributions}
\label{sec:Regularization}
{Like other astrophysical entities, ETIs and technosignatures may be described by quantities with heavy-tailed distributions{. The radiated energy is widely thought to be a possible example \citep{Drake73-L}; others may include the lifespan of societies or transmitters, or the number of societies per metasociety.}} Plausible distributions like shallow power laws can have variances and means dominated by very rare events with extreme properties. These objects may be too uncommon to detect in any practical selection, however, and thus throw off estimates {for the properties of a typical sample}.

When needed, this series {estimates mean and variance} by trimming values of {a} random variable that are far outside the usual range in a typical sample. {This} regularization is done by imposing cutoff values {$\jSingleCUTLO$ and $\jSingleCUTHI$} based on extreme value statistics {(Appendix~\ref{sec:ExtremeValue}). Starting from a random variable $\jjSingleGen$ and trimming with respect to a selection $\Selection{\AltLabel}{{\KMark}}$, the regularized variable ${\jjSingleGenREGjkAlt(\jjTuple; \jkTuple)}$ has a truncated distribution,
\begin{equation}
\label{eqn:RegDist}
\PDF{{\jjSingleGenREGjkAlt}} (\jSingle) \equiv \displaystyle \frac{\PDF{\jjSingleGen} (\jSingle) \cdot \IndicatorOf{\jSingleCUTLO \le \jSingle \le \jSingleCUTHI}}{{\jkPAlt}(\jSingleCUTLO \le \jjSingleGen \le \jSingleCUTHI)} .
\end{equation}
}

{Equation~\ref{eqn:RegDist} also regularizes aggregate variables, {as} they can be treated as just a special case of singleton variables (equation~\ref{eqn:SingleToAgg}). Because aggregate variables are sums, the calculations are more complicated. If the individual contributions have a narrow enough distribution that their full range is expected to be well-sampled in each aggregate, any large fluctuations in the sum must be due to fluctuations in the number of objects included: $\jjkAggGenREGjlAlt \approx \jjkNGenREGjlAlt \jkMeanGen{\jjSingleGen}$. These fluctuations may be relatively small and the regularization has essentially no effect. If the individual contributions have a very broad distribution -- when we are calculating the aggregate luminosity of objects with a shallow power-law distribution, for example -- the sum is almost entirely dominated by the largest contribution in the sample and $\jjkAggGenREGjlAlt \approx \jlMaxAlt{\jjSingleGen}$.}

The cutoff is a statistical quantity, however, and any actual sample may include extreme outliers.  Thus this regularization will underestimate the sample mean and variance some of the time, possibly by a large degree in rare instances.

\subsection{The discreteness criterion}
\label{sec:DiscretenessCriterion}
In order to make a detection of a broadcast {in a survey} {${\SurvLabel}$}, {at least one must be} sampled during a program\footnote{The converse is not always true -- a sampled broadcast may be too faint to detect, or confused with many others.}, and a null result should occur whenever zero broadcasts are sampled. It is possible our observation was too short, too narrowband, or had too small a field to intercept rare broadcasts even when they are extremely bright. This {happens} outside of SETI too: a supernova or a quasar episode can boost a galaxy's luminosity {far} above its typical value {for a short time}. If {a} galaxy's measured luminosity is {fainter than} a supernova or a quasar, that just means one was not active during the observation, not that they never {are found there}. Thus, a null result is compatible with any models {with} a high enough probability of zero broadcasts being detected.

The discreteness {criterion} states that we should only expect a detection {in survey ${\SurvLabel}$ with} model{s having}
\begin{equation}
\label{eqn:DiscreteP}
\fpP(\bNSurv = 0) \le {\fpPBAR}
\end{equation}
{for a suitably conservative false negative threshold $\fpPBAR$ {near $0$}.} Only models that fulfill this criterion are likely to be constrained by observations. To order of magnitude, we expect a {nonempty} sample only if the mean number of objects is $\ga 1$.\footnote{This follows from Markov's inequality, which gives $\fpP({\jjNSurv} \ge 1) \le \Mean{{\jjNSurv}}$ for {the} {nonnegative} ${\jjNSurv}$ \citep{Wasserman04}.} However, no broadcasts will be detected if the sample is missing any of the ancestor nodes on the ``tree'' -- societies, metasocieties, or {galaxies}. {E}quation~\ref{eqn:DiscreteP} implies {the often-simpler formula} 
\begin{equation}
\label{eqn:DiscretenessCriterion}
{\Mean{\jkNSurv}} \ge 1 - {\fpPBAR} \approx 1~{\text{if}~\KMark \in \AncestorTypesOf{\BcMark}~\text{or}~\KMark = \BcMark.}
\end{equation}

\subsection{An example: lifespan bias in time-limited selection}
\label{sec:LifespanBias}
Long-lived objects are disproportionately represented in a field once a population has reached equilibrium. This phenomenon serves as an example of how to make selection-relative calculations.

Suppose we have objects of type $\JMark$ in a {host} {${\KMark}$}. {How is the probability distribution of object lifespan affected by a time-limited selection? To start, {let us} suppose that each such object forms} at time $\jjTStart$ and exists for {duration} $\jjDuration$ without interruption. The distribution function for these objects has the form
\begin{equation}
{\jjkDist (\jjTuple | \jkTuple)} = \frac{d^3 \Mean{{\jjkN}}}{d\jjDuration d\jjTStart d\jjParamsOther} {(\jjTuple | \jkTuple)}
\end{equation}
where $\jjParamsOther$ stands for a tuple of other, irrelevant parameters {to marginalize over}. Now consider a window ${\GenLabel}$ that selects all objects in ${\KMark}$ {active anytime during} the contiguous time interval $[\oTStartGen, \oTStartGen + \oDurationGen]$. Clearly, an object in this population is selected if it either formed before $\oTStartGen$ and lives long enough to overlap the window {($\oTStartGen - \jjDuration \le \jjTStart < \oTStartGen$)} or if it forms within the window {($\oTStartGen \le \jjTStart \le \oTStartGen + \oDurationGen)$}. Thus, if the objects form at a constant rate ${\jjkRateTotal}$ {(equation~\ref{eqn:jRate})},
\begin{align}
\nonumber \frac{d\Mean{{\jjkNGen}}}{d\jjDuration} {(\jDurationCore | \jkTuple)} & = \int_{\oTStartGen - {\jDurationCore}}^{\oTStartGen+\oDurationGen} \int {\jjkDist(\jjTuple | \jkTuple)} d\jjParamsOther d\jjTStart \\
\label{eqn:BiasedDurationNDist}
                                                & = ({\jDurationCore} + \oDurationGen) {\cdot \jjkRateTotal(\jkTuple) \cdot} {\jkPDF{\jjDuration} (\jDurationCore)}
\end{align}
from equation~\ref{eqn:MeanNGenericSelection}, where ${\jkPDF{\jjDuration}}$ is the unbiased lifespan distribution. It follows immediately that ${\Mean{\jjkNGen} = (\jkMean{\jjDuration} + \oDurationGen) \cdot \jjkRateTotal(\jkTuple)}$, giving us {the ${\GenLabel}$-biased duration distribution} (equation~\ref{eqn:BiasedPDF})
\begin{equation}
{\jkPDFGen{\jjDuration} (\jDurationCore)} = \frac{{\jDurationCore} + \oDurationGen}{{\jkMean{\jjDuration}} + \oDurationGen} {\jkPDF{\jjDuration} (\jDurationCore)}.
\end{equation}
Finally, we have (equation~\ref{eqn:BiasMean}),
\begin{equation}
\label{eqn:BiasedDurationMean}
{\jkMeanGen{f(\jjDuration)}} = \frac{{\jkMean{\jjDuration f(\jjDuration)}} + \oDurationGen {\jkMean{f(\jjDuration)}}}{{\jkMean{\jjDuration}} + \oDurationGen} .
\end{equation}
These formulae assume that the rate has been constant infinitely far into the past, at least in that each object's lifespan is shorter than the age of ${\KMark}$. Otherwise, an additional cutoff is needed for the integration in equation~\ref{eqn:BiasedDurationNDist}, as is well known in the field of stellar populations \citep[e.g.,][]{Miller79}. {In the limit that all the objects are more short-lived than the host's age at the start of ${\GenLabel}$, $\oTStartGen - \jkTStart$, equation~\ref{eqn:BiasedDurationMean} still applies. On the other hand, if most all the {$\JMark$ objects} live longer than $\oTStartGen - \jkTStart$, it can be shown the bias disappears: {almost} every object that has ever been born is still around, so we get a fair sample of them.}

\begin{figure*}
\centerline{\includegraphics[width=17cm]{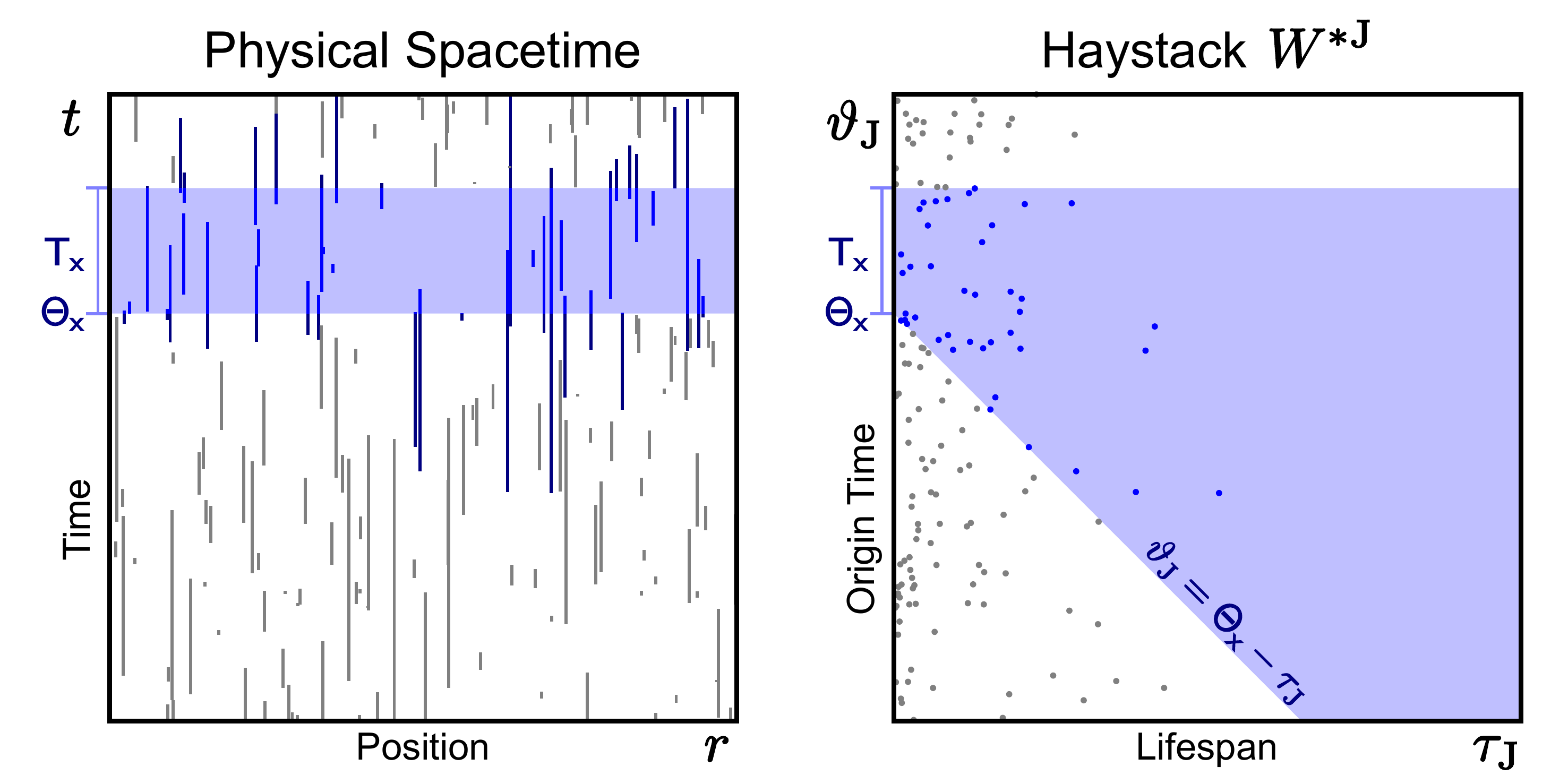}}
\figcaption{{Lifespan bias by a window ${\GenLabel}$ in a population of objects with {exponentially distributed} lifespans. Selected objects are in blue (bright blue within the window). On {the} left, the objects are scattered in spacetime, and their lifespans are intervals in time. Those with longer $\jjDuration$ are more likely to reach the window, which covers a fixed span of real time (blue shading). On {the} right, the objects are represented by points in a simplified haystack (spatial position not included). The window covers a bigger $\jjTStart$ range as $\jjDuration$ increases.} \label{fig:LifespanBias}}
\end{figure*}

When the observational window is short compared to typical lifespans, equation~\ref{eqn:BiasedDurationMean} effectively weights the biased average by another power of $\jjDuration$. Every biased moment depends on the next higher unbiased moment, save the zeroth where a cancellation occurs. For the mean lifespan {in particular}, ${\jkMeanGen{\jjDuration}} = {\jkMean{\jjDuration}} + {\jkVar{\jjDuration}}/({\jkMean{\jjDuration}} + \oDurationGen) \ge {\jkMean{\jjDuration}}$. Hence, one can replicate the result in \citet{Kipping20} that ${\jkMeanGen{\jjDuration}} \approx 2 {\jkMean{\jjDuration}}$ for exponentially distributed $\jjDuration$. Or, one can show that the lifespan{'s selection-relative} mean is ill-behaved for a {power-law} distribution ${\jkPDF{\jjDuration}} \propto \jjDuration^{-\alpha}$ extending to infinity unless $\alpha > 3$.

{Lifespan bias can be thought of in two ways (Figure~\ref{fig:LifespanBias}). In real spacetime, ``bigger'' objects are more likely to cross into a window with fixed duration, having a larger {``cross-section.''} In the haystack, the window itself becomes wider for longer lifespans and thus has a longer reach for these objects. A similar bias applies to band-limited selection by frequency: wideband broadcasts are more likely to be intercepted, a significant fact in the chord model presented later.}

{Figure~\ref{fig:LifespanBias} helps illustrate the different types of means. Each dot represents a specific, realized object; $\Mean{\jjDurationJ}$ is the simple mean of the lifespan for that object ${\JIndex}$, which is entirely determined by the dot's position in the haystack. Then $\Mean{\jjDuration} = \jjMean{\jjDuration} {(\jjTuple)}$ is a \emph{function} ranging over the entire space that returns the average lifespan for an object given its parameters -- it returns the $\jjDuration$ coordinate in the {right-hand} panel. A realized host ${\KIndex}$ has an accompanying distribution of {objects} over the haystack:} {$\MeanK{\jjDuration}$ averages over the entire haystack, while $\MeanGenK{\jjDuration}$ averages only over the {blue-shaded} region.} {One such realization of the distribution is shown in the figure. Likewise, for a random host ${\KMark}$, the distribution itself is a function of $\jkTuple$, as are $\jkMean{\jjDuration}(\jkTuple)$ and $\jkMeanGen{\jjDuration}(\jkTuple)$.}

\section{Assumptions about randomness}
\label{sec:PopAssumptions}
{Technosignatures} are the result of complex {biological,} social{,} and technological phenomena, with a panoply of factors shaping their populations. The full interplay responsible for broadcast characteristics may be {analytically} intractable. Nonetheless, treating broadcasts as the result of a hierarchy of {conditionally} {independent point} processes {illuminates our understanding of the ETI populations.}

{
\subsection{``Random'' ETIs}
\label{sec:RandomBc}
Random variables are a very general concept (Section~\ref{sec:RVBasics}) and can describe deterministic events. Still, the rest of the paper views broadcasts and other objects as unpredictable. This does not mean that the intelligences behind them have no specific motivations. Random processes {can be used to describe human behavior by focusing on statistical trends rather than the unpredictable specific outcomes resulting from complex motivations \citep[e.g.,][]{Jusup22}}; the same seems plausible for ETIs.

The most basic properties of metasocieties and societies are where they are, when they start, and how long they last. The first two of these likely involve stochastic elements. Potential habitats are scattered randomly through the galaxy, and the timing of ETI evolution likely involves some contingency, although subsequent interstellar expansion might not. 

{Broadcasts can have predictable structure, though.} Predictability can be a byproduct of how a transmitter operates. A narrowbeam transmitter on a rotating world sweeps past the Earth at regular intervals, for example \citep{Gray02}. ETIs could even exploit periodicity to encode information or make the artificial nature of their broadcasts obvious \citep[as in][]{Borra12,Harp18}.

{However, although} broadcasts may have predictable features, their properties are not fully determined -- {otherwise,} we would simply aim our telescopes where and when we expect a broadcast instead of sifting through the haystack. Broadcasts plausibly turn on and off at unpredictable times, the result of unknowable social factors. Randomness in the host {galaxy} also induces randomness in broadcasts. Each host star has random light travel delay times, parallaxes, {and} Doppler shifts, and sits in a turbulent interstellar medium with random dispersion and scintillation. A faint periodic broadcast might only be observable during rare moments when scintillation magnifies it to a detectable level \citep{Cordes97}.

The real issue is whether the treatment of broadcasts as independent isolated bursts of radiation is inadequate, and thus the haystack is defined according to the wrong parameters. Introducing dependence is one way to address this: in a periodic train, knowing two subsequent arrivals {allows} us {to} predict the others. Often there is a natural parameterization in which we can treat the entire complex of signals as a single broadcast, however, like defining a periodic train by its period and phase. Although the specific distributions used in this paper would not apply, the general ideas of the analysis still do.
}

\subsection{Independence}
\label{sec:Independence}
I {also} assume the properties of two sibling objects are independent of each other, conditionalized on the shared parent's parameter tuple. Likewise, the number of children objects in a parent is conditionally independent of each child object's parameters. Thus, each broadcast is assumed to be independent of every other, except in the sense they depend on a shared ancestor (society, metasociety, {galaxy, or the universe}); societies are independent of each other, aside from \emph{their} shared {ancestors; and so on.}

Independence is by no means obvious. On the broadcast level, although beacons and {noncommunicative} broadcasts are plausibly completely independent, intragalactic communications are part of a web of messages, providing a natural mechanism for coordination. Broadcasts may be more common if there are more potential receiving societies{, and some may even be direct replies to others}. Metasocieties {can} converge to a common broadcast protocol, like designated frequency channels. {Dependence could extend to other object types -- if interstellar expansion is rampant, the existence of a society around one star strongly suggests the existence of them around nearby stars, and so on.}

{The framework handles all such seeming dependence by moving the shared properties up the tree.} Instead of {viewing sibling objects} as mutually dependent, we instead {postulate that the ``dependent quantities'' are actually properties of the ancestor} itself. Thus, the common properties {all follow} from the {ancestral tuple, which governs the populations within it. \emph{Conditionalized}} on {that tuple}, the {siblings} are independent of each other. The difference is that the population does not respond to {its own} fluctuations.

{Consider: what if} each {galaxy} has exactly one society arising at a random time {since the Big Bang} and the society's lifespan is always one hundred years{?} Without knowing when the society existed, a single broadcast from that society could occur any time {in cosmic history}. {Once} we know when one broadcast occurred, however, any others {had to have been} made to within one hundred years because there was no broadcasting society at other times {-- a failure of independence. {However,} that is only because the broadcast tells us when the society appeared. If we already knew that, learning when one broadcast {happens} does not necessarily give us any new information -- the broadcast times can be independent conditional on societal parameters.}

{The model can be altered to include different levels to allow for still further degrees of dependence, but it nonetheless assumes that everything can be grouped into discrete {objects} sorted into hierarchical levels. Perhaps ETI distributions are more like scale-free fractals with a continuous range of dependence. This would necessitate a different kind of framework.}

{Just because the properties of one object are conditionally independent of another's does not mean that the parameters of a single object are mutually independent. Massive stars are younger on average; metasocieties encompassing more worlds could live longer; ``gregarious'' societies could transmit broadcasts at a higher rate and make them brighter on average; and so on. This just means that the joint distribution is not separable into the product of the PDFs of the variables; {like when} all the objects fall along a line or surface in the haystack. It prevents us from {modeling} the aggregate emission with compound Poisson statistics. The distribution can still be calculated, however, and we still find the mean with Campbell's formula (equation~\ref{eqn:CampbellFormula}). If the societal population is Poissonian, equation~\ref{eqn:CampbellVar} {allows} us {to} find the variance.}

\section{The astrophysical context of ETIs: The universe, galaxies, and stars}
\label{sec:HighLevels}

{
\begin{deluxetable*}{lccccccc}
\tabletypesize{\footnotesize}
\tablecolumns{8}
\tablewidth{0pt}
\tablecaption{{Shared notation for objects \label{table:SharedObjectNotation}}}
\tablehead{\colhead{Quantity} & \colhead{$\JMark$-Type} & \colhead{Universe} & \colhead{Galaxy} & \colhead{Metasociety} & \colhead{Society} & \colhead{Broadcast} & \colhead{Star}}
\startdata 
Random object index      & $\JMark$  & $\UnivMark$  & $\GalMark$  & $\MetaMark$  & $\SocMark$  & $\BcMark$  & $\StarMark$\\
Realized object index    & $\JIndex$ & $\UnivIndex$ & $\GalIndex$ & $\MetaIndex$ & $\SocIndex$ & $\BcIndex$ & $\StarIndex$\\
\hline
Parameter tuple          & $\jjTuple, \jjTupleJ$ 
                         & $\uTuple, \uTupleUniv$ 
												 & $\gTuple, \gTupleGal$ 
												 & $\zTuple, \zTupleMeta$ 
												 & $\aTuple, \aTupleSoc$ 
												 & $\bTuple, \bTupleBc$ 
												 & $\hTuple, \hTupleStar$\\
Position                 & $\jjPosition, \jjPositionJ$   
                         & \nodata        
												 & $\gPosition, \gPositionGal$   
												 & $\zPosition, \zPositionMeta$   
												 & $\aPosition, \aPositionSoc$   
												 & $\bPosition, \bPositionBc$   
												 & $\hPosition, \hPositionStar$\\
Size (volume)            & $\jjSize, \jjSizeJ$       
												 & $\uSize, \uSizeUniv$       
												 & $\gSize, \gSizeGal$       
												 & $\zSize, \zSizeMeta$       
												 & \nodata        & \nodata        & \nodata\\
Starting time            & $\jjTStart, \jjTStartJ$     
                         & $\uTStart, \uTStartUniv$     
												 & $\gTStart, \gTStartGal$     
												 & $\zTStart, \zTStartMeta$     
												 & $\aTStart, \aTStartSoc$     
												 & $\bTStart, \bTStartBc$     
												 & $\hTStart, \hTStartStar$\\
Duration                 & $\jjDuration, \jjDurationJ$   
                         & $\uDuration, \uDurationUniv$   
												 & $\gDuration, \gDurationGal$   
												 & $\zDuration, \zDurationMeta$   
												 & $\aDuration, \aDurationSoc$   
												 & $\bDuration, \bDurationBc$   
												 & $\hDuration, \hDurationStar$\\
Arbitrary singleton variable & $\jjSingle, \jjSingleJ$
                             & $\uSingle, \uSingleUniv$
														 & $\gSingle, \gSingleGal$
														 & $\zSingle, \zSingleMeta$
														 & $\aSingle, \aSingleSoc$
														 & $\bSingle, \bSingleBc$
														 & $\hSingle, \hSingleStar$\\
\hline
Distribution (intensity) & $\jjDist$         & $\uDist$       & $\gDist$       & $\zDist$       & $\aDist$       & $\bDist$         & $\hDist$\\
Number of objects        & $\jjN$            & \nodata        & $\gN$          & $\zN$          & $\aN$          & $\bN$            & $\hN$\\
Haystack (space state)   & $\jjHaystack$     & $\uHaystack$   & $\gHaystack$   & $\zHaystack$   & $\aHaystack$   & $\bHaystack$     & $\hHaystack$\\
Random sample (point process) & $\jjSample$  & $\uSample$     & $\gSample$     & $\zSample$     & $\aSample$     & $\bSample$       & $\hSample$\\
Realized sample          & $\jjSampleReal$   & $\uSampleReal$ & $\gSampleReal$ & $\zSampleReal$ & $\aSampleReal$ & $\bSampleReal$   & $\hSampleReal$\\
Temporal rate per star   & $\jjRate$         & \nodata        & \nodata        & $\zRate$       & $\aRate$       & $\bRate$         & \nodata\\
Abundance per star       & $\jjAbund$        & \nodata        & \nodata        & $\zAbund$      & $\aAbund$      & $\bAbund$        & \nodata\\
Total temporal {(formation)} rate      & $\jjRateTotal$    & \nodata        & \nodata        & $\zRateTotal$  & $\aRateTotal$  & $\bRateTotal$    & {$\gRateStarCore$}\\
Arbitrary aggregate variable & $\jjAgg$
                             & \nodata
														 & $\gAgg$
														 & $\zAgg$
														 & $\aAgg$
														 & $\bAgg$
														 & $\hAgg$\\
\enddata
\tablecomments{{The {$\JMark$ type} objects are generic, used for general formulae, as are the $\KMark$ and $\LMark$ types.\\
Singleton variables (upper half) are here marked with {the uppercase index for a type of object (e.g., $\BcMark$ for broadcasts), if the variable is treated as a general parameter instead of describing a fixed object.} A lowercase index (defaults given for each type) in a singleton variable indicates the variable refers to a realized object with fixed parameters. Other indices may be substituted, however, if they are defined for realized objects. Singleton variables may also specify a quantity window.\\
Aggregate variables (lower half) can also be marked with an object selection, and a quantity window if different from object window. By default, the windows are the {ALL (}$\AllLabel${)} window and the universe (${\UnivMark}$) as a host.}}
\end{deluxetable*}
}

{
\subsection{The universe}
\label{sec:Universe}
The universe is the root node of the tree (Figure~\ref{fig:Tree}), the ``object'' that contains all the other objects as descendants. Even the universe has a parameter tuple, $\uTuple$, and a haystack, the space of all such tuples $\uHaystack$ (Table{~\ref{table:SharedObjectNotation} lists} relevant {notation}.). Cosmological parameters like the Hubble constant can be used to define this haystack, but in this {work,} I assume these are set to fixed values. Instead, the model parameters describe the descendant populations of ETIs and their broadcasts. For instance, suppose we thought that in every transmitter population, the luminosity distribution follows a {power-law} distribution, all with the same slope $\bPowerLaw$. We do not know what the slope is, however. Then we can consider $\buPowerLaw$ as a cosmic-level parameter, one describing all ETI populations in the Universe. In a given model universe ${\UnivIndex}$, it takes on a value $\bPowerLawUniv$, which all the lower-level populations inherit, and then we make calculations for each $\bPowerLawUniv$ variate to compare predictions.

{As discussed in Section~\ref{sec:ModelNature}, no universe distribution is considered here, only the consequences of different $\uTuple$ for observed populations.}
}

{
\subsection{Galaxies as domains for ETIs}
}
\label{sec:Galaxies}
{ETIs can coordinate their properties on large scale through replication and communication, but there must be a limit to how far these processes can operate. We can imagine the Universe ends up broken up into a patchwork of regions over which coordination succeeded, much like the magnetic domains in a piece of iron \citep[see][]{Olson15}. In this series, a} {domain} is an astrophysical {region} so large and isolated that ETIs cannot spread {outside of it}, completely confining any metasocieties. Depending on how difficult space travel is, a {domain} might be as small as a {planetary body} or as large as the entire Hubble volume. {Every domain has sites where individual societies can reside. This work treats} galaxies as the natural scale for {domains}.\footnote{\vphantom{I}{If intergalactic travel or communication is allowed, galactic metasocieties could fuse into an intergalactic metasociety on the domain scale. If cohesive galactic {submetasocieties} persist, then these would be the children of the top-level intergalactic metasocieties, instead of being the parents. We could also consider galaxies analogously to stars in Section~\ref{sec:Stars}.}}

Although they are home to entire populations of discrete stars and worlds, galaxies themselves form a cosmic population with a distribution
\begin{align}
\nonumber {\guDist (\gTuple | \uTuple)} & \equiv \frac{d\Mean{{\guN}}}{d\gTuple} {(\gTuple | \uTuple)} \\
                                                & = \frac{d^3 \Mean{\guN}}{d\hgN d\gPosition d\gParamsOther} {(\gTuple | \uTuple)}.
\end{align}
The two {parameters I single out} are the number of stars in the entire galaxy, $\hgN$, and its {position $\gPosition$}, with $\gParamsOther$ as a catch-all for any remaining parameters. All other things being equal, galaxies with more stars have more habitats in which intelligent life can evolve, increasing the mean galaxy-wide rate {at which} metasocieties appear. It takes time for intelligence to {evolve}, thus {the cosmic epoch of the galaxy matters}; {distance} also directly affects observables. There are many other factors that could affect the evolution and spread of ETIs: star formation history, metallicity distribution, stellar density {and} velocity dispersion, {nuclear activity}, and the nature of the interstellar medium all could be important (e.g., \citealt{Gonzalez01,Lineweaver04,Dayal15,DiStefano16,Gowanlock16}{; \citealt{Balbi17,Lingam19};} \citealt{Lacki21-Traversability}).

It is also possible the development of new metasocieties is affected by {their predecessors}, and thus there are techno-historical parameters for each galaxy as well. The science-fiction trope wherein relics of a {long-dead} ETI yield advanced technologies for practical interstellar travel (e.g., the ``subway'' system in \citealt{Sagan85}) comes to mind -- later ETIs {would} have an easier time spreading than the first ones. Or, early ETIs might {instead} launch self-replicating probes that actively destroy all new ETIs \citep[see][and discussion therein]{Brin83}. Even so, we might regard all these ETIs as being part of a single metasociety, albeit one whose societies are highly clustered in time, their appearances poorly described by a Poisson process (see Section~\ref{sec:MetaScenarios}).

Observational programs only draw a sample of the cosmic population. {This sample is fixed in targeted observations} typical of most extragalactic radio SETI efforts to date \citep{Shostak96,Gray17}. Large-scale cosmic surveys may be viewed as drawing random samples from the Universe {\citep{Martinez02}}. {Random samples of background galaxies are also present in targeted pointings at nearby objects \citep[see][]{Garrett23}.} {When the galaxies being studied are already known,} we conditionalize on the known galaxy sample{; when treating galaxies statistically,} there is a {galactic} contribution to the sampling variance.

{
\subsection{Stars: A tracer of ETIs?}
\label{sec:Stars}

{Planetary systems around s}tars are commonly regarded as the origin sites of ETIs. The properties of a sun can bear on whether a metasociet{y} originate{s} there.\footnote{\vphantom{H}{Hypothetically, ETIs could originate elsewhere; rogue planets detached from any star may even be habitable \citep[e.g.,][]{Stevenson99,Abbot11,Badescu11}. On the whole, most baryonic bodies like planets should trace the stellar mass on large scales, modulated by other quantities, as discussed below.}} The star needs to live long enough for intelligence to evolve, for one {\citep{Huang59,Carter83,Livio99}, and its luminosity should be steady enough that a planet can remain habitable for geological times \citep{Hart79,Kasting93}. There is a continuing debate on whether planets around M dwarfs can develop life, against the backdrop of planetary tidal locking and large stellar flares (\citealt{Shields16} and references therein); the space weather around them might discourage radio communications if a society evolves \citep{Cirkovic20}.} Stellar parameters could also affect whether a planetary system is settled: mass and metallicity are correlated with number and types of attendant planets \citep[e.g.,][]{Johnson10,Mulders15}; exotic types of stars might draw ETIs interested in astroengineering {\citep{Dyson63,Learned08,Chennamangalam15,Semiz15,Osmanov16,Imara18,Lacki20-LensFlare,Lingam20}}; and so on.} {For a more precise, fine-grained treatment, we consider stellar systems instead of stars, with the distribution including quantities describing the architecture of the system: how many stars, in what orbits; the number, sizes, and locations of planets; their satellites; and other variables that could be relevant to whether there are inhabitants \citep[e.g.,][]{Ward00}.}

{S}ocieties and their broadcasts {may be} concentrated in regions of high stellar density, {without} necessarily being hosted by stars. {O}n a galactic scale, we may expect baryonic resources, stellar energy, and interstellar habitats like interstellar objects to trace the stellar mass distribution. {These could be useful for ETIs in astroengineering projects. {However,} ETIs could migrate to the cores or outskirts of galaxies, driven by the possibility of more efficient collaboration \citep{Kardashev85,Smart09}, more optimal thermodynamic environments \citep{Cirkovic06-OuterGalaxy}, or large-scale gradients in habitats. Some possible ``attractor'' objects trace young stars instead of total stellar mass, like high-mass X-ray binaries \citep[as in][]{Vidal11}, maybe even concentrating ETIs into a few clumps while leaving the rest of the galaxy empty. Even if they roughly follow stars, the dependence need not be linear; they might be far more prevalent in regions of higher stellar density because interstellar travel is easier, for example \citep{DiStefano16,Gajjar21,Lacki21-Traversability}. In the absence of secure knowledge of the relevant tradeoffs, I posit that the density of societies in inhabited galaxies follows the stars on a large scale.}

{Although they do not fit neatly into the ``tree'' of objects, it is useful to treat stars as another class of objects with its own point process. The stars are points scattered in a ``stellar haystack'' $\hHaystack$, described by parameter tuples $\hTuple$. Like other types of objects, windows select stars based on position in their haystack. Galaxies host stars, but stars might also be regarded as hosted by metasocieties and societies if their planetary system is home to an ETI.\footnote{\vphantom{F}{Formally, we could use projections of the position and time of the ``host'' (meta)society to pick out stars (Section~\ref{sec:Windows}). In any case, b}ecause societies are localized in single planetary systems {or interstellar space}, $\haN$ is zero, one, or perhaps up to a few in multiple star systems, depending on whether we are considering stars \emph{per se} or stellar systems.} Thus, we can consider the stellar sample ${\hjjSampleGen}$ drawn by a selection $\Selection{\GenLabel}{{\JMark}}$, consisting of ${\hjjNGen}$ stars. 

The stellar properties considered here are current position $\hPosition$, birth time $\hTStart$, and initial mass $\hM$. Lifespan $\hDuration$ is regarded as a function of $\hM$. The stars then have a distribution function
\begin{equation}
\label{eqn:StellarDistribution}
{\hjjDist (\hTuple | \jjTuple)} = \frac{d^4 \Mean{{\hjjN}}}{d\hPosition d\hTStart d\hM d\hParamsOther} {(\hTuple | \jjTuple)},
\end{equation}
a joint stellar density, initial mass function, and star formation history. Any other parameters are absorbed into $\hParamsOther$. This distribution can be used as a term in metasocietal, societal, or broadcast distribution to introduce a dependence on stellar properties.}

\section{\texorpdfstring{The ETI\lowercase{s} themselves: metasocieties and societies}{The ETIs themselves: metasocieties and societies}}
\label{sec:ETILevels}

{
\begin{deluxetable}{cp{7cm}}
\tabletypesize{\footnotesize}
\tablecolumns{2}
\tablewidth{0pt}
\tablecaption{Other variable notation for objects \label{table:OtherObjectNotation}}
\tablehead{\colhead{Notation} & \colhead{Explanation}}
\startdata
\cutinhead{Galaxies and stars}
{$\hM$}                           & \vphantom{M}{Mass of individual star}\\
{$\hgMAggTime$}                   & \vphantom{T}{Total stellar mass of galaxy {at one epoch}}\\
\cutinhead{Metasocieties and societies}
$\zFC$                                    & Fraction of societies in a metasociety {${\MetaMark}$} that are communicative (broadcasting)\\
$\hParamsMetaOrigin$                      & Parameter tuple for {the} star from which the metasociety first emerged\\
{$\zhRateTotal$}                   & \vphantom{M}{Mean rate of metasocietal origin events around star}\\
{$\ahRateTotal$}                   & \vphantom{M}{Mean rate of societal origin events around star}\\
\cutinhead{Broadcasts}
$\bPolQuantity$						                & Quantity describing polarization of broadcast; in {general, it} is a vector or matrix\\
$\bFPol(\PolVar)$                         & Fraction of broadcast emission in polarization $\PolVar$\\
$\bFPolGen$                               & Sum of $\bFPol$ over all polarizations in $\oPolSetGen$\\
$\bSolidAngle$                            & Solid angle that broadcast is emitted into\\
$\bBandwidth$                             & Total bandwidth of broadcast in {the} source frame, frequency span over {the} entire duration \\
$\bBandwidthTime$                         & Instantaneous bandwidth of broadcast in {the} source frame\\
$\bBandwidthTimeOFGen$                    & Bandwidth covered by broadcast over {the} duration of ${\GenLabel}$\\
$\bNuMid$                                 & Central frequency of broadcast in {the} source frame\\
$\bDriftRate$						                  & Drift rate of broadcast in {the} source frame\\
${\bjjRatenu}$                    & \vphantom{M}{Mean temporal rate of broadcasts in host ${{\JMark}}$ per unit frequency per star}\\
${\bjjAbundnu}$                   & \vphantom{M}{Mean number of broadcasts in host ${{\JMark}}$ per unit frequency per star}\\
${\bjjRatenuTotal}$               & \vphantom{M}{Mean total temporal rate of broadcasts in host ${{\JMark}}$ per unit frequency}\\
${\bjjAbundnuTotal}$              & \vphantom{M}{Mean total number of broadcasts in host ${{\JMark}}$ per unit frequency}\\
$^{\EarthFrameDecor}$                     & Superscript indicating {observer-frame} quantity \\
\enddata
\end{deluxetable}
}

\subsection{Metasocieties}
\label{sec:Metasocieties}

Intelligences are organized into metasocieties and societies. {A}n inhabited galaxy contains one or more metasocieties at the present moment: ${\zgNTime(\TimeVar)} \ge 1$. A metasociety is a collection of interacting ETI societies shaped by a shared history. A society and all of its ``offspring'' always are part of the same metasociety {for our purposes here}. The technosignatures of a metasociety {have a shared} influence {from} a common origin. Metasocieties, unlike individual societies, can be extended {over} galactic scales. 

The nature and evolution of metasocieties is one of the most speculative areas of SETI, with direct bearing on the Fermi Paradox and expected technosignatures. A metasociety might consist of only one planet-confined society, a loose galactic network of worlds, an expanding interstellar settlement front, {or} a dense equilibrium state, {among} other possibilit{ies} (section~\ref{sec:MetaScenarios}).

Given the diversity of hypotheses, huge uncertainties, and the possible amplification of stochastic effects by initial exponential growth, there is no one obviously correct single set of parameters to describe them. {I characterize them with these parameters:}
\begin{itemize}
\item All metasocieties are regarded as having a single origin, {arising} at a {single well-defined} time $\zTStart$, at a single location. {T}he origin {site} is indicated by {a parameter tuple $\hParamsMetaOrigin$ describing} the first inhabited host star {(see section~\ref{sec:Stars})}. {The full distribution includes a} rate of metasociety appearance per star in an empty {galaxy}, even allowing us to establish a functional dependence on stellar properties. However, the number of current metasocieties does not necessarily scale with {$\Mean{\hgNTime}$} if they are sufficiently common and cover the entire {galaxy} (section~\ref{sec:MetaScenarios}). Actual metasocieties may form as a fusion of smaller metasocieties \citep{Kardashev85,Forgan17}; the ``origin'' might be for the first or perhaps a dominant component, but maybe the origins of the others affect the result as well.
\item All metasocieties cover a finite volume $\zSize$ containing a finite number of astrophysical objects {centered at position $\zPosition$. N}ot all regions with the same spatial volume are equivalent -- {maybe there are} more societies in a volume that includes the core or disk of a galaxy than one that only includes the distant halo. {An alternative is the mean number of stars covered by a metasociety, {$\Mean{\hzN}$}. In the (perhaps simplistic) scenarios I discuss, however, the metasociety is either restricted to one stellar system with $\zSize \to 0$ or pervades the entire {galaxy}.} Conceivably, multiple metasocieties might overlap, {perhaps inhabiting} different habitats (e.g., planetary surfaces versus deep space versus compact objects).
\item The lifespan of the metasociety, $\zDuration$. Metasocieties may have effectively undefined lifespans, however, especially if {they} cover entire {galaxies}, continuously regenerating through internal replication.\footnote{Of course, in the very long run, all metasocieties will perish in our current understanding of cosmology, but there will be hydrogen-burning stars to host them for {$10$} trillion years \citep{Laughlin97}, much less more exotic remnants \citep{Adams97}.}
\item The mean fraction of societies within the metasociety that are communicative, $\zFC$.
\item Other parameters describing its internal distribution of communicative societies, $\zComSocDistParams$, and metasociety-wide properties of broadcasts $\zBcDistParams$. Most obviously, this can include something that governs the number of communicative societies, like the rate of appearance of new societies per star. A metasociety necessarily has at least one society {over its history}, but the number of communicative societies may be zero if most societies are unable or unwilling to make broadcasts. For broadcasts, this could include a typical energy scale or typical rate per society. {These are all lumped under a residual tuple, $\zParamsOther$.}
\end{itemize}

{Under the assumption that metasocieties {are defined by these specific} properties, the adopted} metasocietal distribution {has} the form:
{
\begin{equation}
{\zgDist (\zTuple | \gTuple)} = \frac{d^7 \Mean{{\zgN}}}{d\zTStart d\zDuration d\zPosition d\zSize d\zFC d\hParamsMetaOrigin d\zParamsOther} {(\zTuple | \gTuple)}.
\end{equation}}
Depending on scenario, metasocieties may interact in complicated ways, so the metasocietal point process ${\zgSample(\gTuple)}$ can be far from Poissonian {(see discussion in Section~\ref{sec:MetaScenarios})}.

{Although most of these parameters are regarded as single numbers, the framework is sufficiently general that they can be replaced with {subtuples}, expanding the haystack with more dimensions. Metasocietal size is treated here as a fixed value, if only as a placeholder because the volume does not come into play in this paper. While interstellar expansion can be quick on cosmological timescales, it still takes time and we could catch it in progress \citep{Zackrisson15}. To account for that, we could replace ``size'' with parameters describing the velocity of the expansion front \citep{Jones81}, stellar diffusion parameters \citep{CarrollNellenback19}, initial low filling factors due to percolation \citep{Landis98}, and so on. The origin star's parameter tuple could be expanded to model the star's trajectory through the galaxy, including its initial velocity and perhaps artificial propulsion \citep{Badescu06}, if we regard its position as remaining important. The communicative fraction $\zFC$ could be replaced with parameters for a function describing communicativeness as a function of societal age \citep{Sagan73}. Many such complications can be postulated, and the distribution altered to treat them.}

\subsection{Societies}
\label{sec:Societies}

In this series, a society is considered a localized, independent entity with its infrastructure.\footnote{\vphantom{T}{Though ``society'' may suggest distinct individuals cooperating {toward} common goals, the concept applies even if these (inter)planetary entities are solitary beings, group minds, ecospheres, or even large collections of {noninteracting} intelligences.}} A communicating society is one that emits broadcasts. In this view, a ``society'' does not just include sentient beings, but the technology they produce. Thus, a society may produce broadcasts long after its inhabitants have perished; as many have noted, the longevity of technosignatures themselves determines the number of detectable ETIs (e.g., \citealt{Carrigan12,Balbi21}{; \citealt{Wright22}}). Conversely, a communicating society may effectively disappear when it loses interest in broadcasting, even while the society itself thrives (\citealt{Sagan73}; {\citealt{Smart09}}).

Classically, SETI considers single planetary systems or worlds as unitary bodies, naturally imposing a discreteness. {The} society {level could} be much finer, {though,} including individual interstellar vehicles and inhabited interstellar objects{, or distinct organizations on a single world.} The most relevant factor {here} is the {induced clumping} of broadcasts; a vast number of interstellar transmitting entities could reduce the Poisson fluctuations of inhabited {galaxies}. {However,} there could plausibly be more {``levels,''} each imposing {its} discreteness effects, requiring more complex models.\footnote{{W}hat are called ``civilizations'' or ``societies'' in SETI are planetary metasocieties; all the diverse cultures, institutions, and peoples of a planet are amalgamated into one entity. This single entity could have a much longer lifespan than any society as sociologically defined. But a single world {or planetary system} can be far more tightly integrated than an entire galaxy{, so there is motivation to treat them as one unit}.}

Each society starts at time $\aTStart$ and {survives for} a finite duration $\aDuration${.} {It is located at $\aPosition$; when considering very long timescales, we can instead use parameters describing its trajectory.} Thus {I adopt 
\begin{equation}
{\azDist (\aTuple | \zTuple)} = \frac{d^4 \Mean{{\azN}}}{d\aTStart d\aDuration d\aPosition d\aParamsOther} {(\aTuple | \zTuple)}
\end{equation}
as the societal distribution for a metasociety ${\MetaMark}$.} {Any other parameters, including those describing a society's broadcast population, are collected into $\aParamsOther$.} {As with metasocieties, individual parameters listed here can be replaced with sets of parameters (e.g., describing the society's trajectory instead of just its position).} If a metasociety has more than one society, the appearance of those societies is assumed to be Poissonian. Furthermore, all the properties of the societies are assumed to be independent of one another, including their broadcast distributions.

Long-lived interstellar metasocieties are plausibly in an equilibrium state, motivating a single $\azAbund$ for each one. If ETIs {tend to fill} available habitats, then they can do so in much less than 1 Gyr. Sustainability arguments suggest that metasocieties necessarily have to self-regulate to avoid resource depletion \citep{Fogg87,HaqqMisra09}. The reality may be more complicated than a simple Poissonian equilibrium process, however, with many hierarchical levels of organization that this model neglects. Perhaps metasocieties are highly chaotic, with a turbulent series of spikes, booms, plateaus, and crashes. Metasocieties that consist of several {submetasocieties} knit together by messages in relics could be extremely intermittent. Percolation hypotheses suggest that metasocieties may be spatially inhomogeneous as well \citep{Landis98}, {albeit} subject to stellar mixing \citep{Wright14-Paradox} {and societal turnover \citep{Wiley11}}. Without interstellar travel, even a multitude of independent transmitting programs is ``clumped'' onto the rare broadcasting host worlds that presumably dominate the Poissonian fluctuations. More detailed models are necessary to treat effects like these.

\subsection{Scenarios for metasocietal evolution}
\label{sec:MetaScenarios}

\subsubsection{The ``classical'' scenario: metasocieties as societies}
The default scenario in SETI theory is that societies are {confined to one solar system}, surviving for a continuous span of time before disappearing forever without replication \citep[e.g.,][]{Oliver71}. Thus, each metasociety covers just one {stellar system} and contains just one society, which may or may not be communicative. Each (meta)society is regarded as appearing and behaving independently of the others. The adopted metasocietal distribution function is
{
\begin{multline}
\label{sec:MetaDistClassic}
{\zgDist (\zTuple | \gTuple)} = \fDirac(\zPosition - \hPositionMetaOrigin) \fDirac(\zSize) {\gPDF{\zDuration, \zFC, \zParamsOther}} \\
\cdot~  \zhRateTotal(\zTStart | \hParamsMetaOrigin) {\hgDist}(\hParamsMetaOrigin {| \gTuple})  .
\end{multline}
The metasociety thus has zero size ($\zSize = 0$) and its position is identical to that of the star it originated from. The number of metasocieties is proportional to the stellar distribution, but it is modulated by the number rate of metasocieties appearing around each star {at} a given time after its birth. That term includes dependence on the stellar parameter tuple -- factors like the star's position in a galaxy, mass, and metallicity might all affect the prevalence of metasocieties. Note also that the {delay-time} distribution does not generally integrate to $1$; it is vastly less if ETIs are rare, but can be more than $1$ if technological societies evolve several times in a given planetary system \citep{Wright18-PITS,Schmidt19}.\footnote{A simple {societal} intensity distribution does not contain enough information to model the clustering of metasocieties that happens when they evolve multiple times on rare worlds (e.g., those with complex multicellular life). {An additional level for inhabited worlds can be added to the tree to address that.} Nor does equation~\ref{sec:MetaDistClassic} account for what happens if subsequent metasocieties around the same star overlap; it is presumed that their {lifespans are} much shorter than the gap time between them.}}

{With only one society per metasociety, there is no coordination or dependence between different societies.} Whether or not a metasociety{'s single society} is communicative is random:
\begin{equation}
{\azN(\zTuple)} \sim \Bernoulli({\zFC}) ,
\end{equation}
{with a probability ${\zFC}$ {determined by $\zTuple$}, the ``communicative'' factor in the Drake equation.} {The distribution of metasocieties (and societies) is entirely set by galactic properties. For a realized galaxy they are treated as a Poissonian process because of this independence} \citep[as in][]{Glade12}: 
\begin{gather}
\nonumber {\zgNGen(\gTuple)} \sim \Poisson(\Mean{{\zgNGen}})\\
{\agNGen(\gTuple)} \sim \Poisson(\Mean{{\agNGen}}) \sim \Poisson({\gMeanGen{\zFC}} \Mean{{\zgNGen}}) ,
\end{gather}
{where the marking and thinning of societies by their communicativeness preserves the Poisson character.} Communicative societies inherit the lifespan and origin point of their parent metasociety, leading to a societal distribution 
{
\begin{equation}
{\azDist (\aTuple | \zTuple)} = {\zFC} \fDirac(\aTStart - {\zTStart}) \fDirac(\aDuration - {\zDuration}) {\fDirac(\aPosition - {\zPosition})} {\zPDF{\aParamsOther}} .
\end{equation}
}

As long as ${\gMean{\zDuration}}$ is much shorter than the {galaxy}'s evolutionary timescales,
\begin{gather}
\nonumber \Mean{{\zgNTime} (\TimeVar)} \approx {\zgRate}(\TimeVar) {\gMean{\zDuration}} {\Mean{\hgNTime(\TimeVar)}}\\
\Mean{{\agNTime} (\TimeVar)} \approx {\zgRate}(\TimeVar) {\gMean{\zDuration} \gMean{\zFC} \Mean{\hgNTime (\TimeVar)}}  .
\end{gather} 
Thus, ${\zgAbund \approx \zgRate \gMean{\zDuration}}$. {The rate and abundance themselves depend on factors like {star formation} rate and the delay-time distribution (Section~\ref{sec:Drake}).}

\subsubsection{Galactic clubs as metasocieties}
In the galactic club scenario, the first communicative society successfully contacts the succeeding ones through either remote transmissions or {nonreplicating} probes, establishing norms and protocols (\citealt{Bracewell75}). The protocols become locked in, with each new communicative society conforming to the precedent set by the previous ones. The implicit coordination results in a single galaxy-spanning metasociety, even though the societies themselves remain planetbound, aside possibly from some automated probes. {T}he {classical} scenario {is a special kind of galactic club} scenario; it simply interprets ``metasociety'' differently.

{Much of} the metasocietal distribution function is a formality. A metasociety's size and position are simply those of the host galaxy. A galactic club is born with the first society, its origin star is the site of that society, and it endures for $\zDurationInfty$, a{n arbitrary} timescale longer than the current age of the {galaxy}. The number of metasocieties {in a given host galaxy} is a Bernoulli random variable, {simply determined by the societies:} $\fpP({\zgNTime(\TimeVar)} = 1) = \fpP({\agNHist(\TimeVar)} \ge 1)$. {The way the metasociety influences its children is through a single metasociety-wide $\zFC$, and the distribution of any additional parameters in $\zParamsOther$ describing shared societal and broadcast properties.}\footnote{\vphantom{F}{For example, perhaps each society broadcasts only at a frequency assigned by the metasociety according to some distribution. Or perhaps the ``culture'' of the galactic club can bias the lifespans of its member societies.}}

The galactic club metasociety is {therefore largely} an emergent phenomenon, {considered here} only as a mechanism to coordinate societal and broadcast properties. Societies are the fundamental entities {driving the} galactic club. {Despite the seeming circular dependence between the metasociety and its hosted societies, in practice, the dependence is one-way for any single parameter. The births of societies are entirely independent of each other, so if we restrict the societal haystack to just $\aPosition$ and $\aTStart$, a realized host galaxy's population is a Poisson point process. Meanwhile, the metasociety's $\zFC$ and $\zParamsOther$ are chosen and then used to mark each society with its other parameters. The societal haystack is then expanded into the remaining dimensions, and thinned according to which ones are communicative. Finally, the origin of the metasociety is identified with the first (communicative) society.}

{When we consider a host galaxy with fixed properties and conditionalize on any metasocietal parameters governing societal/broadcast distributions, the societies are modeled with a Poisson point process:}
\begin{equation}
{\agNGen(\gTuple | \zParamsOther)} \sim \Poisson(\Mean{{\agNGen | \zParamsOther}}) .
\end{equation}
{The societal distribution includes a dependence on the stellar distribution, because they originate around stars:}
\begin{multline}
\label{eqn:GalacticClubSocDist}
{\agDist (\aTuple | \gTuple, \zParamsOther)} = {\gPDF{\aDuration, \aParamsOther | \zParamsOther}} \\
\cdot~\int_{\hHaystack} \fDirac(\aPosition - {\hPosition}) \ahRateTotal (\aTStart | {\hTuple, \zParamsOther}) {\hgDist}({\hTuple | \gTuple, \zParamsOther}) d\hTuple.
\end{multline}

{Actual galactic metasocieties may be more complicated, even in the absence of interstellar travel. \citet{Forgan17} argues that instead multiple ``galactic cliques'' would arise, although they may in turn contact each other and fuse through a long period of inter-negotiation. There may in fact be a patchwork of different technosignature footprints in a galaxy, on intermediate spatial scales.}\footnote{This might also be true of expansive metasocieties{.}}

\subsubsection{Expansive metasocieties}
Common interstellar travel allows for galaxy-spanning metasocieties. In an expansive metasociety, the metasociety spreads throughout a galaxy in a {(cosmologically)} {minuscule} time, treated as instantaneous here.\footnote{\vphantom{I}{If a galaxy is seeded with life by direct panspermia \citep{Crick73}, in turn evolving ETIs, it could be viewed as hosting an expansive metasociety that takes billions of years to develop.}} The technosignature population of the galaxy effectively undergoes a ``phase transition'' \citep[see][]{Cirkovic08}. Only one metasociety is allowed at a time in a galaxy. {Like the galactic club, the size and position of the metasociety is that of the galaxy itself.} A metasociety may persist once established as internal migration reestablishes societies in locations that have fallen, in which case we can set ${\gPDF{\zDuration}} = \fDirac(\zDuration - \zDurationInfty)$. {Or}, perhaps some internal process causes them to vanish in mere millions of years or less \citep[as in][]{HaqqMisra09,Prantzos20}. Note this lifespan {is likely} far greater than that of individual societies; a galaxy-wide catastrophe is required for a galaxy-spanning metasociety to collapse.

{Each metasociety arises from a single stellar system, but because only one metasociety exists at a time, the realized rate can have a complicated dependence. If a new society evolves independently in an already established metasociety, there are many possible outcomes: it could be assimilated into it without affecting its properties appreciably, it could ``rejuvenate'' a metasociety and extend its lifespan, or perhaps it could subsume the extant metasociety with its own. There could be different metasocieties coexisting, or aggressive metasocieties could inhibit the evolution of intelligence on all other planets in the galaxy, and so on. Each of these affects the metasocietal distribution in different ways. The key point is that this interaction induces a \emph{dependent} thinning on the metasocietal point process.}

{A fairly simple subscenario is one where metasocieties are exclusive and inhibitory: once a metasociety appears, it persists without interference for its entire lifespan, then vanishes. No other metasociety can arise during its reign. This means ${\zgNTime(\zTStart, \gTuple)}$ is a Bernoulli variable, $1$ if the galaxy has a metasociety at {$\TimeVar = \zTStart$} and $0$ if it is empty.} To model {these metasocieties}, define an auxiliary distribution {describing their appearance rate in the absence of inhibition, if ${\zgNTime(\zTStart)}$ just happened to be $0$:
\begin{multline}
\label{eqn:ExpansiveDist}
{\zgDistEMPTY (\zTuple | \gTuple)} = \fDirac(\zPosition - {\gPosition}) \fDirac(\zSize - {\gSize}) {\gPDF{\zDuration, \zFC, \zParamsOther}} \\
\cdot~\zhRateTotalEMPTY(\zTStart | \hParamsMetaOrigin) {\hgDist}(\hParamsMetaOrigin {| \gTuple})  .
\end{multline}
with $\zhRateTotalEMPTY = d{\gMean{\zhN | \zgNTime(\zTStart) = 0}}/d\zTStart$. In an unpopulated galaxy, metasocieties originate around stars much like they do in the classical scenario. {Without any interaction, their population is Poisson given the galaxy's properties.} Hence, there is again a stellar distribution function and a rate per star. This counterfactual distribution can be marginalized to get an effective rate of appearance ${\zgRateTotalEMPTY} \equiv d\Mean{{\zgN | \zgNTime (\zTStart) = 0}}/d\zTStart$. If extant metasocieties have lifespans unaffected by any independent ``second origins'' in their domains, then the distribution function is
\begin{equation}
{\zgDist (\zTuple | \gTuple)} = (1 - \Mean{{\zgNTime (\zTStart)}}) {\zgDistEMPTY(\zTuple | \gTuple)} .
\end{equation}
It is found by integrating ${\zgDist}$ with origin time $\zTStart$ restricted by duration $\zDuration$ ($\TimeVar - \zDuration \le \zTStart \le \TimeVar$; section~\ref{sec:LifespanBias}). Differentiating by $\zTStart$ turns this into a partial differential equation.}

{A few simple results are evident. When every expansive metasociety is persistent, $\zDuration \to \zDurationInfty$. A galaxy then has an expansive metasociety if and only if one has ever appeared {any time} during its history. The only way it lacks one is if no society develops during any interval of time. Because the first society {in a realized galaxy} arises according to a Poisson process, the time until one appears has an exponential distribution governed by ${\zgRateEMPTY \equiv \zgRateTotalEMPTY / \Mean{\hgNTime(\zTStart)}}$. The mean number of metasocieties is
\begin{equation}
\label{eqn:zNSysExpansiveApprox}
\Mean{{\zgNTime}(\TimeVar)} = 1 - \exp\left(-\Mean{{\hgNTime}(\TimeVar)} {\zgAbundEMPTY} (\TimeVar {| \gTuple}) \right) ,
\end{equation}
defined in terms of the effective abundance per star
\begin{equation}
{\zgAbundEMPTY} (\TimeVar {| \gTuple}) = \frac{1}{\Mean{{\hgNTime}(\TimeVar)}} \int_{{\gTStart}}^{\TimeVar} {\zgRateTotalEMPTY}(\zTStart {| \gTuple}) d\zTStart,
\end{equation}
in turn using the effective rate found by marginalizing ${\zgDistEMPTY}$.}

{But what if metasocieties are short-lived compared to the galaxy's age? If ${\zgRateEMPTY}(\zTStart)$ is roughly constant {given a $\gTuple$}, then the evolution should be described by 
\begin{equation}
\frac{d\Mean{{\zgNTime}{(\TimeVar)}}}{d\TimeVar} \approx {\zgRateEMPTY} \Mean{{\hgNTime}} (1 - \Mean{{\zgNTime (\TimeVar)}}) - \frac{\Mean{{\zgNTime(\TimeVar)}}}{{\gMean{\zDuration}}},
\end{equation}
with the first term on the right side describing the modified appearance rate, and the second term describing their disappearance. The natural equilibrium number of metasocieties is
\begin{equation}
\label{eqn:zNSysExpansiveDutyCycle}
\Mean{{\zgNTime(\TimeVar)}} = [1 + ({\zgRateEMPTY} \Mean{{\hgNTime (\TimeVar)}} {\gMean{\zDuration}})^{-1}]^{-1} .
\end{equation}
}

{The convergence to $1$ in equation~\ref{eqn:zNSysExpansiveDutyCycle} is slower because all galaxies start with ${\zgNTime(\TimeVar)} = 0$. If ${\gMean{\zDuration}}$ is long on cosmological scales, then there has not been time to reach the second ``gap'' between metasocieties -- each galaxy has either avoided having any or is in the era of its first metasociety. In contrast, if ${\gMean{\zDuration}}$ is short, then each galaxy has already had a random sequence of metasocieties and gaps. Equation~\ref{eqn:zNSysExpansiveDutyCycle} thus describes the duty cycle of expansive metasocieties once the transient initial condition has vanished. I adopt equation~\ref{eqn:zNSysExpansiveApprox} unless otherwise stated, but to {order of magnitude} the abundance of metasocieties should be the same.}

Expansive metasocieties differ from galactic clubs in that the number of communicative societies saturates at a very high equilibrium level once they become established -- potentially in the billions or more. We have a metasociety-level societal distribution of 
\begin{multline}
{\azDist(\aTuple | \zTuple) = \azRate(\aTStart | \zTuple)} \zPDF{\aDuration, \aParamsOther} \\
{\frac{d\Mean{{\hzNTime(\aTStart)}}}{d\hPosition} (\aPosition {| \zTuple}).}
\end{multline}
Furthermore, ${\azRate}$ is not only expected to be larger than but \emph{independent} of ${\zgRateEMPTY}$. The number of communicative societies in an expansive metasociety is treated separately. {They form a Poisson point process, dependent on the metasociety's properties:}
\begin{equation}
{\azNGen(\zTuple)} \sim \Poisson(\Mean{{\azNGen}})
\end{equation}
with $\Mean{{\azNGen}} \approx {\azRate \zMean{\aDuration} \Mean{\hgNGen}}$ when $\oDurationGen \ll \aDuration$. This means ${\agNTime (\TimeVar)}$ has a mixture distribution, with a $(1 - \Mean{{\zgNTime{(\TimeVar)}}})$ probability of being $0$ and a $\Mean{{\zgNTime(\TimeVar)}}$ probability of being ${\azNTime (\TimeVar)}$, yielding:
\begin{equation}
\fpP({\agNTime (\TimeVar)} \ge 1) = \Mean{{\zgNTime (\TimeVar)}}[1 - \exp(-{\gMean{\azNTime(\TimeVar)}})] . 
\end{equation}
The independence results in interesting effects: in particular, when $\Mean{{\zgNTime (\TimeVar)}} \ll 1$, the probability of detecting a broadcast can vary quadratically with the number of stars in a {galaxy} (Section~\ref{sec:WhichHostGalaxies}).

{The qualitative characteristics of galactic clubs and persistent expansive metasocieties converge when $\Mean{{\zgN}}$ approaches $1$. In both cases, metasociety-wide traits can be treated as galactic traits and societies form a Poisson point process in {a realized galaxy}. A galactic club can even be regarded as an expansive metasociety that spreads through telecommunications instead of settlement, albeit with much lower population densities in the absence of replication. A single metasociety approximation, with ${\zgN} = 1$, serves to work for both scenarios when ETIs are not very rare.}

\subsection{The Drake Equation in the formalism}
\label{sec:Drake}
The Drake equation applies to the classical scenario when the galaxy is in {a steady state} \citep{Drake65,Glade12}. New stars are formed at a constant rate, and as they age {over} billions of years, some fraction of those host single ETI societies of limited lifespan. To implement it simplistically, if an ETI evolves, it {must} appear {exactly} $\hTstarBAR$ after the star's birth: the {delay-time} distribution is 
\begin{equation}
\zhRateTotal = \hFeti \fDirac((\zTStart - \hTStart) - \hTstarBAR) .
\end{equation}
Stars need to maintain a stable luminosity long enough for intelligence to evolve on Earth-like planets, excluding massive stars. {However,} let us imagine for simplicity that all stars with $\hM \le \hMCUTHI$ are equally likely to host an ETI. This probability $\hFeti$ includes all terms in the Drake equation relating to the number of planets, habitability, and the evolution of life and intelligence. Then we can apply the above to the classical metasociety distribution (equation~\ref{eqn:ExpansiveDist}).

We want to calculate the number of communicative societies existing at any one time in the {galaxy}. First, {let us} calculate the {galactic}-level societal distribution, using equation~\ref{eqn:AncestorRelDist}. We convolve $\azDist (\aTuple | \zTuple)$ with ${\zgDist} (\zTuple {| \gTuple})$. Directly plugging these into that equation yields a formidable integral over all stellar and metasocietal parameters. Actually, most of the variables are effectively nuisance parameters, and the expression is simplified by the many delta functions and the separable integrals. In the end, we are left with
\begin{multline}
{\agDist (\aTuple | \gTuple)} = {\gMean{\hFeti}} {\gMean{\zFC} \gPDF{\hPositionMetaOrigin}(\aPosition) \gPDF{\zDuration}(\aDuration)} \\
\cdot~{\gPDF{\aParamsOther}}  {\gRateStar(\zTStart - \hTstarBAR)} \CDF{\hMMetaOrigin}(\hMCUTHI) .
\end{multline}
In other words, the communicative societies trace the stars in space, each existing for the lifespan of its metasociety, and they appear at a rate proportional to the {star formation} rate $\hTstarBAR$ ago, including only stars that last long enough for ETIs to evolve.

The {expected} number of currently active communicative societies {$\Mean{{\agNTime} (\TimeVar)}$} is {the integral of ${\agDist}$ with $\TimeVar - \aDuration \le \aTStart \le \TimeVar$ (section~\ref{sec:LifespanBias}). Now let us suppose that the {star formation} rate of the galaxy has been constant, that (meta)societies are always much younger than the galaxy, and that $\CDF{\hMMetaOrigin}(\hMCUTHI) \approx 1$ because most stars are low mass and likely to survive for the required billions of years. Then we finally get,}
\begin{equation}
\label{eqn:DrakeFormal}
\Mean{{\agNTime} (\TimeVar)} \approx {\gRateStar} {\gMean{\hFeti}} {\gMean{\zFC} \gMean{\zDuration}} .
\end{equation}
This is the Drake equation rewritten in the formalism.

{Including different {delay-time} distributions, galactic habitability, dependence on stellar mass, and other effects is fairly straightforward. We simply change the $d{\gMean{\zNStar}}/d\ztStarDelay$ distribution to include dependencies on these parameters of the host star. A spread in delay times is necessary} when treating quiescent galaxies where the star formation rate {was} quenched about {$10$} billion years ago.

\section{Broadcasts: General considerations}
\label{sec:BroadcastConsiderations}

{A} broadcast {is} an artificial release of energy at a discrete site through some specified time and frequency range (see Table{s~\ref{table:SharedObjectNotation}{--}\ref{table:EmissionNotation}} for notation). Not every technosignature is a broadcast: unpowered artifacts in the {solar system} (\citealt{Freitas85}; {\citealt{Rose04};} \citealt{Davies13}) and anomalous atmospheric compositions resulting from industrial pollution \citep{Whitmire80,Lin14} are not broadcasts. On the other hand, not every broadcast need be an attempt at communication -- {they include} directed energy transmission {for} power beaming \citep{Inoue11,Benford16}, propulsion \citep{Lingam17}{,} or remote sensing \citep{Scheffer14}. Broadly speaking, waste heat from megastructures \citep{Dyson60} and exhaust radiation from vehicles \citep{Harris86} {are broadcasts too}. 

{The modulation of starlight by astronomical-scale megastructures (\citealt{Arnold05,Chennamangalam15}; {\citealt{Wright16};} \citealt{Zackrisson18}{; \citealt{Lacki19-Sunscreen,Suazo22}}) is an interesting case -- here, the technosignature is the blocking of an energy release that normally would happen. The{se} might be viewed as negative energy broadcasts. Searches for galactic-scale obscuration of starlight \citep{Annis99,Zackrisson15} {use} a collective bound as in Paper II, much as waste heat searches look for the positive energy release from the {reprocessing} of this missing starlight.}

\subsection{Basic parameters to describe broadcasts}
\label{sec:BcParameters}
{The} basic considerations {for} whether a {{broadcast is detectable} are where it is, how bright it becomes, when it happens, and how it behaves. The first question is answered by the broadcast position $\bPosition$.}

Fundamentally, the brightness is controlled by the total energy released $\bEactual$ and its distribution into the transmitter's sky {over solid angle $\bSolidAngle$ (refer to Table~\ref{table:EmissionNotation} for emission-related variables)}. These can be considered separately, but I use the total effective isotropic energy $\bEiso \equiv 4 \pi d\bEactual/d\bSolidAngle$ {of} the broadcast to describe the brightness{, evaluated toward the direction of the observer}. In the absence of beaming, $\bEiso = \bEactual$. When broadcasts are beamed, the $\bEiso$ distribution has a large peak near zero for off-axis broadcasts. Yet even on-axis beamed broadcasts may have a wide range of $\bEiso$ because of different beaming angles or simply different $\bEactual$.  Previous SETI works have essentially considered two basic classes of $\bEiso$ distributions -- the monoenergetic $\fDirac$-distribution with a characteristic $\bEiso$ and {power-law} distributions (e.g., {\citealt{Drake73-L};} \citealt{Gulkis85,Dreher04}).  Both narrow and broad distributions seem plausible at this point, with the former favored by possible engineering constraints on deliberate transmissions and the latter {motivated} by the possible diversity of ETI capabilities and goals. Each broadcast's emission may {also} have distinct degrees and states of polarization. The {polarization} properties {are} encapsulated in $\bPolQuantity$, which should be understood as a Stokes vector or similar representation.

Surveys are sensitive to particular kinds of time-frequency properties -- spectral lines or pulses, for example. Although the time-frequency behavior may be endlessly complex, I model {it with} five basic parameters. Broadcasts are limited in time, with a source-frame duration $\bDuration$, and {in} frequency, with a {constant} source-frame instantaneous bandwidth $\bBandwidthTime$. Broadcast{s} also start at times $\bTStart$ and are centered on source-frame frequencies $\bNuMid$. Finally, a drift rate $\bDriftRate$ describes the {skewness} of the broadcast in {the} time-frequency space. The broadcast is entirely contained in the time range $[\bTStart, \bTStart + \bDuration]$ and frequency range $[\bNuMid - \bBandwidth/2, \bNuMid + \bBandwidth/2]${, where the total bandwidth $\bBandwidth$ is derived from $\bBandwidthTime$ and $\bDriftRate$ {($\bBandwidth = \bBandwidthTime + \bDuration |\bDriftRate|$ when they are constant)}.}

{These quantities define a haystack in which broadcasts are scattered. Most of them match the quantities defining haystacks in \citet{Tarter07} and \citet{Wright18}. A separate modulation parameter is missing here, not being directly relevant for energy or photon detection, but bandwidth and drift rate both can be considered kinds of modulation. The main differences are in the treatment of time, reduced to a single quantity in previous works. \citet{Wright18} {interpret} this as a repetition period, while the duty cycle and the longevity of the transmitter are ignored. Including periodicity would necessitate adding at least one more dimension to the ones here (see discussion in Section~\ref{sec:RandomBc}). Bursts repeating at random intervals \citep{Kipping22}, however, require no new dimensions -- the mean rate is simply considered a property of the hosting society.}

\subsection{Selection of broadcasts}

{When is a broadcast sampled? Primary considerations are whether a broadcast ``touches'' the window in physical spacetime (Section~\ref{sec:LifespanBias}) and frequency, with a ``cross section'' set by a duration and bandwidth.} A broadcast has an effective bandwidth $\bBandwidthTimeOFGen$ during window ${\GenLabel}$, a quantity {accounting for a broadcast's spectral} evolution. In some cases, the {broadcast's spectrum is unchanging} ({like} the box model; section~\ref{sec:BoxModelIntro}), in which case we can simply use $\bBandwidthTimeOFGen = {\bBandwidthTime}$. {For the purposes of selection,} I adopt
\begin{equation}
\label{eqn:bBandwidthEff}
\bBandwidthTimeOFGen = \diam\left\{\FreqVar \Big| {\bDerivativeEisoGen} (\TimeVar, \FreqVar | \bTuple) > 0 \right\}
\end{equation}
where $\diam \fpSubset = {\max[}|a_1 - a_2|\,| a_1, a_2 \in \fpSubset{]}$.\footnote{$\bBandwidthTimeOFGen$ and $\bBandwidthGen$ are distinct -- the latter {accounts for the} frequency, sky field, and so on as well in restricting the bandwidth. A broadcast at a frequency never {covered} by ${\GenLabel}$ has $\bBandwidthGen {= 0}$ and a drifting signal that just barely ``touches'' ${\GenLabel}$ has {$\bBandwidthGen \ll \oBandwidthGen$}, but neither situation affects $\bBandwidthTimeOFGen$. {In all cases, $\bBandwidth \ge \bBandwidthTimeOFGen \ge \bBandwidthGen$.}}

{I regard the selection of a broadcast by window ${\GenLabel}$ as a binary decision based on its time and frequency properties:
\begin{multline}
\label{eqn:BcSelectionCriteria}
\obPGen(\bTuple) = \IndicatorOf{-\bDuration \le \bTStart - \oTStartGen \le \oDurationGen}\\
                  \cdot~\IndicatorOf{|\bNuMid - \oNuMidGen| \le (1/2) (\bBandwidthTimeOFGen + \oBandwidthGen)}
\end{multline}
The presumption is that the broadcast's ancestors also are chosen by ${\GenLabel}$, as required (Section~\ref{sec:Samples}), because if the broadcast is active at some time and place, so are all its hosts -- so ${\GenLabel}$ must not impose any filters on characteristics specific to society, metasociety, or {galaxy}. Polarization characteristics are not considered, because of the likely {crosstalk} between the broadcast's polarization and the observed polarizations, at least for linear polarization.}

\subsection{Energy and photons emitted by broadcasts}

{A broadcast is detected through its emission (Table~\ref{table:EmissionNotation}). This can be measured in various forms, including energy and photons.} Each broadcast has an (effective isotropic) energy output per polarization per unit time and frequency
\begin{equation}
\label{eqn:LnuSingle}
\bLnupoliso (\TimeVar, \FreqVar, \PolVar) \equiv \bDerivativeEisoPol (\TimeVar, \FreqVar, \PolVar) = \bFPol(\PolVar) \bLnuiso (\TimeVar, \FreqVar) .
\end{equation}
I assume that the degree and state of polarization {are} independent of time and frequency. Thus, I separate $\bLnupoliso$ into the fraction of energy emitted in polarization $\PolVar$, $\bFPol(\PolVar)$, and the spectral luminosity, $\bLnuiso$.\footnote{For example, a broadcast with linear polarization fraction $\bPolDegreeLin$ has $\bFPol(\PolVarLin) = (1/2) (1 - \bPolDegreeLin) + \bPolDegreeLin \cos^2 {\phi_{\PolVarLin; \BcMark}}$, where ${\phi_{\PolVarLin; \BcMark}}$ is the relative angle between broadcast's polarization angle and $\PolVarLin$.} {The same basic ideas can be applied to broadcasts in other messengers, like neutrinos and gravity waves.}

Of course, we do not collect all of the energy emitted by a broadcast, only that which falls on our instruments within a time and frequency window in set of observed polarizations. Out of a total effective isotropic emission $\bEmissioniso$, the broadcast releases only a limited amount $\bEmissionisoGen$ within these constraints, zero if the broadcast is {not in} the sample. The amount of energy released within the time, frequency, and polarization constraints for ${\GenLabel}$ is
\begin{equation}
\label{eqn:EnergyGeneric}
\bEisoGen = \int_{\oNuMidGen - \oBandwidthGen/2}^{\oNuMidGen + \oBandwidthGen/2} \int_{\oTStartGen}^{\oTStartGen + \oDurationGen} \sum_{\PolVar \in \oPolSetGen} \bLnupoliso (\TimeVar, \FreqVar, \PolVar) d\TimeVar d\FreqVar
\end{equation}
{When the polarization properties are independent of time and frequency,} the sum over polarizations {is replaced} by a single factor {$\bFPolGen$, such that}
\begin{equation}
{\bFPolGen = \sum_{\PolVar \in \oPolSetGen} \bFPol(\PolVar | \bPolQuantity) .}
\end{equation}
Note that the frequency and time variables are all in the {\emph{source frame}}.\footnote{{An object at redshift $\yRedshift$ is observed at Earth to have} $\oNuMidGen = \oeNuMidGen (1 + \yRedshift)$, $\oBandwidthGen = \oeBandwidthGen (1 + \yRedshift)$, and $\oDurationGen = \oeDurationGen / (1 + \yRedshift)$.} 

When dealing with photon detectors, the relevant quantity is the number of photons emitted within the observational window:
\begin{equation}
\label{eqn:PhotonGeneric} 
\bPhotonisoGen \equiv \int_{\oNuMidGen - \oBandwidthGen/2}^{\oNuMidGen + \oBandwidthGen/2} \int_{\oTStartGen}^{\oTStartGen + \oDurationGen} \sum_{\PolVar \in \oPolSetGen} \frac{\bLnupoliso(\TimeVar, \FreqVar, \PolVar)}{h \FreqVar} d\TimeVar d\FreqVar,
\end{equation}
where $h$ is Planck's constant. 

Even though the time domain and frequency domain amplitudes are related by the Fourier transform, $\bLnuiso$ can have extremely complex dependence on time and frequency.\footnote{{Formally,} $\bBandwidth$ and $\bDuration$ {cannot both be} finite; applying a rectangular window to {one} domain convolves the {other} by a $\sinc$ function that is {nonzero} over an infinite range. Many observations integrate over $\oDurationObs \oBandwidthObs \gg 1$, {resulting in negligible} leakage. {These include all practical observations in the optical {or higher energies}, and} ``filterbank'' data products in radio SETI {which also} employ polyphase filterbank techniques to reduce sidebands \citep{Price16}.} This motivates the later use of the box and chord models as simplifications.

{
\begin{deluxetable*}{lccccc}
\tabletypesize{\footnotesize}
\tablecolumns{6}
\tablewidth{0pt}
\tablecaption{{Notation for emission quantities \label{table:EmissionNotation}}}
\tablehead{\colhead{Quantity} & \colhead{Arbitrary} & \colhead{Energy} & \colhead{Power} & \colhead{Photons} & \colhead{Note}}
\startdata 
Emission from broadcast                                          & $\bEmission$         & $\bEactual$           & $\bLactual$        & $\bPhotonactual$      & a, b\\
Effective isotropic emission from broadcast                      & $\bEmissioniso$      & $\bEiso$              & $\bLiso$           & $\bPhotoniso$         & b\\
\ldots per unit frequency                                        & $\bEmissionnuiso$    & $\bEnuiso$            & $\bLnuiso$         & $\bPhotonnuiso$       & c\\
\ldots per polarization $\PolVar$                                & $\bEmissionpoliso$   & $\bEpoliso$           & $\bLpoliso$        & $\bPhotonpoliso$      & c\\
\ldots per unit frequency per polarization $\PolVar$             & $\bEmissionnupoliso$ & $\bEnupoliso$         & $\bLnupoliso$      & $\bPhotonnupoliso$    & c\\
\ldots {occurring} during window ${\GenLabel}$ & $\bEmissionisoGen$   & $\bEisoGen$           & $\bLisoGen$        & $\bPhotonisoGen$      & c\\
Effective isotropic aggregate emission from broadcast population selected by $\Selection{\GenLabel}{{\JMark}}$ 
                                                                 & ${\bjjAggRisoGen}$      & ${\bjjAggEisoGen}$       & ${\bjjAggLisoGen}$    & ${\bjjAggQisoGen}$       & b\\
Effective isotropic background emission in window ${\GenLabel}$    
                                                                 & $\rAggRisoGen$       & $\rAggEisoGen$        & $\rAggLisoGen$     & $\rAggQisoGen$        & b\\
Effective isotropic total emission in window ${\GenLabel}$ from object ${{\JMark}}$
                                                                 & ${\tjjAggRisoGen}$      & ${\tjjAggEisoGen}$       & ${\tjjAggLisoGen}$    & ${\tjjAggQisoGen}$       & b\\
\hline
Emission distance                                                & $\ybDistanceR$       & $\ybDistanceE$        & $\ybDistanceL$     & $\ybDistanceQ$        & \\
Dilution factor for broadcast                                    & $\lDilutionR$        & $\lDilutionE$         & $\lDilutionL$      & $\lDilutionQ$         & \\
{Transmittance factor within window ${\GenLabel}$}          
                                                                 & {$\lTransmittanceRGen$} & {$\lTransmittanceEGen$} & {($\lTransmittanceEGen$)} & {$\lTransmittanceQGen$}\\
\hline
Fluence (flux) from single broadcast in window ${\GenLabel}$       
                                                                 & $\lFluenceGen$       & $\lFluenceEGen$       & $\lFluxEGen$       & $\lFluenceQGen$       & d\\
Aggregate fluence (flux) from population of broadcasts selected by $\Selection{\GenLabel}{{\JMark}}$ 
                                                                 & ${\mjjFluenceGen}$      & ${\mjjFluenceEGen}$      & ${\mjjFluxEGen}$      & ${\mjjFluenceQGen}$      & d\\
Background fluence (flux) in window ${\GenLabel}$       & $\kFluenceGen$       & $\kFluenceEGen$       & $\kFluxEGen$       & $\kFluenceQGen$       & d\\
Total fluence (flux) including background in window ${\GenLabel}$ from object ${{\JMark}}$     
                                                                 & ${\qjjFluenceGen}$      & ${\qjjFluenceEGen}$      & ${\qjjFluxEGen}$      & ${\qjjFluenceQGen}$      & d\\
\hline
Instrument-measured quantity for single broadcast in window ${\GenLabel}$               
                                                                 & $\lMeasureGen$       & $\lEnergyGen$         & $\lPowerGen$       & $\lPhotonGen$         & \\
Aggregate measured quantity from population of broadcasts selected by $\Selection{\GenLabel}{{\JMark}}$        
                                                                 & ${\mjjMeasureGen}$      & ${\mjjEnergyGen}$        & ${\mjjPowerGen}$      & ${\mjjPhotonGen}$        & \\
Background measured quantity in window ${\GenLabel}$    & $\kMeasureGen$       & $\kEnergyGen$         & $\kPowerGen$       & $\kPhotonGen$         & \\
Total instrument-measured quantity in window ${\GenLabel}$
                                                                 & $\qMeasureGen$       & $\qEnergyGen$         & $\qPowerGen$       & $\qPhotonGen$         & \\
\enddata
\tablenotemark{a}{When no quantity window is given, the ALL window is assumed to apply; so ${\bEiso}$ is the total effective isotropic energy released over the entire lifespan, frequency range, and all polarizations from broadcast ${{\BcMark}}$.}
\tablenotetext{b}{The variable listed under power is the effective isotropic radiated power (EIRP).}
\tablenotetext{c}{Any emission variable, aggregate or {singleton}, can be substituted with these modifiers to the same effect. Thus, ${\bjjAggLpolisoGen}$ is the aggregate single-polarization EIRP from the broadcasts in ${\bjjSampleGen}$; $\lFluxEnu$ is the energy flux per unit frequency of a single random broadcast ${\BcMark}$, unfiltered by any window.}
\tablenotetext{d}{The variable listed under power is energy flux; all others are types of fluence.}
\end{deluxetable*}
}

\subsection{Distances, fluences, and dilutions}
Observable quantities depend not just on the intrinsic properties of the broadcast, but where it is {-- its distance ($\ybDistance$) and position {in} the sky ($\lSkyLocation$).} The difference in redshift between broadcasts in the same galaxy is insignificant{, so I set them equal ($\ybRedshift = \ygRedshift$). Variation in broadcast distances} is important {for} the Milky Way, which after all is another {galaxy} with some population of broadcasts (even if {it is} empty aside from our own). For extragalactic systems that are not Galactic satellites, however, the size of the target {galaxy} is negligible compared to its distance, and we can approximate the distance as that to the {galaxy} itself {($\ybDistance = \ygDistance$)}.

Aside from the cuts imposed by the selections, only the emission incident on our detector can be measured. The fluence $\lFluenceGen$ of a broadcast is the amount of emission per unit area integrated over {all combinations of time, frequency, and polarization} {covered by window ${\GenLabel}$}. Specific types of fluence include the energy fluence $\lFluenceEGen$ where the emission is quantified as energy and the photon fluence $\lFluenceQGen$ which counts the expected number of photons (Table~\ref{table:EmissionNotation}). 

The distance determines the flux and fluence of a broadcast with a given amount of emission. The energy fluence can be found using the ``energy distance'' $\yDistanceE = \sqrt{1+\yRedshift} \yDistanceM$ which accounts for redshift, correcting the transverse angular distance $\yDistanceM$ (see \citealt{Hogg99} for how to calculate $\yDistanceM$). Likewise, the photon fluence uses a ``photon distance'' $\yDistanceQ$ equal to $\yDistanceM$.

 {The observed emission is also partly blocked by opacity from dust and gas on the sightline, plus the Earth's atmosphere, a range-limiting issue in optical SETI \citep{Howard04}. Only a fraction $\lTransmittanceRGen$ of $\bEmissionCore$ emission remains after absorption and scattering, the transmittance factor, which can vary with both position and frequency. The frequency dependence means that windows at different frequencies have different transmittances. At the usual radio frequencies observed in SETI, however, $\lTransmittanceRGen$ is very near $1$.} 

It is convenient to express the inverse-square law and redshift effects as a ``dilution {factor,}'' $\lDilutionR = 1/(4\pi {\ybDistanceR}^2)$. For emission of type $\SingleEmissionCore$,
\begin{equation}
\label{eqn:EmissionToFluence}
\lFluenceGen = \frac{\bEmissionisoGen {\lTransmittanceRGen}}{4 \pi {\ybDistanceR}^2} = \bEmissionisoGen {\lTransmittanceRGen} \lDilutionR .
\end{equation}
In principle, the $\bEmissionisoGen$ and $\lDilutionR$ are not independent -- although the intrinsic properties of the broadcast are modeled as independent, the cuts depend on $\oDurationGen$ and $\oBandwidthGen$, which are affected by redshift. Even the Doppler shifts of broadcasts should have a slight effect on $\bEmissionisoGen$. {Transmittance, when it is not near $1$, is likely to be much lower in regions of high obscuration and thus is certainly not independent of $\lDilutionR$.} {However,} for practical purposes, {with} the emission and dilution independent{,}
\begin{equation}
{\jjMeanGen{{\lFluenceGen}^n} = \jjMeanGen{{\bEmissionisoGen}\vphantom{\bEmissionisoGen}^n} \jjMean{{{\lTransmittanceRGen}^n} {\lDilutionR}^n}}
\end{equation}
{for any arbitrary exponent $n$}.

\section{Populations of broadcasts}
\label{sec:BroadcastPopulations}

\subsection{The broadcast distribution function}

In this series, {the broadcast distribution function takes the form}
\begin{equation}
\label{eqn:pDistJ}
{\bjjDist (\bTuple | \jjTuple)} = \frac{d^8 \Mean{{\bjjN}}}{d\bEiso d\bPolQuantity d\bDuration d\bTStart d\bBandwidthTime d\bNuMid d\bDriftRate {d\bPosition}} {(\bTuple | \jjTuple)},
\end{equation}
{using the variables identified in Section~\ref{sec:BcParameters}.} {The broadcast population is a Cox point process, and the broadcast population of a realized society in particular is a Poisson point process.} The instantaneous rate of broadcasts per frequency {per star} is especially useful:
\begin{equation}
{\bjjRatenu} (\TimeVar, \FreqVar) \equiv {\frac{1}{\Mean{{\hjjNTime} (\TimeVar)}} \frac{d^2 \Mean{{\bjjN}}}{d\bTStart d\bNuMid}  (\bTStart = \TimeVar, \bNuMid = \FreqVar)} {,}
\end{equation}
{as is} the frequency abundance {per star}, 
\begin{equation}
{\bjjAbundnu} (\TimeVar, \FreqVar) = {\frac{1}{\Mean{{\hjjNTime} (\TimeVar)}} \frac{d\Mean{{\bjjNTime} (\TimeVar)}}{d\bNuMid} (\bNuMid = \FreqVar) .}
\end{equation}
The {stellar} rate ${\bjjRate}$ and the {stellar} abundance ${\bjjAbund}$ {are additionally marginalized over $\bNuMid$, as defined as in Section~\ref{sec:Marginalization}. It is more convenient to use the total abundances and rates, the derivatives without dependence on $\Mean{{\hjjNTime(\FreqVar)}}$, when considering individual societies around single star systems: ${\baRatenuTotal} \equiv d^2 \Mean{{\baN}}/d\bTStart d\bNuMid$, ${\baAbundnuTotal} \equiv d\Mean{{\baNTime (\TimeVar)}}/d\bNuMid$, ${\baRateTotal} \equiv d\Mean{{\baNFreq(\FreqVar)}}/d\bTStart$, ${\baAbundTotal} \equiv \Mean{{\baNTimeFreq(\TimeVar, \FreqVar)}}$.}

{Specific examples of the broadcast distribution are given in Sections~\ref{sec:BoxModelIntro}~and~\ref{sec:ChordModelIntro}.}

{
\subsection{Interchangeability}
When I compute properties of aggregate observables in this paper, I apply results for compound Poisson random variables. {However,} to do this, the variables being summed need to be identically distributed. Basically, e}ach society is assumed to have the ``same'' broadcast distribution as the others, an ``average'' distribution. Likewise, each metasociety has the ``same'' {societal and broadcast} distribution{s} as any other in the {galaxy}. {{The \emph{interchangeability} assumption is that all distributions of an object type are the same, except for translations in spacetime (because objects are located in different places and start at different times).} The broadcast distribution in every society (regardless of metasociety) {in galaxy ${\GalMark}$} has the form:
\begin{multline}
\label{eqn:InterchangeableBcDist}
\baDist {(\bTuple | \aTuple)} = f_{{\GalMark}}^{\BcMark} (\bEiso, \bPolQuantity, \bDuration, \bBandwidthTime, \bNuMid, \bDriftRate) \\
\cdot \fDirac(\bPosition - \aPosition) \cdot~\IndicatorOf{0 \le \bTStart - \aTStart \le \aDuration},
\end{multline}
and every metasociety has a societal distribution
\begin{equation}
\azDist {(\aTuple | \zTuple)} = f_{{\GalMark}}^{\SocMark}  (\aDuration) \cdot \IndicatorOf{0 \le \bTStart - \aTStart \le \aDuration},
\end{equation}
with fixed functions $f_{{\GalMark}}^{\BcMark}$ and $f_{{\GalMark}}^{\SocMark}$.}

{S}ocieties are selected at different points of their lifecycles{, but for the compound Poisson distribution to be applicable,} the societal broadcast distribution must be {time}-stationary. If, for example, societies lose interest in communicating as they age, then young societies have a higher ${\baRatenu}$ than older ones. {Interchangeability} does not allow this. {Thus there can be no further dependence on $\bTStart - \aTStart$ in equation~\ref{eqn:InterchangeableBcDist}, as long as the broadcast is within bounds. Likewise, the societal distribution is stationary. More generally, there cannot be any edge effects wherein societies or metasocieties spend different amounts of time in the selection window; thus {all} lifespans must either be much longer or much shorter than the window's duration.}

Real societies and metasocieties may be incredibly diverse, each with its own unique array of technosignatures. {Societies and broadcasts are represented by Cox processes (Section~\ref{sec:OtherPointProcesses}). Because these are generally {nonergodic}, a single society or metasociety is a very poor representative of the ensemble properties of the underlying distribution -- we only see one realization of the distribution, which itself is a random variable. If half of societies broadcast in radio and the rest in {gamma rays}, and there is only a single society {that} happens to be radio-broadcasting, we will not be able to ``see'' the true diversity in broadcasts.} The observed sample is {thus} subjected to stochasticity {when there are only a few societies. A} large sample of societies {can sample the} entire gamut of broadcasts. {In that sense, all the broadcasts can} be pool{ed} together into a {galactic} broadcast distribution{, although the effects of the ``clumping'' remain underestimated by the compound Poisson distribution.}

{There are expressions for the mean and variances when dealing with diverse populations, but they are very complicated because of the nesting selection-relative averages and variances (equation~\ref{eqn:RecursiveAverageInequality}). It is more insightful to work directly with the distributions themselves when they are available (section~\ref{sec:NonInterchangeableExample}).} 

Table~\ref{table:ScenarioEqns} lists equations for some important quantities characterizing broadcast populations, given the assumptions of {independence} and {interchangeability}.

\subsection{The number of broadcasts}
The number of broadcasts intercepted by ${\GenLabel}$ from a society ${\SocMark} \in {\ajjSampleGen}$ is {$\baNGen(\aTuple)$, a Poisson random variable given $\aTuple$}.  Its mean is given by the number of broadcasts that ``touch'' the ${\GenLabel}$ window:
\begin{multline}
\label{eqn:MeanNBTouchSociety}
\Mean{{\baNGen}} = \idotsint_{\oNuMidGen-(\bBandwidthTimeOFGen+\oBandwidthGen)/2}^{\oNuMidGen+(\bBandwidthTimeOFGen+\oBandwidthGen)/2} \int_{\oTStartGen - \bDuration}^{\oTStartGen + \oDurationGen }\\
{\baDist} (\bTuple {| \aTuple}) d\bTStart d\bNuMid d\bEiso d\bPolQuantity d\bDuration d\bBandwidthTime d\bPosition d\bDriftRate .
\end{multline}
If {societies are long-lived compared to $\bDuration$, and if} ${\baRatenu}$ varies slowly enough in time and frequency compared to common values of $\bDuration$ and $\bBandwidthTimeOFGen$,
\begin{equation}
\label{eqn:MeanNBSoc}
\Mean{{\baNGen}} \approx {\baRatenuTotal}(\oTStartGen, \oNuMidGen) {\cdot} (\oBandwidthGen + {\aMean{\bBandwidthTimeOFGen}}) (\oDurationGen + {\aMean{\bDuration}}) {.}
\end{equation}

\begin{deluxetable*}{l|c|c}
\tabletypesize{\footnotesize}
\tablecolumns{3}
\tablewidth{0pt}
\tablecaption{Equations for aggregate variables under standard assumptions \label{table:ScenarioEqns}}
\tablehead{\colhead{Variable} & \colhead{Galactic Club} & \colhead{Expansive Metasociety}}
\startdata 
\cutinhead{Societal variables}
$\fpP({\baNGen} \ge 1)$       & $1 - \exp\left(-\Mean{{\baNGen}}\right)$      & $1 - \exp\left(-\Mean{{\baNGen}}\right)$\\
$\Var{{\baNGen}}$             & $\Mean{{\baNGen}}$                            & $\Mean{{\baNGen}}$\\
$\Mean{{\baAggGen}}$          & $\Mean{{\baNGen}} {\aMean{\bSingleGen}}$        & $\Mean{{\baNGen}} {\aMean{\bSingleGen}}$\\
$\Var{{\baAggGen}}$           & $\Mean{{\baNGen}} {\aMean{\bSingleGen^2}}$      & $\Mean{{\baNGen}} {\aMean{\bSingleGen^2}}$\\
\cutinhead{Metasocietal variables}
{$\fpP({\azNGen} \ge 1)$}      & {\nodata}                                       & {$1 - \exp\left(-\Mean{{\azNGen}}\right)$}\\
{$\Var{{\azNGen}}$}            & {\nodata}                                       & {$\Mean{{\azNGen}}$}\\
{$\Mean{{\azAggGen}}$}         & {\nodata}                                       & {$\Mean{{\azNGen}} \Mean{\aSingleGen}$}\\
{$\Var{{\azAggGen}}$}          & {\nodata}                                     & {$\Mean{{\azNGen}} \Mean{\aSingleGen^2}$}\\
\hline
$\fpP({\bzNGen} \ge 1)$      & \nodata                                       & $1 - \exp\left[-\Mean{{\azNGen}} \left(1 - e^{-\Mean{\baNGen}}\right)\right]$\\
$\Mean{{\bzNGen}}$           & \nodata                                       & $\Mean{{\azNGen}} \Mean{\baNGen}$\\
$\Var{{\bzNGen}}$            & \nodata                                       & $\Mean{{\bzNGen}} \left[1 + \Mean{\baNGen}\right]$\\
$\Mean{{\bzAggGen}}$         & \nodata                                       & $\Mean{{\bzNGen}} \aMean{\bSingleGen}$\\
$\Var{{\bzAggGen}}$          & \nodata                                       & $\Mean{{\bzNGen}} \left[\aMean{\bSingleGen^2} + \Mean{\baNGen} \aMean{\bSingleGen}^2\right]$\\
\cutinhead{Galactic variables}
{$\fpP({\agNGen} \ge 1)$}       & {$1 - \exp\left(-\Mean{{\agNGen}}\right)$}      & {$\Mean{{\zgNGen}} \left[1 - \exp\left(-\Mean{{\agNGen}}\right)\right]$}\\
{$\Mean{{\agNGen}}$ }           & {$\Mean{{\agNGen}}$}                            & {$\Mean{{\zgNGen}} \Mean{\azNGen}$}\\
{$\Var{{\agNGen}}$}             & {$\Mean{{\agNGen}}$}                            & {$\Mean{{\agNGen}} \left[1 + \Mean{\azNGen} \left(1 - \Mean{\azNGen}\right)\right]$}\\
{$\Mean{{\agAggGen}}$}          & {$\Mean{{\agNGen}} \Mean{\aSingleGen}$}         & {$\Mean{{\zgNGen}} \Mean{\azNGen} \Mean{\aSingleGen}$}\\
{$\Var{{\agAggGen}}$}           & {$\Mean{{\agNGen}} \Mean{\aSingleGen^2}$}       & {$\Mean{{\agNGen}} \left[\Mean{\aSingleGen^2} + \Mean{{\zgNGen}} (1 - \Mean{{\zgNGen}}) \Mean{\aSingleGen}^2\right]$}\\
\hline
$\fpP({\bgNGen} \ge 1)$       & $1 - \exp\left[-\Mean{{\agNGen}} \left(1 - e^{-\Mean{\baNGen}}\right)\right]$ 
                                      & $\Mean{{\zgNGen}} \left[1 - \exp\left(-\Mean{\azNGen} \left(1 - e^{-\Mean{\baNGen}}\right)\right)\right]$\\
$\Mean{{\bgNGen}}$            & $\Mean{{\agNGen}} \Mean{\baNGen}$
                                      & $\Mean{{\zgNGen}} \Mean{\azNGen} \Mean{\baNGen}$\\
$\Var{{\bgNGen}}$             & $\Mean{{\bgNGen}} \left[1 + \Mean{\baNGen}\right]$
                                      & $\Mean{{\bgNGen}} \left[1 + \left(\Mean{\baNGen} + (1 - \Mean{{\zgNGen}}) \Mean{\bzNGen}\right)\right]$\\
$\Mean{{\bgAggGen}}$          & $\Mean{{\bgNGen}} \aMean{\bSingleGen}$ 
                                      & $\Mean{{\bgNGen}} \aMean{\bSingleGen}$\\
$\Var{{\bgAggGen}}$           & $\Mean{{\bgNGen}} \left[\aMean{\bSingleGen^2} + \Mean{\baNGen} \aMean{\bSingleGen}^2 \right]$
                                      & $\Mean{{\bgNGen}} \left[\aMean{\bSingleGen^2} + \aMean{\bSingleGen}^2 \left(\Mean{\baNGen} + \left(1 - \Mean{{\zgNGen}}\right) \Mean{\bzNGen}\right)\right]$\\
\enddata
\tablecomments{{These expressions are calculated under the assumptions of independence and interchangeability. Note that ${\jlMean{\jjkNGen}} \equiv {\jlMeanGen{\jjkNGen}}$ and ${\jkMean{\jjSingleGen}} \equiv {\jkMeanGen{\jjSingleGen}}$ by convention (section~\ref{sec:SelRelOperationRules}).}}
\end{deluxetable*}

The number of broadcasts is a compound {random sum}, with means and variances given in the table. As long as $\bDuration \ll \aDuration, \zDuration$ and ${\bgRatenu}$ is slowly varying, 
\begin{multline}
\label{eqn:MeanNBSys}
\Mean{{\bgNGen}} \approx {\bgRatenu}(\oTStartGen, \oNuMidGen) {\cdot \Mean{\hgNGen}} (\oBandwidthGen + {\gMean{\bBandwidthTimeOFGen}}) \\
\cdot (\oDurationGen + {\gMean{\bDuration}}) .
\end{multline}
It is scaled by the mean rate that broadcasts occur per unit frequency per star, which is 
\begin{equation}
{\bgRatenu} = \begin{cases}
                {\agAbund \gMean{\baRatenuTotal}}                          & \text{(Galactic club)}\\
								{\Mean{\zgNGen} \gMean{\azAbund} \gMean{\baRatenuTotal}}   & \text{(Expansive interstellar)}
								\end{cases}
\end{equation}
This abundance is a key quantity in SETI. The ``clumping'' into societies and metasocieties adds additional sampling variance, more so in the expansive interstellar scenario where ${\zgNGen}$ is independent of $\azNGen$. When fewer than one broadcast is typically {selected} per society, the probability that one broadcast is intercepted is
\begin{multline}
\label{eqn:PInterceptionRare}
P({\bgNGen} \ge 1) \approx \begin{cases}
                1 - \exp(-\Mean{{\bgNGen}})                     \\
								\hfill \text{(Galactic club)}                  \\
								\Mean{{\zgNGen}} [1 - \exp(-\Mean{\bzNGen})] \\
								\hfill \text{(Expansive interstellar)}
								\end{cases} .
\end{multline}
{In the opposite limit of many selected broadcasts per society, this probability is simply the probability that one communicative society is selected. The variance in the number of broadcasts per metasociety is then mainly from the Poissonian clumping into societies, approaching $\Mean{{\azNGen}} \Mean{\baNGen}^2$ (full expressions are given in Table~\ref{table:ScenarioEqns}).}

\subsection{Aggregate emission and fluence}
\label{sec:AggregateEmission}
The total single-polarization spectral luminosity from the broadcasts in a sample ${\bjjSampleGen}$ is
\begin{equation}
{\bjjAggLnupolisoGen} (\TimeVar, \FreqVar, \PolVar) = {\jjSumBcGen} \bLnupoliso (\TimeVar, \FreqVar, \PolVar)
\end{equation}
Of course, ${\bjjSampleGen}$ itself is random{, even when ${\JMark}$ itself is specified}. {The usual conditions for broadcast selection by the window ${\GenLabel}$ (equation~\ref{eqn:BcSelectionCriteria})} define {the} bounds of integration {to} be used in equation~\ref{eqn:BasicAggMean}. Under the {interchangeability}-independen{ce} assumptions,
\begin{multline}
\label{eqn:MeanLnuETI}
\Mean{{\bjjAggLnupolisoGen} (\TimeVar, \FreqVar, \PolVar)} \approx {\bjjRatenu}(\TimeVar, \FreqVar) {\cdot \Mean{\hjjNGen}} {\jjMeanGen{\bLnupoliso(\TimeVar, \FreqVar, \PolVar)}} \\
\cdot~(\oBandwidthGen + {\jjMean{\bBandwidthTimeOFGen}}) (\oDurationGen + {\jjMean{\bDuration}}) .
\end{multline}
{Strictly speaking, this is not necessarily a compound Poisson variable because ${\zgNGen(\gTuple)}$ is Bernoulli in the expansive metasociety scenario. It reaches the compound Poisson limit when $\Mean{{\zgNGen}} = 1$ and the galaxy as a whole can be identified with the metasociety (Table~\ref{table:ScenarioEqns}).}  

The aggregate fluence {is a similar compound sum} under these assumptions. A fluence ${\mjjFluenceGen}$ of any type from the sample is found by summing fluences $\lFluenceGen$ of the same type from individual broadcasts in a sample. Examples include the total energy fluence ${\mjjFluenceEGen}$ and total photon fluence ${\mjjFluenceQGen}$ of a sample, composed of the energy fluences $\lFluenceEGen$ and photon fluences $\lFluenceQGen$ of individual broadcasts. We have:
\begin{multline}
\Mean{{\mjjFluenceGen}}  \approx {\bjjRatenu (\oTStartGen, \oNuMidGen)} {\cdot \Mean{\hjjNGen}} (\oBandwidthGen + {\jjMean{\bBandwidthTimeOFGen}})  \\
 {\cdot (\oDurationGen + {\jjMean{\bDuration}}) \jjMean{\lFluenceGen}}.
\end{multline}

{The clumping of broadcasts into societies can have a big effect on the variance of ${\mjjFluenceGen}$ (Table~\ref{table:ScenarioEqns}). When a window typically samples $\ll 1$ broadcast per society or the variance in $\lFluenceGen$ is sufficiently large, the variance in an inhabited galaxy ${\GalMark}$ with ${\zgNGen} = 1$ approaches what we would expect if ${\bgNGen}$ were Poissonian. In the opposite limit, $\Var{{\mgFluenceGen}} \approx \Mean{{\bgNGen}} \Mean{\baNGen} \aMean{\lFluenceGen}^2 = \Mean{{\agNGen}} \Mean{\maFluenceGen}^2$. The societies become a population of ``standard {candles,}'' and the fluctuations are just shot noise in their number. This limit requires many broadcasts per society just within a window, all becoming confused, which can be a tall order (Paper II).}

\subsubsection{What do we expect the aggregate emission to look like?}
Although we usually envision broadcasts to resemble isolated ``spikes'' in $\tAggLnuiso$, on coarse enough time-frequency scales, this {roughness blurs out}.  The mean spectrum, as shown by equation~\ref{eqn:MeanLnuETI}, is proportional to ${\bgRatenu}$ itself, as long as it varies slowly enough and a mean $\bLnupoliso$ exists.  If there is no ``magic frequency'' \citep[unlike][]{Drake73-Freq,Blair93} or ``magic time'' \citep[unlike][]{Pace75,Makovetskii77,Corbet03,Nishino18} that ETIs in the target {galaxy} prefer to broadcast at, ${\bgRatenu}$ is most likely a smooth function{, {perhaps} a {power law}}.  

Supposing that the $\bEiso$ distribution has a mean, over large bandwidths and long durations the aggregate emission converge{s} to a constant luminosity source with a smooth spectrum. In other words, it will resemble diffuse nonthermal emission commonly seen in objects like active galactic nuclei. {Like} any population of discrete sources {\citep[e.g.,][]{Tonry88}, there are Poissonian fluctuations in the emission, far greater than those from the immense number of particles contributing to natural diffuse emission.} With enough broadcasts being added together, the fluctuations also become Gaussian by the central limit theorem.  This motivates use of the total emission of galaxies to set constraints on broadcasts, as in Paper II.

If the $\bEiso$ distribution has no mean, or if broadcasts are so rare that confusion does not occur for any feasible sample, the aggregate emission {remains} {``spiky,''} with strongly non-Gaussian fluctuations that could be picked out by conventional search strategies.  Nonetheless, there still could be a background of fainter broadcasts for which aggregate constraints still apply.

Even if there is a magic frequency, the broadcasts {from an extended metasociety} may be smeared out in received frequency because of velocity differences between the {different transmitter sites}.  Then the aggregate emission will appear much like natural line emission, although possibly at a frequency corresponding to no natural transition.  The luminosity of candidate ``lines'' in the galaxy's emission sets constraints on such broadcast populations.

\subsection{The diffuse approximation}
{Because societies are discrete, they clump broadcasts. {However,} including the societal level in the tree greatly complicates {the} analysis, compounding the Poissonian character of the number of broadcasts in collections of societies. The} \emph{diffuse approximation} ignor{es} the discreteness of societies{, instead imagining the transmitters being spread diffusely across the galaxy.} The variance in ${\bgNGen}$ is thus underestimated.  How large this correction is depends on the nature of the discretization, but the diffuse approximation gives good results when there are many societies, few of which emit a detectable broadcast {(${\jjMean{\baNGen}} \ll 1$)}. {This is appropriate when using very fine observations (e.g., narrow channels for narrowband lines) except in the most extreme cases -- far, far into the confusion regime for heavily populated galaxies.}

{Practically speaking, the diffuse approximation treats the broadcasts as the children of metasocieties (expansive metasociety scenario) or galaxies (classical and Galactic club scenario). Thus, the diffuse approximation works directly with the metasociety's or galaxy's broadcast distribution instead of building those up from societies' broadcast distributions. Just as the societal distribution posits that societies trace stars (Section~\ref{sec:Stars} \&~\ref{sec:MetaScenarios}), in the diffuse approximation, that dependence is shifted down to the broadcast distribution. In the end, the broadcast distribution for the effective parent ${{\JMark}}$ in the diffuse approximation is found by substituting
\begin{equation}
\label{eqn:DiffuseSubst}
\baRatenuTotal \fDirac(\bPosition - \aPosition) \to {\bjjRatenu} \frac{d\Mean{{\hjjNTime}}}{d\hPosition}(\bPosition);
\end{equation}
the substitution can also apply to the broadcast abundance per frequency variable.}

{
\subsection{Aggregate luminosities in diverse, noninterchangeable hosts: an example}
\label{sec:NonInterchangeableExample}
Diversity among host objects increases the sampling variance of aggregate quantities, and the variance increases as it manifests in {higher-level} ancestors. This section presents a simple example: how would the cosmic-relative variance in an aggregate metasocietal luminosity $\bzAggLiso$ change if the luminosities were shared among different types of hosts? To keep the example focused, the number distributions $\baN{(\aTuple)} \sim \Poisson(\Mean{\baN})$ and $\azN{(\zTuple)} \sim \Poisson(\Mean{\azN})$ are strictly the same among different societies and metasocieties. In {c}ase 1, every metasociety and society is interchangeable, with 
\begin{equation}
\aPDF{\bLiso}(\ell) = \exp(-\ell / \bLisoBAR) / \bLisoBAR .
\end{equation}
In {c}ase 2, all the broadcasts in an individual society have the same luminosity, but this characteristic luminosity is exponentially distributed between societies:
\begin{equation}
\aPDF{\bLiso}(\ell) = \fDirac(\ell - \bLisoBAR_{\SocMark})~\text{and}~\zPDF{\bLisoBAR_{\SocMark}}(\ell) = \exp(-\ell / \bLisoBAR) / \bLisoBAR .
\end{equation}
In case 3, each metasociety has decreed all broadcasts have the same luminosity, but different metasocieties decide on different luminosities, with an exponential distribution:
\begin{equation}
\zPDF{\bLiso}(\ell) = \fDirac(\ell - \bLisoBAR_{\MetaMark})~\text{and}~\uPDF{\bLisoBAR_{\MetaMark}}(\ell) = \exp(-\ell / \bLisoBAR) / \bLisoBAR .
\end{equation}
These can be compared with case 0, in which all broadcasts have the same luminosity: $\aPDF{\bLiso}(\ell) = \fDirac(\ell - \bLisoBAR)$. {In case 0, broadcasts, societies, and metasocieties are all interchangeable.}

There are slightly different approaches to calculating the variance in each case, but it involves building up from broadcast-level averages to the cosmic level. All the broadcasts in a society are drawn from the same {luminosity distribution} -- a degenerate one in cases 0, 2, and 3, and an exponential one in case 1. Thus, the aggregate luminosity of broadcasts in a {realized} society is a compound Poisson variable. In cases 0, 1, and 3, the societies are interchangeable, so we can then apply a compound Poisson distribution again. However, in case 2, the societies have different luminosity distributions depending on their $\bLisoBAR_{\SocMark}$. To advance to the metasocietal-level variables, we have to integrate over $\zPDF{\bLisoBAR_{\SocMark}}$, which is a sort of marginalized societal haystack (equation~\ref{eqn:CampbellVar}). Finally, {in} cases 0--2, all metasocieties are interchangeable, and so $\Mean{\bzAggLiso} = \uMean{\bzAggLiso}$ and $\Var{\bzAggLiso} = \uVar{\bzAggLiso}$. {However, in} case 3, we need to find $\uVar{\bzAggLiso} = \uMean{(\bzAggLiso)^2} - \uMean{\bzAggLiso}^2$ by averaging over $\uPDF{\bLisoBAR_{\SocMark}}$ (equation~\ref{eqn:BiasMean}), effectively a metasocietal haystack. 

In all four cases, the mean aggregate luminosity is the same, $\uMean{\bzAggLiso} = \Mean{\bzN} \bLisoBAR$, because the mean number of broadcasts and the mean broadcast luminosity are the same. The variance, however, increases from one case to the next:
\begin{equation}
\uVar{\bzAggLiso} = \begin{cases}
                 2 \bLisoBAR^2 \Mean{\bzN} (\frac{1}{2} + \frac{1}{2} \Mean{\baN})     & (\text{Case 0})\\
                 2 \bLisoBAR^2 \Mean{\bzN} (1 + \frac{1}{2} \Mean{\baN})               & (\text{Case 1})\\
                 2 \bLisoBAR^2 \Mean{\bzN} (1 + \Mean{\baN})                           & (\text{Case 2})\\
                 2 \bLisoBAR^2 \Mean{\bzN} (1 + \Mean{\baN} + \frac{1}{2} \Mean{\bzN}) & (\text{Case 3})
		\end{cases}
\end{equation}
In case 0, the variance is totally due to {compound} Poissonian {fluctuations} in the number of broadcasts, supplemented by intrinsic variance in the luminosity in case 1. All the broadcasts in a society share the same luminosity in cases 2 and 3, amplifying the fluctuations; in case 3, as the shared luminosity $\bLisoBAR_{\MetaMark}$ varies, the entire aggregate luminosity does as well. 

This example only scratches the surface of the subject, but the methods in sections~\ref{sec:ProbabilityBack}~and~\ref{sec:SelectionOverview} can be applied to these problems even when the compound Poisson distribution does not apply.
}

\section{The box model}
\label{sec:BoxModelIntro}
\begin{figure}
\centerline{\includegraphics[width=8.5cm]{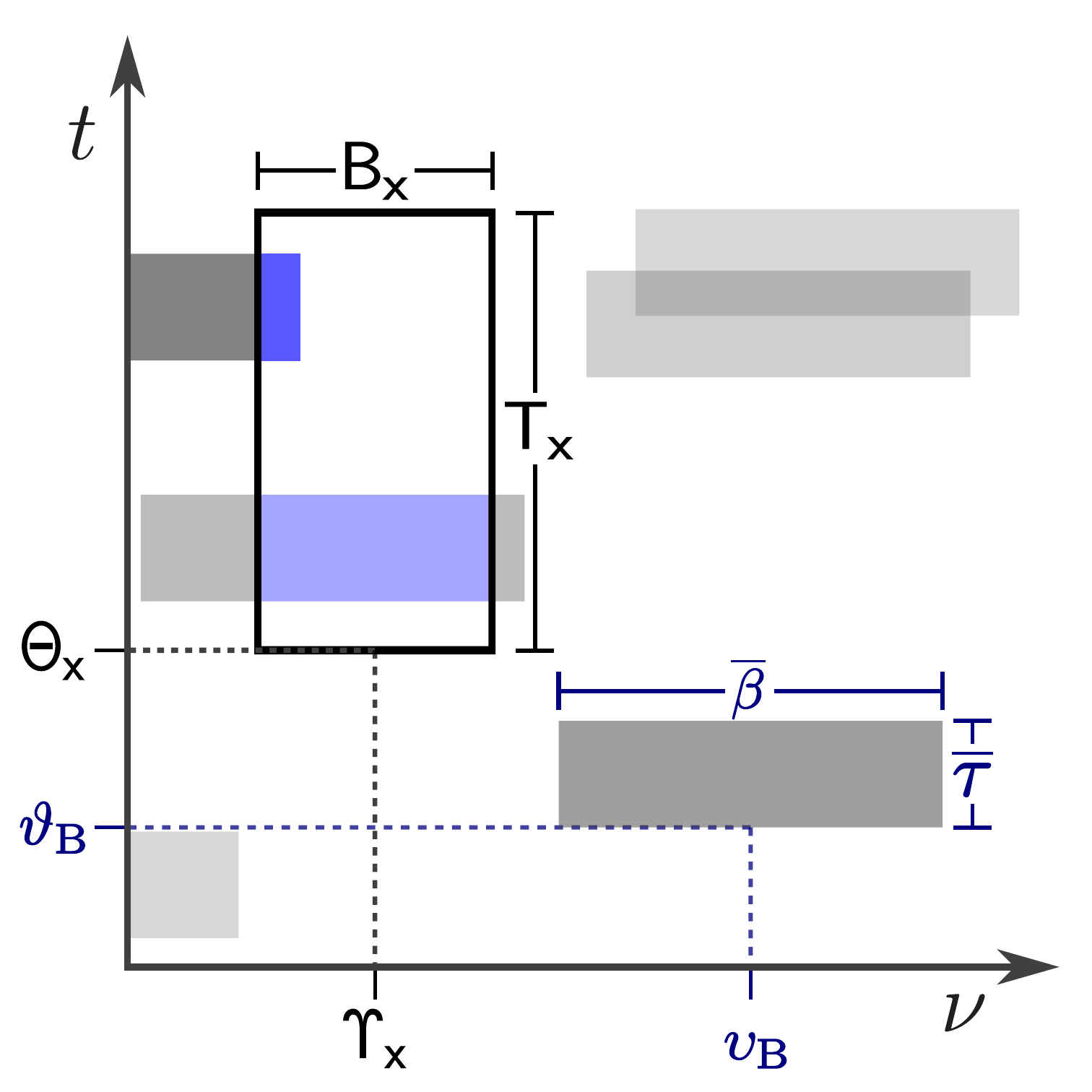}}
\figcaption{{Sketch} of a spectrogram in the box model.  All broadcasts have equal duration $\bDurationBox$ and bandwidth $\bBandwidthBox$, and {$\bLnuiso$} is constant within each ``box'' although the total $\bEiso$ may vary (filled boxes of different shades).  The selection also is a window of duration $\oDurationGen$ and bandwidth observation $\oBandwidthGen$ at all {times} and frequencies within those ranges. Only emission within the selection (blue shading) is observed. \label{fig:BoxModel}}
\end{figure}

In the box model, broadcasts are simple contiguous ``boxes'' in time-frequency space. Each box covers the frequency range $|\FreqVar - \bNuMid| \le \bBandwidthBox / 2$ and time range $0 \le \TimeVar - \bTStart \le \bDurationBox$. Because these boxes are not skewed, the drift rate $\bDriftRate$ is $0$. Furthermore, all the boxes are assumed to have identical bandwidths $\bBandwidthBox$ and durations $\bDurationBox$ (Figure~\ref{fig:BoxModel}). The distribution is assumed to be stationary, in that $\bTStart$ and $\bNuMid$ have uniform distributions over any observable range, with
\begin{equation}
{\baDist} = {\baRatenuTotal} \aPDF{\bEiso, \bPolQuantity} \fDirac(\bDuration - \bDurationBox) \fDirac (\bBandwidthTime - \bBandwidthBox) \fDirac(\bDriftRate) {\fDirac(\bPosition - \aPosition) .}
\end{equation}
In the box model, $\bBandwidthTimeOFGen = \bBandwidthBox$ (equations~\ref{eqn:bBandwidthEff} and~\ref{eqn:MeanNBTouchSociety}).

The selection itself is also a contiguous box in time-frequency space, spanning the frequency range $|\FreqVar - \oNuMidGen| \le \oBandwidthGen/2$ and time range $0 \le \TimeVar - \oTStartGen \le \oDurationGen$.  Sample properties in the box model depend on the number and amount of overlap between broadcast boxes and the sample box in time-frequency space (Figure~\ref{fig:BoxModel}). 

Consider a sample ${\bjjSampleGen}$, consisting of broadcasts within a field of ${\hjjNGen}$ stars with a time and frequency response described by a box.  The number of broadcasts expected from a host ${{\JMark}}$ is {proportional to the rate per star per unit frequency, number of stars, and the temporal and frequency ``cross sections'' resulting from the combined width of the window and the broadcasts:}
\begin{equation}
\label{eqn:NSampleBox}
\Mean{{\bjjNGen}} = {\bjjRatenu} {\Mean{\hjjNGen}} (\oDurationGen + \bDurationBox) (\oBandwidthGen + \bBandwidthBox) .
\end{equation}

Observables in the box model tie to an overlap quantity $\bOverlapGen$, the fraction of the broadcast's time-frequency ``area'' within the sample box. Appendix~\ref{sec:BoxModelDetails} provides calculations relating to this quantity.

As uniform ``boxes'' in time-frequency space, broadcasts in the box model have steady luminosities and flat spectra:
\begin{equation}
\label{eqn:dEdnudtBox}
\bLnupoliso (\TimeVar, \FreqVar, \PolVar) = \begin{cases}
\displaystyle \bFPol(\PolVar) \frac{\bEiso}{\bBandwidthBox \bDurationBox}   &  \begin{array}{l}
                                                                                    \text{if}~0 \le \TimeVar - \bTStart \le \bDurationBox~\\
																																										\text{and}~|\FreqVar - \bNuMid| \le \bBandwidthBox/2
																																								\end{array}\\
        0                                                                   & \text{otherwise}
        \end{cases} .
\end{equation}
Although the broadcasts may share the same degree and type of polarization, the polarization state is random. The mean luminosity spectrum of a sample is
\begin{equation}
\Mean{{\bjjAggLnupolisoGen} (\TimeVar, \FreqVar, \PolVar)} = \frac{1}{2} {\Mean{\hjjNGen}} {\bjjRatenu} {\jjMeanGen{\bEiso}} .
\end{equation}
{with a } mean {energy} fluence {of}
\begin{equation}
\Mean{{\mjjFluenceEGen}} = \frac{|\oPolSetGen|}{2} {\bjjRatenu \Mean{\hjjNGen} \jjMeanGen{\bEiso} \jjMeanGen{\lDilutionE {\lTransmittanceEGen}}}  \oDurationGen \oBandwidthGen.
\end{equation}

Many observational quantities simplify when we assume that $\bDurationBox \nsim \oDurationGen$ and $\bBandwidthBox \nsim \oBandwidthGen$.  The box model {has four natural limits}:
\begin{itemize}
\item \emph{Lines} -- Long-lasting, narrowband broadcasts with $\bDurationBox \gg \oDurationGen$ but $\bBandwidthBox \ll \oBandwidthGen$. These include the classic carrier wave ``beacons'' of radio SETI (\citealt{Drake61}; see also \citealt{Enriquez17} and references therein) and artificial laser lines in optical \citep{Schwartz61}.  Instead of $\bEiso$ and ${\bjjRatenu}$, it is most natural to consider the isotropic luminosity $\bLiso \equiv \bEiso / \bDurationBox$. As the lifetime is unknown and largely irrelevant, the observable rate is ${\bjjAbundnu} = {\bjjRatenu} {\jjMean{\bDuration}} = {\bjjRatenu} \bDurationBox$.
\item \emph{Pulses} -- Short-lasting, wideband broadcasts with $\bDurationBox \ll \oDurationGen$ and $\bBandwidthBox \gg \oBandwidthGen$. Several SETI surveys have sought pulses (e.g., \citealt{Shvartsman93}; {\citealt{Howard04};} \citealt{Siemion10,Maire19}). Instead of $\bEiso$, it is most natural to consider the isotropic energy spectrum $\bEnuiso \equiv \bEiso / \bBandwidthBox$. The observable rate is ${\bjjRate} = {\bjjRatenu} {\jjMean{\bBandwidth}} = {\bjjRatenu} \bBandwidthBox$.
\item \emph{Blips} -- Short-lasting, narrowband broadcasts with $\bDurationBox \ll \oDurationGen$ and $\bBandwidthBox \ll \oBandwidthGen$. Blips can form the basis of wideband communication systems \citep{Messerschmitt15}, or result from transient broadcasts like our radars or narrowbeam beacons on rotating worlds \citep{Gray02}.
\item \emph{Hisses} -- Long-lasting, wideband broadcasts with $\bDurationBox \gg \oDurationGen$ and $\bBandwidthBox \gg \oBandwidthGen$, essentially noise. Thermal waste heat {\citep{Dyson60}} is effectively a hiss, {as is exhaust radiation \citep{Harris86},} and {a typical random waveform is a hiss} similar to white noise. Long-lasting continuum sources emit hisses. Instead of $\bEiso$, it is most natural to consider the isotropic luminosity spectrum $\bLnuiso \equiv \bEiso / (\bDurationBox \bBandwidthBox)$. The observable abundance is ${\bjjAbund} = {\bjjRatenu} {\jjMean{\bDuration} \jjMean{\bBandwidth}} = {\bjjRatenu} \bDurationBox \bBandwidthBox$.
\end{itemize} 
Table \ref{table:BoxRegimes} summarizes key quantities for each of these regimes.

\begin{deluxetable*}{lcccc}
\tabletypesize{\footnotesize}
\tablecolumns{5}
\tablewidth{0pt}
\tablecaption{Quantities in the {extreme regimes of the} box {model} \label{table:BoxRegimes}}
\tablehead{\colhead{Quantity} & \colhead{Line} & \colhead{Pulse} & \colhead{Blip} & \colhead{Hiss}}
\startdata 
$\bDurationBox$             & $\gg \oDurationGen$         & $\ll \oDurationGen$          & $\ll \oDurationGen$          & $\gg \oDurationGen$\\
$\bBandwidthBox$            & $\ll \oBandwidthGen$        & $\gg \oBandwidthGen$         & $\ll \oBandwidthGen$         & $\gg \oBandwidthGen$\\
Natural energy variable     & $\bLiso$                    & $\bEnuiso$                   & $\bEiso$                     & $\bLnuiso$\\ 
Natural rate variables                                                                & 
  ${\bjjAbundnu} = {\bjjRatenu} \bDurationBox$                                              &
	${\bjjRate} = {\bjjRatenu} \bBandwidthBox$                                                &
	${\bjjRatenu}$                                                                         & 
	${\bjjAbund} = {\bjjRatenu} \bBandwidthBox \bDurationBox$                                  \\
$\bOverlapGen$                                                                        &
	$\oDurationGen / \bDurationBox$                                                     &
	$\oBandwidthGen / \bBandwidthBox$                                                   &
	$1$                                                                                 &
	$\oDurationGen \oBandwidthGen / (\bDurationBox \bBandwidthBox)$                      \\
$\Mean{{\bjjNGen}}$                                                                      & 
	${\bjjAbundnu} {\Mean{\hjjNGen}} \oBandwidthGen$                                                 & 
	${\bjjRate} {\Mean{\hjjNGen}} \oDurationGen$                                                     & 
	${\bjjRatenu} {\Mean{\hjjNGen}} \oBandwidthGen \oDurationGen$                                    & 
	${\bjjAbund} {\Mean{\hjjNGen}}$                                                                   \\
${\jjMean{\bEisoGen}}$                                                                   &
	$(\oNumPolGen/2) {\jjMean{\bLiso}} \oDurationGen$                                      &
	$(\oNumPolGen/2) {\jjMean{\bEnuiso}} \oBandwidthGen$                                   &
	$(\oNumPolGen/2) {\jjMean{\bEiso}}$                                                    &
	$(\oNumPolGen/2) {\jjMean{\bLnuiso}} \oBandwidthGen \oDurationGen$                      \\
${\jjMean{\bEisoGen^2}}$                                                                 &
	${\jjMean{{\bFPolGen}^2}} {\jjMean{{\bLiso}^2}} \oDurationGen^2$                              &
	${\jjMean{{\bFPolGen}^2}} {\jjMean{{\bEnuiso}^2}} \oBandwidthGen^2$                           &
	${\jjMean{{\bFPolGen}^2}} {\jjMean{{\bEiso}^2}}$                                              &
	${\jjMean{{\bFPolGen}^2}} {\jjMean{{\bLnuiso}^2}} \oBandwidthGen^2 \oDurationGen^2$            \\
$\Mean{{\bjjAggLnuisoGen}}$                                                                 & 
	$\displaystyle {\frac{\oNumPolGen}{2} \bjjAbundnu} {\Mean{\hjjNGen}} {\jjMean{\bLiso}}$                                                 & 
	$\displaystyle {\frac{\oNumPolGen}{2} \bjjRate} {\Mean{\hjjNGen}} {\jjMean{\bEnuiso}}$                                                  & 
	$\displaystyle {\frac{\oNumPolGen}{2} \bjjRatenu} {\Mean{\hjjNGen}} {\jjMean{\bEiso}}$                                                  & 
	$\displaystyle {\frac{\oNumPolGen}{2} \bjjAbund} {\Mean{\hjjNGen}} {\jjMean{\bLnuiso}}$                                                  \\
${\jjMean{\lFluenceEGen} (\text{distant host})}$                                                                                                        &
  $\displaystyle \frac{\oNumPolGen}{2} \frac{{\jjMean{\bLiso}} \oDurationGen {\jjMean{\lTransmittanceGen}}}{4 \pi (1 + \yRedshift) {{\yjjDistanceM}}^2}$                   &
	$\displaystyle \frac{\oNumPolGen}{2} \frac{{\jjMean{\bEnuiso}} \oBandwidthGen{\jjMean{\lTransmittanceGen}}}{4 \pi (1 + \yRedshift) {{\yjjDistanceM}}^2}$                &
	$\displaystyle \frac{\oNumPolGen}{2} \frac{{\jjMean{\bEiso}} {\jjMean{\lTransmittanceGen}}}{4 \pi (1 + \yRedshift) {{\yjjDistanceM}}^2}$                                 &
	$\displaystyle \frac{\oNumPolGen}{2} \frac{{\jjMean{\bLnuiso}} \oBandwidthGen \oDurationGen {\jjMean{\lTransmittanceGen}}}{4 \pi (1 + \yRedshift) {{\yjjDistanceM}}^2}$
\enddata
\tablecomments{Quantities in {the} source frame: broadcast rates and abundances (${\bjjAbundnu}$, ${\bjjRate}$, ${\bjjRatenu}$, ${\bjjAbund}$), window definition quantities ($\oBandwidthGen$, $\oDurationGen$), broadcast durations and bandwidths ($\bBandwidthBox$, $\bDurationBox$), broadcast emission ($\bLiso$, $\bEnuiso$, $\bEiso$, $\bLnuiso$, $\bEisoGen$, ${\bjjAggLnuisoGen}$).\\
Quantities in observer frame: fluence  ($\lFluenceEGen$).\\
{When considering a fixed sample of stars, the known number of stars $\hjjNGen$ can be substituted for $\Mean{\hjjNGen}$.}}
\end{deluxetable*}

Often there are multiple time and frequency scales involved in a survey.  A broadcast might be longer than a few seconds long exposure but multiple pointings of the same sky region could occur over years.  The approximations for the four quadrants apply when the inequalities hold for all relevant time and frequency scales.

\section{The chord model for frequency-drifting lines}
\label{sec:ChordModelIntro}

The broadcasts in the box model have no skew. Although a satisfactory approximation when the time or frequency resolution is coarse, both line and pulse searches use fine sampling windows sensitive to frequency drifts {-- lines drift} because of the changing Doppler shifts of an accelerating source \citep{Sheikh19}, while pulses drift because of dispersion induced by the interstellar and intergalactic media \citep{Siemion10}. In fact, lines without any drift or dispersion are generally attributable to anthropogenic radio frequency interference, since the rotation of the Earth will add a location-dependent frequency drift \citep{Enriquez17,Sheikh21}. Individual broadcasts can be dedrifted or dedispersed by shifting the spectrum by different delays in different channels to maximize signal-to-noise {(see Paper II)}. Nonetheless, a population of broadcasts form a background with a range of drift rates that cannot be dedrifted.

\begin{figure}
\centerline{\includegraphics[width=8.5cm]{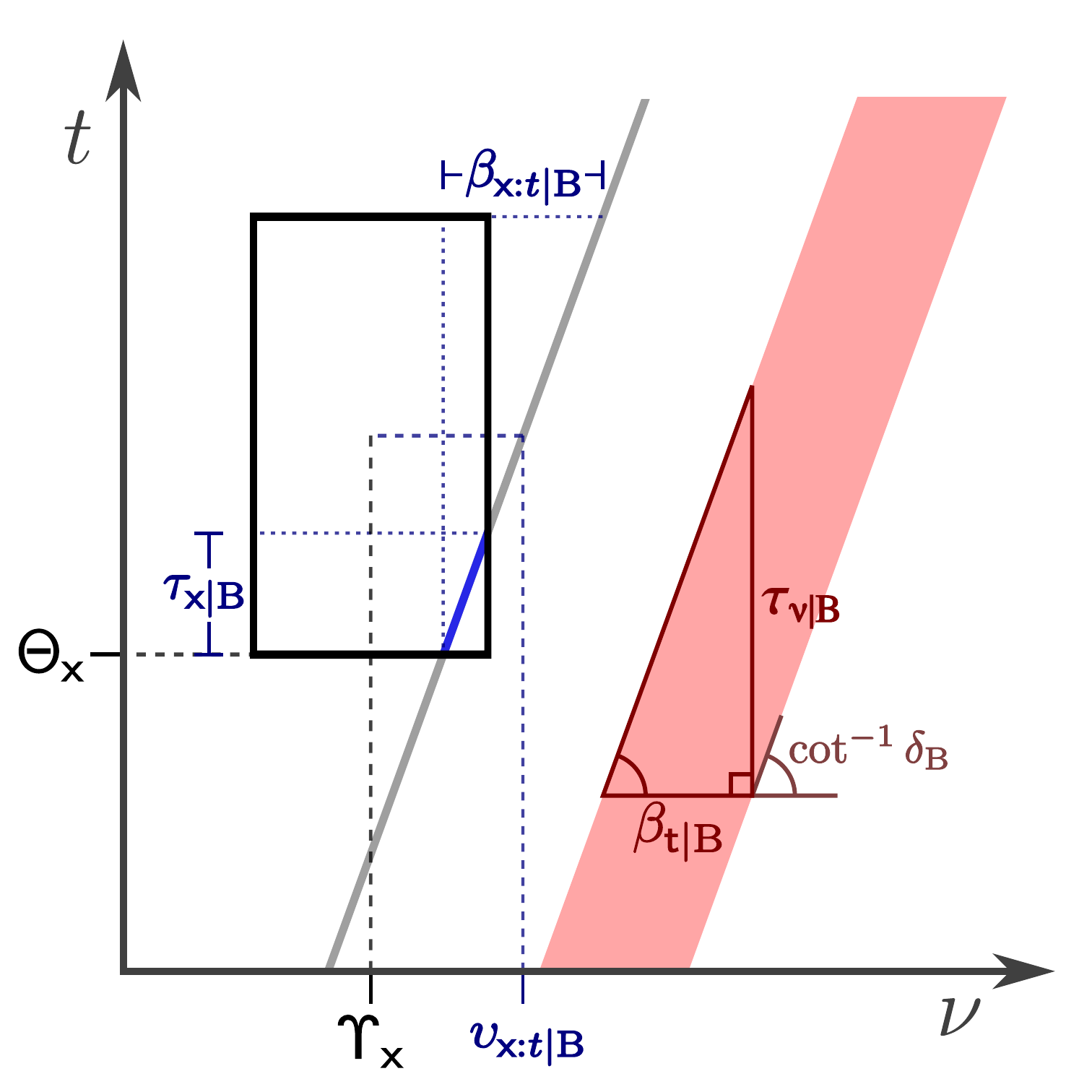}}
\figcaption{{Sketch} of a spectrogram in the chord model.  Drifting signals could resemble a band (red) on the diagram. In the chord approximation, the band is assumed to have negligible bandwidth and looks like a skewed line ({gray}). A {nonzero} fluence is intercepted by {the window} only if the line cuts across the {window} box in a chord (highlighted in blue).\label{fig:ChordModel}}
\end{figure}

An archetypal drifting line never ``ends'' in the middle of an observational sample window, because $\oDurationGen \ll \bDuration$. To avoid edge effects, these broadcasts can be considered as thin bands with an instantaneous bandwidth $\bBandwidthTime$ and a duration at fixed frequency $\bDurationFreq$ (Figure~\ref{fig:ChordModel}). The ratio of these quantities defines the slope or drift rate
\begin{equation}
|\bDriftRate| \equiv \frac{\bBandwidthTime}{\bDurationFreq} .
\end{equation}
The drift rate is a signed quantity that is positive if the broadcast drifts to high frequencies at later times and negative if it drifts to low frequencies. In the chord model, $\bDriftRate$ does not vary with time -- broadcasts have linear drift in frequency. As in the box model, the selection window is a contiguous box in time-frequency space.

The chord model simplifies further by taking these bands to have negligible $\bBandwidthTime$ and $\bDurationFreq$, even as $\bDriftRate$ itself is fixed. Because the band has infinitesimal {band}width, the broadcast is either entirely within the sample or not at any given time  ({gray}-blue line in Figure~\ref{fig:ChordModel}). A broadcast is part of the sample if the band ever crosses the box, appearing as a chord in time-frequency diagrams. The duration of the chord's time in the sample box is $\bDurationGen$, and it sets the amount of emission received. The effective bandwidth used in calculating the expected number of broadcasts in the sample is $\bBandwidthTimeOFGen = |\bDriftRate| \bDuration$ (equations~\ref{eqn:MeanNBTouchSociety} and \ref{eqn:bBandwidthEff}). The chord approximation breaks down if $\bDurationFreq \ga \min(\oDurationGen, \oBandwidthGen / |\bDriftRate|)$.

I adopt a distribution for drifting broadcasts similar to that for lines in the box model, with a frequency abundance of {$\baAbundnuTotal$ per society}:
\begin{multline}
{\baDist} = {\baAbundnuTotal} \fDirac(\bDuration - \bDurationInfty) \fDirac (\bBandwidthTime) \fDirac(\bDriftRate)  {\fDirac(\bPosition - \aPosition)} {\fDirac(\bTStart)}\\
{\cdot~\aPDF{\bEiso, \bPolQuantity, \bDriftRate} .}
\end{multline}
{The origin time of each broadcast is set to $\TimeVar = 0$ while the duration is set to an arbitrarily long $\bDurationInfty$; the distribution is then assumed to extend to all negative and positive frequencies with the understanding that emission at frequencies and times outside any window being used is ignored. This allows for a direct relation between $\bEiso$ and the more relevant effective isotropic luminosity $\bLiso = \bEiso / \bDurationInfty$, with $\aPDF{\bLiso} = \bDurationInfty \aPDF{\bEiso}$.} The mean number of broadcasts within ${\bjjSampleGen}$ is
\begin{equation}
\Mean{{\bjjNGen}} = {\bjjAbundnu} {\Mean{\hjjNGen}} (\oBandwidthGen + \oDurationGen {\jjMean{|\bDriftRate|}}) ,
\end{equation}
with {an average time spent within ${\GenLabel}$ of}
\begin{equation}
{\jjMean{\bDurationGen}} = \frac{\oDurationGen}{1 + \oDurationGen {\jjMean{|\bDriftRate|}} / \oBandwidthGen} .
\end{equation}

The drift rate distribution is not evenly sampled. Broadcasts with high ${|\bDriftRate|}$ are more likely to cut through the box simply because they cover more frequency, although they spend less time on average within the observational window. Calculations of statistical quantities need to take this into account (see Appendix~\ref{sec:ChordModelDetails}, equation~\ref{eqn:BoxChordMean}). Although the biasing can be extreme for individual observations in single channels, modern radio SETI surveys cover hundreds of MHz or more and thus should sample drift rates more or less fairly.

A drift rate distribution is necessary to calculate the higher moments. In some cases, all $\bDriftRate$ may have the same value, if all broadcasts are coming from transmitters at a single location within a narrow frequency range. A galactic population of line transmitters may contain a panoply of sites with different accelerations with a potentially vast $\bDriftRate$ range \citep{Sheikh19}. In this series, I adopt a uniform drift rate distribution:
\begin{equation}
\label{eqn:PDFUniformDrift}
{\jjPDF{\bDriftRate}} = \IndicatorOf{|\bDriftRate - {\bjjDriftRateMid}| \le {\bjjDriftRateBAR}}/(2{\bjjDriftRateBAR})
\end{equation}
where ${\bjjDriftRateMid} \equiv {\jjMean{\bDriftRate}}$ is the ``center'' of the distribution, which may be {nonzero} because of Doppler effects imposed by the Earth's rotation and revolution. A canonical value for the drift rate scale is ${\bjjDriftRateBAR} = 1\ \Hz\,\sec^{-1} (\FreqVar/\GHz)$ \citep{Oliver71}, though there could be a tail extending to much higher $\bDriftRate$ \citep{Sheikh19}. {Note this is the unbiased drift rate distribution.}

The energy emission in this model is treated as being defined by a {steady} luminosity:
\begin{equation}
\label{eqn:dEdnudtChord}
\bLnupoliso (\TimeVar, \FreqVar, \PolVar) = \bFPol(\PolVar) \bLiso \fDirac(\FreqVar - \bNuMidTime(\TimeVar)) ,
\end{equation}
where $\bNuMidTime(\TimeVar)$ is the frequency of the chord at time $\TimeVar$. Taking the limit of broadcasts with small $\bBandwidthTime$ allows us to calculate 
\begin{equation}
\Mean{{\bjjAggLnuisoGen} (\TimeVar, \FreqVar)} = \frac{\oNumPolGen}{2} {\Mean{\hjjNGen}} {\bjjAbundnu} {\jjMeanGen{\bLiso}}.
\end{equation}
This is a smooth function although any realized ${\bjjAggLnuisoGen}$ itself is very {``spiky.''} Windows integrate over these ``spikes'' with their nonzero bandwidth, yielding a mean fluence of
\begin{equation}
\Mean{{\mjjFluenceEGen}} = \frac{\oNumPolGen}{2} {\bjjAbundnu} {\Mean{\hjjNGen}} {\jjMeanGen{\bLiso}} {\jjMeanGen{\lDilutionE {\lTransmittanceEGen}}} \oDurationGen \oBandwidthGen.
\end{equation}

Further details and formulae are given in Appendix~\ref{sec:ChordModelDetails}.

\section{Measurements and noise}
\label{sec:Measurements}

The final step from populations to {observables} is the measurement itself. {A full discussion of measurements and instrumental effects is deferred {(see Paper II)}, but it is worth noting that measurements are random variables that depend on the sample. They generally depend on a detector response $\iResponse$ that varies with position, but this distribution is predictable for each point on the haystack, and the observables are tractable with the point process framework.} An archetypal observable is an aggregate variable summing a quantity $\lMeasureGen$ {(with $\Mean{\lMeasureGen} \propto \lFluenceGen$)} for every broadcast in a sample mixed with some kind of background. Measurements can also include derived quantities that describe the statistics of integrated observables. A simple example is the derived signal-to-noise ratio, which is discussed in detail in Paper II. Others include cross-correlation statistics to detect large populations of faint signals \citep{Drake65}. {Another major} type {is} counts of {objects} fulfilling a particular criterion. 

Actual measurements are subject to noise, microscopic fluctuations in the instrument, background radiation, or the broadcast radiation itself. The noise increases the variance in observables beyond the sampling variance, making individual broadcasts harder to detect.  {If we are doing a measurement on a broadcast sample $\bSampleGen$, the variance in that measurement is (equation~\ref{eqn:TotalVariance})
\begin{equation}
\Var{\lMeasureGen} = \Mean{\Var{\lMeasureGen | \bSampleGen}} + \Var{\Mean{\lMeasureGen | \bSampleGen}},
\end{equation}
the first term being the mean noise variance and the second representing sample variance.}

\section{Where should we look for extragalactic broadcasts?}
\label{sec:WhichHostGalaxies}
The idea that starfaring societies replicate until they pervade a galaxy suggests large galaxies as disproportionately favorable targets for SETI.

It is plausible that spreading metasocieties are very rare, with $\ll 1$ {on average} per Milky Way-sized galaxy. The lack of evidence for an omnipresent metasociety in our own Galaxy is suggestive, although perhaps that is the result of anthropic selection effects \citep{Hanson21} {or active measures to remain hidden \citep{Ball73}}. {Rarity} is also consistent with the lack of evidence for Kardashev Type III {meta}societies \citep{Annis99,Garrett15,Griffith15,Lacki16-K3,Chen21} {and} negative extragalactic SETI results (\citealt{Horowitz93,Shostak96,Gray17}). With the framework developed, we now consider how many broadcasts {to} expect from galaxies of differing sizes, contrasting galactic clubs and spreading metasocieties.

\begin{figure}
\centerline{\includegraphics[width=8.5cm]{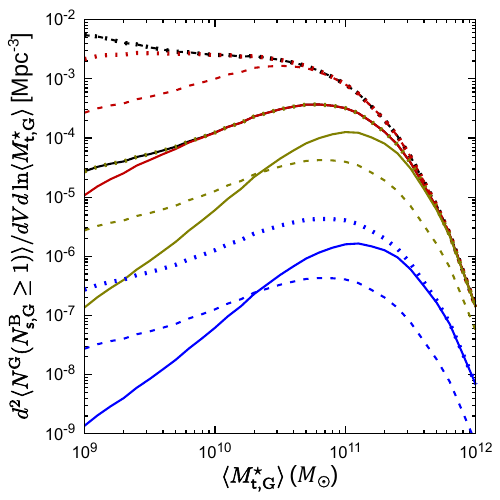}}
\figcaption{Number density of galaxies with at least one {detected beacon} in galactic clubs (dashed{, $\agAbund = 0.1$}) and expansive metasocieties (solid {for $\zgAbundEMPTY = 10^{-12}$ and dotted for $\zgAbundEMPTY = 1$, both with $\azAbund = 1$}). {Different colors are used for different $\Mean{\baNSurv}$: $100$ (black), $10^{-10}$ (red), $10^{-12}$ (gold), and $10^{-14}$ (blue).} {The fraction of galaxies with detectable beacons can be near $1$ (common broadcasts and societies), increase linearly with stellar mass (galactic clubs or common expansive metasocieties), or increase quadratically with stellar mass (rare expansive metasocieties.)}  \label{fig:BroadcastsVsMStarSys}}
\end{figure}

Suppose, under the {interchangeability} and independence assumptions, that societies in populated galaxies create ultranarrowband beacons{, all very bright and easily detected out to cosmological distances.} In the line regime of the box model (or equivalently, the chord model with ${\bgDriftRateBAR} = 0$), {$\Mean{\baNSurv} = \baAbundnuTotal \oBandwidthSurv$ for survey ${\SurvLabel}$}.

What is the probability that ${\SurvLabel}$ intercepts at least one {such} broadcast {in a comoving volume}? {And do we expect them in the more numerous small galaxies, or in the rarer large galaxies? If these beacons are sufficiently rare,} most observed societies have none active within the band observed by the survey ($\Mean{\baNSurv} \ll 1$), meaning equation~\ref{eqn:PInterceptionRare} is a good estimate of the probability that at least one broadcast passing the luminosity cut is intercepted. The stellar mass density distribution of galaxies with a detectable broadcast is given by
\begin{equation}
\frac{d^2 \Mean{\gN (\bgNSurv \ge 1)}}{d\VolVar d\ln {\Mean{\hgMAggTime}}} = \fpP(\bgNSurv \ge 1) \frac{d^2 \Mean{\gN}}{d\VolVar d\ln {\Mean{\hgMAggTime}}} .
\end{equation}
{To find this mass function of galaxies with detected beacons,} I adopt the $z \sim 0.1$ galaxy mass distributions from \citet{Moustakas13}, and assume $\Mean{\hM} = 0.2\ \Msun$ motivated by \citet{Chabrier03}. 

I compare the expansive metasociety and galactic club scenario in Figure~\ref{fig:BroadcastsVsMStarSys}. Expansive metasocieties are assumed {by default} to be fairly rare ($\zgAbundEMPTY = 10^{-12}$) but {they} plant one society around every star ($\azAbund = 1$), so that $\Mean{\agNSurv} \approx 10^{10}$ {for $\Mean{\hgNSurv} = \Mean{\hgNTime} = 10^{11}$, a moderate sized galaxy}. The {comparable} galactic club scenario has $\agAbund = 0.1${, so that the mean number of societies in a galaxy of the same size is nearly equal, $\Mean{\agNSurv} = 10^{10}$ for $\Mean{\hgNTime} = 10^{11}$}. This is likely extremely generous {to the galactic club models because abundances this high} require a mean societal lifespan of order {$1\ \Gyr$}. {Note that in the galactic club scenario, the societies are basically spread evenly between galaxies, while in the expansive metasociety scenario, only about $10\%$ of $\Mean{\hgNTime} = 10^{11}$ galaxies are inhabited, but these have {$10$} times the {mean} number of societies.}

As seen in Figure~\ref{fig:BroadcastsVsMStarSys}, there are two regimes of behavior: an asymptotic density distribution when {$\Mean{\bgNSurv} \gg \Mean{\hgNTime}^{-1}$}, and a steeper form when {$\Mean{\bgNSurv} \ll \Mean{\hgNTime}^{-1}$}. In the latter case, broadcasts are rare, and thus bigger galaxies are more likely to have one. Detections are biased to more massive galaxies in the expansive metasociety scenario{, where} $\fpP({\bgNSurv} \ge 1)$ has a quadratic dependence on {the expected number of stars at the present,} $\Mean{\hgNTime}$.

The reason for the quadratic dependence is simple: both $\Mean{\zgNTime}$ and $\Mean{{\azNSurv}}$ independently grow with stellar mass. When expansive metasocieties are rare, larger galaxies are proportionately more likely to host one simply because they have more stars for them to evolve around. Then, when one does appear and spread across the galaxy, there are more stars to populate, resulting in more societies, and thus more broadcasts. In contrast, while more stars {mean} more societies in the galactic club scenario, each society is confined to one star regardless of galaxy mass.\footnote{There are additional reasons to favor larger galaxies: their stellar populations tend to be older {\citep[e.g.,][]{Gallazzi05,Conroy14}}, allowing for more time for ETIs to evolve, and they tend to have higher metallicities according to the mass-metallicity relationship, suggesting more planets may be around \citep{Dayal15}.} {This effect applies only when expansive metasocieties are rare; when they are extremely common (dotted lines in the figure), the linear dependence on mass is restored -- though with a higher constant of proportionality, because we expect many more societies in an expansive metasociety.}

This demonstration suggests that if interstellar migration is a credible possibility, extragalactic SETI {c}ould do well to focus on {high-mass} galaxies. Breakthrough Listen's nearby galaxy survey observes several large galaxies, including ellipticals in the Virgo cluster \citep{Isaacson17}. It also has commensal access to MeerKAT, which will conduct deep observations of the Fornax cluster, which also is home to many large galaxies \citep{Czech21}.

\section{Conclusion}
\label{sec:Conclusion}

If interstellar travel and migration {are} indeed possible, then ETIs are unlike known astrophysical phenomena in that they can reproduce. Replication can amplify quirks of history onto galactic scales. Thus, supposing that starfaring ETIs are rare, {one} galaxy could have no ETIs while {another, astrophysically} indistinguishable{,} could have billions of inhabited worlds. This motivates the use of a {probabilistic} treatment of the observable technosignatures of galaxies, wherein different {galaxies} can have wildly different broadcast distributions.

This work introduces a framework that walks step by step from this population to measurements. Galaxies are treated as isolated systems {within the universe}. Within them, ETIs are congregated into localized societies. Entire lineages of societies are aggregated into metasocieties, which can cover the entire galaxy. Broadcasts are produced by societies. {These objects are organized in a random tree structure, where objects can have ancestor hosts and in turn host descendants.}

{Each of these objects is described by a parameter tuple}, points in their respective ``haystacks'' \citep[c.f.,][]{Wright18}. {Their populations are random point {processes} in the haystacks,} one process defined for every possible host. Every point {process} has a distribution (intensity), describing the mean number of objects in different parts of the haystack. {A selection {is} a thinning according to position in the haystack, within some window, for some host object.} {The result is a biased random set, which is realized as a sample.} {The Poisson point process is a simple description of populations {in realized hosts} when objects} are independent {and interchangeable}. {Observables can then be calculated using compound Poisson distributions.} {Given a distribution of random objects, the population is described by a mixture of Poisson point processes called a Cox point process.} The box model and the chord models specifically describe the observables of broadcasts. 

I present several examples and models to {show} the scope of the framework{:}
\begin{itemize}
\item Lifespan bias{, in which longer-lived objects have a larger temporal ``cross {section,}'' is viewed as the time-limited window reaching farther into the haystack as lifespan increases} {(section~\ref{sec:LifespanBias}; Figure~\ref{fig:LifespanBias}). When sampling with a window ${\GenLabel}$ with duration $\oDurationGen$, the duration mean is biased as
$$\jkMeanGen{f(\jjDuration)} = \frac{{\jkMean{\jjDuration f(\jjDuration)}} + \oDurationGen {\jkMean{f(\jjDuration)}}}{{\jkMean{\jjDuration}} + \oDurationGen},$$
with each biased moment of object lifespan $\jjDuration$ for short windows depending on the next unbiased moment. Selection bias also shows up in the chord model, where broadcasts with high drift rates are more likely to be sampled.}
\item {T}he formalism can be used to rederive {the} Drake equation for classical assumptions about the lack of significant interstellar travel {(section~\ref{sec:Drake})}. {It follows from integrating the societal and metasocietal distribution functions in the classical scenario. This work's version says that the instantaneous mean number of communicative societies in a galaxy is
$$\Mean{\agNTime (\TimeVar)} \approx \gRateStar \gMean{\hFeti} \gMean{\zFC} \gMean{\zDuration},$$
the product of a constant {star formation} rate, the mean fraction of stars hosting ETIs, the mean fraction of societies becoming communicative, and the mean lifespan of societies.}
\item {Expressions for the mean and variance of aggregate random variables are presented for ``galactic club'' and ``expansive metasociety'' scenarios are presented (Table~\ref{table:ScenarioEqns}), under the interchangeability assumption. They apply to the total intercepted emission from all broadcasts in a galaxy. The fluence of the population is proportional to the number of stars, rate/abundance of broadcasts, mean fluence of individual broadcasts, and additional factors related to the bandwidth and duration of the broadcasts and windows (section~\ref{sec:AggregateEmission}):
\begin{multline}
\nonumber \Mean{\mjjFluenceGen}  \approx \bjjRatenu \Mean{\hjjNGen} (\oBandwidthGen + \jjMean{\bBandwidthTimeOFGen}) (\oDurationGen + \jjMean{\bDuration}) \\
\cdot \jjMean{\lFluenceGen} .
\end{multline}
Variance in the aggregate variable increases as the interchangeability assumption is relaxed (Section~\ref{sec:NonInterchangeableExample}).}
\item {T}he presence of expansive metasocieties can actually result in the number of detected broadcasts depending quadratically on stellar mass (section~\ref{sec:WhichHostGalaxies}). {This is because the number of opportunities for an interstellar metasociety to arise in the first place is proportional to the number of stars, and then independently, the number of sites for independent broadcasting societies is again proportional to the number of stars. Larger galaxies may then be disproportionately likely to host detectable technosignatures {(Figure~\ref{fig:BroadcastsVsMStarSys})}.}
\end{itemize}

Future papers of the series will consider limits on broadcast populations from total radio emission and individual searches in face of confusion (Paper II), as well as constraints on populations of broadcasting galaxies from source counts {and commensal searches of background galaxies} (Paper III). 

{T}he ideas behind the framework can be applied to other phenomena, though with a different hierarchy of objects. {Transients like {f}ast {r}adio {b}ursts are generated by discrete objects like neutron stars, each of which could yield a whole population of events. In turn, some kinds of sites may be found only in globular clusters or other {subgalactic} environments \citep[e.g.,][]{Kirsten22}. Thus, like broadcasts, natural transients can be clustered into multiple levels of hosts, which may be described by point processes.}

\acknowledgments
{I thank the Breakthrough Listen program for their support. Funding for \emph{Breakthrough Listen} research is sponsored by the Breakthrough Prize Foundation (\url{https://breakthroughprize.org/}).  In addition, I acknowledge the use of NASA's Astrophysics Data System and arXiv for this research. {I am grateful to the referee for {their dedicated work in reviewing Papers I and II, including} thorough reading{s;} and I thank the statistics editor for guiding me to useful references.}}

\appendix

\section{Extreme value theory and regularization}
\label{sec:ExtremeValue}

\subsection{Review of basic extreme value theory}
\label{sec:ExtremeValueBasics}
Suppose we have a {collection} of $n$ random variables $X_i$ (with $i$ ranging from $1$ to $n$), all {i.i.d.}.  We can sort the values assigned to the $X_i$ and define the order statistic $X_{(i)}$ as the {$i$th} smallest value.  The order statistics themselves are random variables, each with their own well-defined probability distribution ({e.g.,} \citealt{Gumbel58}; {\citealt{Coles01,Castillo05,Embrechts13}}).  Most noteworthy are the minimum $\XOrderMin = \min X_i$ and the maximum $\XOrderMax = \max X_i$.  Their probability distributions are easy to calculate, since it is just the probability that every $X_i$ independently is greater or smaller than $x$.  Given the CDF $\CDF{X}$ and the complementary cumulative distribution function (CCDF) $\CCDF{X} \equiv 1 - \CDF{X}$, the maximum's CDF is 
\begin{equation}
\label{eqn:CDFXMax_FixedN}
{\CDF{\XOrderMax | n}(x) = \prod_{i = 1}^n \CDF{X_i}(x) = [\CDF{X_i}(x)]^n}
\end{equation}
and the minimum's CCDF is 
\begin{equation}
{\CCDF{\XOrderMin}(x) = \prod_{i = 1}^n \CCDF{X_i}(x) = [\CCDF{X_i}(x)]^n} .
\end{equation}  
These random variables generally converge to one of the three types of extreme value distributions -- the Gumbel distribution if $X_i$ {has} an exponential-like PDF tail (including the normal distribution), the Fr\'echet distribution when $X_i$ {has} a {power-law-like} PDF tail, and the Weibull distribution when {they have} finite support \citep[e.g.,][]{Gumbel58,Castillo05}.

{However,} when observing a population of astronomical objects in a field, the actual number sampled is a random variable $N$.  For Poissonian $N$, $\XOrderMin = \min_{1 \le i \le N} X_i$ and $\XOrderMaxNRV = \max_{1 \le i \le N} X_i$ each have a mixture probability distribution, a weighted sum of the probability distributions conditionalized on each value of $n$:
\begin{align}
\nonumber \CDF{\XOrderMaxNRV}(x) & = \frac{\sum_{n = 1}^{\infty} \fpP(N = n) \CDF{\XOrderMaxNRV | N}(x | n)}{1 - \fpP(N = 0)} = \displaystyle \frac{e^{\Mean{N} \CDF{X}(x)} - 1}{e^{\Mean{N}} - 1} \\
\label{eqn:CDFXMax_PoissonN}
\CCDF{\XOrderMin}(x) & = \frac{\sum_{n = 1}^{\infty} \fpP(N = n) \CCDF{\XOrderMin | N}(x|n)}{1 - \fpP(N = 0)} = \displaystyle \frac{e^{\Mean{N} \CCDF{X}(x)} - 1}{e^{\Mean{N}} - 1}
\end{align}
{\citep{Castillo05}.} Note that case where $N = 0$ is specifically excluded because the minimum and maximum $X_i$ in a sample with zero members is not defined.  This exclusion is necessary to ensure {the minimum and maximum} exist.

\subsection{Regularization using median values for minimum and maximum values in a Poissonian sample}
\label{sec:RegularizationDetails}
In this series, regularization is accomplished by introducing cutoffs in the values of $X$. I use quartiles for the minimum and maximum $X$ in a sample, with a lower bound given by {$\CCDF{\XOrderMin}(\XLo) = 3/4$} and an upper bound of {$\CDF{\XOrderMaxNRV}(\XHi) = 3/4$}. Equation~\ref{eqn:CDFXMax_PoissonN} gives us
\begin{align}
\nonumber \CDF{X}(\XHi) & = \frac{1}{\Mean{N}} \ln \left[\frac{3}{4} e^{\Mean{N}} + \frac{1}{4}\right] \\
\label{eqn:XEstimatedMax}
\CDF{X}(\XLo) & = 1 - \frac{1}{\Mean{N}} \ln \left[\frac{3}{4} e^{\Mean{N}} + \frac{1}{4}\right] .
\end{align}

When $\Mean{N} \gg 1$, the great majority of the probability mass is sampled.  As long as we have the inverse CDF of $X$, the bounds are approximated as $\XHi \approx \CDF{X}^{-1} (1 + \ln (3/4) / \Mean{N})$ and $\XLo \approx \CDF{X}^{-1} (-\ln (3/4) / \Mean{N})$. 

When $0 < \Mean{N} \ll 1$, $\CDF{X}(\XLo)$ and $\CDF{X}(\XHi)$ converge to $1/4$ and $3/4$, respectively. Roughly, the spread among all $X_i$ should be the spread in a single $X_i$, because virtually all observations have no events and almost all of the rest have $N = 1$. We could also use other probability thresholds for $\XLo$ and $\XHi$. Using the median $\XOrderMin$ and $\XOrderMaxNRV$ is an intuitive choice when $\Mean{N} \ga 1$, but for $\Mean{N} \ll 1$, it would imply {that the regularized variable has a} variance approaching zero, and is thus unsatisfactory.

\section{Further details for the box model}
\label{sec:BoxModelDetails}
{The distributions and averages I present here are calculated assuming a uniform spread in time and frequency (uniform ${\bjjRatenu}$) and that $(\bBandwidthBox + \oBandwidthGen)/2 \le \oNuMidGen$.}

\subsection{The overlap fraction}
If we regard time and frequency as a plane, the overlap fraction $\bOverlapGen$ is the area of intersection between two contiguous boxes, scaled to the area of the broadcast's box:
\begin{equation}
\label{eqn:OverlapBox}
\bOverlapGen \equiv \frac{\bDurationGen \bBandwidthGen}{\bDurationBox \bBandwidthBox} = \frac{\fRamp[\min(\bTStart + \bDurationBox, \oTStartGen + \oDurationGen) - \max(\bTStart, \oTStartGen)]}{\bDurationBox} \cdot~\frac{\fRamp\left[\min\left(\bNuMid + \frac{\bBandwidthBox}{2}, \oNuMidGen + \frac{\oBandwidthGen}{2}\right) - \max\left(\bNuMid - \frac{\bBandwidthBox}{2}, \oNuMidGen - \frac{\oBandwidthGen}{2}\right)\right]}{\bBandwidthBox} ,
\end{equation}
using the ramp function $\fRamp(x) = \max(0, x)$. The overlap can be decomposed into the intersection between the time ranges spanned by the sample window and the broadcast and the intersection of the frequency ranges spanned by the sample and broadcast. It reaches a maximum value when one of the time ranges is fully inside the other and likewise one frequency range contains the other, although it is possible for the sample window to contain the broadcast for the one and the broadcast to contain the {window} for the other. This maximal value is
\begin{equation}
\label{eqn:OverlapBoxMax}
\bOverlapGenMAX = \min\left(1, \frac{\oDurationGen}{\bDurationBox}\right)\min\left(1, \frac{\oBandwidthGen}{\bBandwidthBox}\right) .
\end{equation}
Among broadcasts within the sample (with a nonzero $\bOverlapGen$), the probability this value is attained is
\begin{equation}
\label{eqn:POverlapBoxMax}
\fpP (\bOverlapGen = \bOverlapGenMAX) = \frac{|\bDurationBox - \oDurationGen| |\bBandwidthBox - \oBandwidthGen|}{(\bDurationBox + \oDurationGen) (\bBandwidthBox + \oBandwidthGen)} .
\end{equation}
This probability only diverges significantly from $1$ if $\bDurationBox \approx \oDurationGen$ or $\bBandwidthBox \approx \oBandwidthGen$. \emph{A priori} we may expect that ETIs and our own efforts are uncoordinated and thus this is unlikely to happen, but there are natural lower limits to bandwidth set by interstellar scattering \citep{Cordes97}, and thus our observations may try to match the broadened bandwidth of a line. In any case, only in rare cases will the partial overlap result in $0 < \bOverlapGen \ll \bOverlapGenMAX$. The probability {density} for this regime is
\begin{equation}
\label{eqn:PDFOverlapBoxLTMax}
\PDFGen{\bOverlapGen} (\bOverlapCore | 0 < \bOverlapCore < \bOverlapGenMAX) = \frac{2}{(\bDurationBox + \oDurationGen) (\bBandwidthBox + \oBandwidthGen)} \left[\max\left(\frac{\oDurationGen}{\bDurationBox}, \frac{\bDurationBox}{\oDurationGen}\right) + \max\left(\frac{\oBandwidthGen}{\bBandwidthBox}, \frac{\bBandwidthBox}{\oBandwidthGen}\right) - 2  - 2 \ln \left(\frac{\bOverlapCore}{\bOverlapGenMAX}\right) \right]
\end{equation}
{The means used in the next section can be calculated from equations~\ref{eqn:OverlapBoxMax}--\ref{eqn:PDFOverlapBoxLTMax}.}

\subsection{Emission and selection windows}
\label{sec:BoxModelEmission}

The emission that falls into the selection box is equal to the total emission ever released in the broadcast multiplied by the overlap fraction and the fraction in observed polarizations: $\bEisoGen = \bEiso \bOverlapGen \bFPolGen$. Thus the mean restricted energy is
\begin{equation}
{\jjMean{\bEisoGen}} = \frac{\oNumPolGen}{2} {\jjMeanGen{\bEiso}} \frac{\oDurationGen \oBandwidthGen}{(\oDurationGen + \bDurationBox) (\oBandwidthGen + \bBandwidthBox)} .
\end{equation}
The variance is found using
\begin{equation}
{\jjMean{\bEisoGen^2}} = {\jjMean{\bFPolGen^2}} {\jjMeanGen{\bEiso^2}} \frac{\oDurationGen \oBandwidthGen}{(\oDurationGen + \bDurationBox) (\oBandwidthGen + \bBandwidthBox)} \max \left[1 - \frac{\bDurationBox}{3\oDurationGen}, \frac{\oDurationGen}{\bDurationBox} - \frac{\oDurationGen^2}{3\bDurationBox^2}\right] \max \left[1 - \frac{\bBandwidthBox}{3\oBandwidthGen}, \frac{\oBandwidthGen}{\bBandwidthBox} - \frac{\oBandwidthGen^2}{3\bBandwidthBox^2}\right] .
\end{equation}

Equation~\ref{eqn:dEdnudtBox} implies the number of photons per unit frequency falls as $\FreqVar^{-1}$. I find
\begin{equation}
{\jjMean{\bPhotonisoGen}} = \frac{\oNumPolGen}{2} \frac{{\jjMeanGen{\bEiso}}}{h} \frac{\oDurationGen}{(\oDurationGen + \bDurationBox) (\oBandwidthGen + \bBandwidthBox)} \ln\left(\frac{1 + \oBandwidthGen/(2 \oNuMidGen)}{1 - \oBandwidthGen/(2 \oNuMidGen)}\right),
\end{equation}
approaching $\Mean{\bEisoGen} / (h \oNuMid)$ when $\oBandwidthGen \ll \oNuMidGen$.  Furthermore, 
\begin{multline}
{\jjMean{\bPhotonisoGen^2}} = {\jjMean{\bFPolGen^2}} \frac{{\jjMeanGen{\bEiso^2}}}{h} \frac{\oDurationGen}{(\oDurationGen + \bDurationBox) (\oBandwidthGen + \bBandwidthBox)} \max \left[1 - \frac{\bDurationBox}{3\oDurationGen}, \frac{\oDurationGen}{\bDurationBox} - \frac{\oDurationGen^2}{3\bDurationBox^2}\right] \frac{1}{\bBandwidthBox^2}\\
{
 \cdot~\begin{cases}
 \displaystyle \left[4 \oBandwidthGen + 4 \oNuMidGen \ln \frac{1 - \oBandwidthGen/(2\oNuMidGen)}{1 + \oBandwidthGen/(2\oNuMidGen)} + \bBandwidthBox \left[\ln \frac{1 - \oBandwidthGen/(2\oNuMidGen)}{1 + \oBandwidthGen/(2\oNuMidGen)}\right]^2\right] & \text{if}~\bBandwidthBox \ge \oBandwidthGen\\
	\begin{aligned}
		2 \left[\bBandwidthBox\left(2 + \fDilog \frac{\bBandwidthBox}{\oNuMidGen - \oBandwidthGen/2 + \bBandwidthBox} - \fDilog \frac{\bBandwidthBox}{\oNuMidGen + \oBandwidthGen/2} - \frac{1}{2}\left[\ln \left(1 + \frac{\bBandwidthBox}{\oNuMidGen - \bBandwidthBox/2}\right)\right]^2\right) \right. \\
		+ \left(\oNuMidGen - \oBandwidthGen/2 + \bBandwidthBox\right) \ln\left(1 + \frac{\bBandwidthBox}{\oNuMidGen - \oBandwidthGen/2}\right) \left[\ln\left(1 + \frac{\bBandwidthBox}{\oNuMidGen - \oBandwidthGen/2}\right) - 1\right] \\
		\left. - \left(\oNuMidGen + \oBandwidthGen/2 - \bBandwidthBox\right) \ln\left(1 {-} \frac{\bBandwidthBox}{\oNuMidGen + \oBandwidthGen/2}\right) \left[\ln\left(1 {-} \frac{\bBandwidthBox}{\oNuMidGen + \oBandwidthGen/2}\right) - 1\right]\right]
	\end{aligned}
	& \text{if}~\bBandwidthBox \le \oBandwidthGen
\end{cases}
}
\end{multline}
with $\fDilog(x)$ being the dilogarithm (Spence's function). If $\oBandwidthGen \ll \oNuMidGen$, ${\jjMean{\bPhotonisoGen^2}} \approx {{\jjMean{\bFPolGen^2}} {\jjMean{\bEisoGen^2}}} / (h \oNuMidGen)^2$.

Mean fluences may be found by multiplying by the mean of the appropriate {transmittance/}dilution factor, ${\jjMeanGen{{\lTransmittanceR^n} \lDilutionR^n}}$ for ${\jjMeanGen{\bEmissionisoGen^n}}$ (equation~\ref{eqn:EmissionToFluence}).

Approximations when $\bDurationBox \nsim \oDurationGen$ and $\bBandwidthBox \nsim \oBandwidthGen$ are given in Table~\ref{table:BoxRegimes}.

\section{Further details for the chord model}
\label{sec:ChordModelDetails}
{The results presented here assume a constant ${\bjjAbundnu}$.}

\subsection{Selection of chords in a contiguous selection window}
\label{sec:ChordSelection}
In the chord model, the window on selected times and frequency is a contiguous {``box,''} as in the box model. Because we only constrain the behavior of the broadcast within the sample window, calculations are simplified when we ``cut'' the chord, only considering its behavior at times between $\oTStartGen$ and $\oTStartGen + \oDurationGen$. This imposes $\bTStartTimeOFGen = \oTStartGen$, $\bDurationTimeOFGen = \oDurationGen$, and $\bBandwidthTimeOFGen = |\bDriftRate| \oDurationGen$. I also use $\bNuMidTimeOFGen$, the frequency during the mid-time of the window. At other times, the band is centered on a frequency $\bNuMidTime (\TimeVar) = \bNuMidTimeOFGen + (\TimeVar - \oTStartGen - \oDurationGen/2) \bDriftRate$. At any given frequency, the center of the band crosses at time $\bTStartFreq (\FreqVar) = (\oTStartGen + \oDurationGen/2) + (\FreqVar - \bNuMidTimeOFGen) / \bDriftRate$. Note that $\oDurationGen = \oeDurationGen / (1 + \yRedshift)$, $\oBandwidthGen = \oeBandwidthGen (1 + \yRedshift)$, and $\bDriftRate = \beDriftRate (1 + \yRedshift)^2$ are in the source frame.

A broadcast crosses the selection window if $|\oNuMidGen - \bNuMidTimeOFGen| \le (\oBandwidthGen + \oDurationGen |\bDriftRate|)/2$. The mean number of broadcasts intercepted is thus {given by}
\begin{equation}
\frac{d\Mean{{\bjjNGen}}}{d\bDriftRate} = {\Mean{\hjjNGen}} {\bjjAbundnu} (\oBandwidthGen + \oDurationGen |\bDriftRate|) {\jjPDF{\bDriftRate}},
\end{equation}
noting that ${\jjPDF{\bDriftRate}}$ is the unbiased drift rate distribution. It immediately follows that $\Mean{{\bjjNGen}} = {\Mean{\hjjNGen}} {\bjjAbundnu} (\oBandwidthGen + \oDurationGen {\jjMean{|\bDriftRate|}})$, regardless of distribution, where ${\jjMean{|\bDriftRate|}}$ is the unbiased mean drift rate magnitude. The biased drift rate distribution, that is, the distribution sampled by the window, favors high drift rate broadcasts:
\begin{equation}
\label{eqn:BiasedDriftRateDist}
{\jjPDFGen{\bDriftRate}}{(\bDriftRateCore)} = \frac{1}{\Mean{{\bjjNGen}}} \frac{d\Mean{{\bjjNGen}}}{d\bDriftRate}{(\bDriftRateCore)}  = \frac{\oBandwidthGen + \oDurationGen |{\bDriftRateCore}|}{\oBandwidthGen + \oDurationGen {\jjMean{|\bDriftRate|}}} {\jjPDF{\bDriftRate}}{(\bDriftRateCore)},
\end{equation}
 (see {equation}~\ref{eqn:BiasedPDF}). 

When expressing calculations, it is convenient to define dimensionless variables, placing the drift rates in ``natural'' units defined by the window: $\bDriftNatGen \equiv \bDriftRate \oDurationGen / \oBandwidthGen$ and ${\bjjDriftNatMidGen} = {\bjjDriftRateMid} \oDurationGen / \oBandwidthGen$. Chords with $|\bDriftNatGen| \ll 1$ behave like lines in the box model.

According to the uniform drift rate distribution model, the drift rates have a uniform unbiased distribution (equation~\ref{eqn:PDFUniformDrift}) spanning the range given by $|{\bjjDriftRateMid} - \bDriftRate| \le {\bjjDriftRateBAR}$. We have
\begin{equation}
{\jjMean{|\bDriftRate|}} = \begin{cases}
                                   {\bjjDriftRateBAR}/2 {\cdot} (1 + ({\bjjDriftRateMid}/{\bjjDriftRateBAR})^2) & {\text{if}~|{\bjjDriftRateMid}| \le {\bjjDriftRateBAR}}\\
																	|{\bjjDriftRateMid}| & {\text{if}~|{\bjjDriftRateMid}| \ge {\bjjDriftRateBAR}}
																	\end{cases}
\end{equation}
and ${\jjMean{\bDriftRate^2}} = {\bjjDriftRateBAR}^2/3 + {{\bjjDriftRateMid}}^2$. {T}he mean drift rate of a sampled broadcast {comes} from equation~\ref{eqn:BiasedDriftRateDist}:
\begin{equation}
{\jjMeanGen{|\bDriftRate|}} = \begin{cases}
                             \displaystyle \frac{{\bjjDriftRateBAR}}{2} \frac{1 + (2/3) {\bjjDriftNatBARGen} + 2 ({\bjjDriftNatMidGen})^2/{\bjjDriftNatBARGen} + ({\bjjDriftNatMidGen}/{\bjjDriftNatBARGen})^2}{1 + (1/2)({\bjjDriftNatBARGen} + ({\bjjDriftNatMidGen})^2/{\bjjDriftNatBARGen})} & \text{if}~|{\bjjDriftRateMid}| \le {\bjjDriftRateBAR}\\
														 \displaystyle \frac{\oBandwidthGen}{\oDurationGen} \frac{{(1/3)}{\bjjDriftNatBARGen}^2 + |{\bjjDriftNatMidGen}| + ({\bjjDriftNatMidGen})^2}{1 + |{\bjjDriftNatMidGen}|} & \text{if}~|{\bjjDriftRateMid}|~{\ge} {\bjjDriftRateBAR}
														\end{cases},
\end{equation}
with ${\bjjDriftNatBARGen} \equiv {\bjjDriftRateBAR} \oDurationGen / \oBandwidthGen$. 

Furthermore, although the unbiased drift rate magnitude CDF is {$\jjCDF{|\bDriftRate|}(\bDriftRateCore) = \jjCDF{|\bDriftRate|}(\bDriftRateCore) - \jjCDF{|\bDriftRateCore|}(-\bDriftRateCore)$}, the biased CDF is
\begin{equation}
{\jjCDFGen{|\bDriftRate|}}({\bDriftNatGen}) = \begin{cases}
           \displaystyle \frac{{\bDriftNatGen}}{{\bjjDriftNatBARGen}} \frac{1 + (1/2) {\bDriftNatGen}}{1 + (1/2)({\bjjDriftNatBARGen} + ({\bjjDriftNatMidGen})^2/{\bjjDriftNatBARGen})} & \text{if}~|{\bjjDriftRateMid}| \le {\bjjDriftRateBAR}~\text{and}~{\bjjDriftRateBAR} - |{\bjjDriftRateMid}| \ge \bDriftRateCore\\
					 \displaystyle \frac{{\bDriftNatGen}^2 + 2 {\bDriftNatGen} + 2 ({\bjjDriftNatBARGen} - |{\bjjDriftNatMidGen}|) + ({\bjjDriftNatBARGen} - |{\bjjDriftNatMidGen}|)^2}{4{\bDriftNatGen} [1 + (1/2)({\bjjDriftNatBARGen} + ({\bjjDriftNatMidGen})^2/{\bjjDriftNatBARGen})]} & \text{if}~|{\bjjDriftRateMid}| \le {\bjjDriftRateBAR}~\text{and}~{\bjjDriftRateBAR} - |{\bjjDriftRateMid}| \le \bDriftRateCore \le {\bjjDriftRateBAR} + |{\bjjDriftRateMid}|\\
           0 & \text{if}~|{\bjjDriftRateMid}| \ge {\bjjDriftRateBAR}~\text{and}~|{\bjjDriftRateMid}| - {\bjjDriftRateBAR} \ge \bDriftRateCore\\
					 \displaystyle \frac{{\bDriftNatGen}^2 + 2 {\bDriftNatGen} - 2 (|{\bjjDriftNatMidGen}| - {\bjjDriftNatBARGen}) {{-}} (|{\bjjDriftNatMidGen}| - {\bjjDriftNatBARGen})^2}{4{\bDriftNatGen} [1 + |{\bjjDriftNatMidGen}|]} & \text{if}~|{\bjjDriftRateMid}| \ge {\bjjDriftRateBAR}~\text{and}~|{\bjjDriftRateMid}| - {\bjjDriftRateBAR} \le \bDriftRateCore \le |{\bjjDriftRateMid}| + {\bjjDriftRateBAR}\\
					 1 & \text{if}~{\bjjDriftRateBAR} + |{\bjjDriftRateMid}| \le \bDriftRateCore
					\end{cases},
\end{equation}
with ${\bDriftNatGen} \equiv \bDriftRate \oDurationGen / \oBandwidthGen$. When the drift rates span a large range (${\bjjDriftRateBAR} \gg \oBandwidthGen / \oDurationGen$ and $\ge |{\bjjDriftRateMid}|$), covering the entire range of integration, the biased probability has a quadratic dependence on $\bDriftRateCore/{\bjjDriftRateBAR}$.

\subsection{Statistics of chord duration in sample}

The central quantity in the chord model is $\bDurationGen$, the time it takes the chord to cross the window:
\begin{equation}
\bDurationGen = \min[\bTStartFreq (\oNuMidGen + \sign(\bDriftRate) \oBandwidthGen/2), \oTStartGen + \oDurationGen] - \max[\bTStartFreq (\oNuMidGen - \sign(\bDriftRate) \oBandwidthGen/2), \oTStartGen]
\end{equation} 
It can be shown that 
\begin{equation}
\bDurationGen = \begin{cases}
									\displaystyle \min\left(\frac{\oBandwidthGen}{|\bDriftRate|}, \oDurationGen\right) & \displaystyle \text{if}~|\oNuMidGen - \bNuMidTimeOFGen| \le \frac{|\oDurationGen|\bDriftRate| - \oBandwidthGen|}{2}\\
									\displaystyle \frac{\oDurationGen + \oBandwidthGen/|\bDriftRate|}{2} - \frac{|\oNuMidGen - \bNuMidTimeOFGen|}{|\bDriftRate|} & \displaystyle \text{if}~\frac{|\oDurationGen|\bDriftRate| - \oBandwidthGen|}{2} \le |\oNuMidGen - \bNuMidTimeOFGen| \le \frac{|\oDurationGen|\bDriftRate| + \oBandwidthGen|}{2}\\
                  0   & \displaystyle \text{if}~|\oNuMidGen - \bNuMidTimeOFGen| \ge \frac{|\oDurationGen|\bDriftRate| + \oBandwidthGen|}{2} .
									\end{cases}
\end{equation}

{For} a {population in a host ${{\JMark}}$ with a uniform $\bNuMid$ distribution}, {sampled} broadcasts {with} a fixed drift rate (necessarily with nonzero $\bDurationGen$) have:
\begin{equation}
{{\jjMeanGen{\bDurationGen^n | \bDriftRate}} = \frac{\oDurationGen^n}{1 + |\bDriftNatGen|} 
	\begin{cases}
	\displaystyle 1 - \frac{n - 1}{n + 1} |\bDriftNatGen|                                          & \text{if}~|\bDriftNatGen| \le 1~\text{and}~n \ne -1\\
	  \displaystyle \frac{1}{|\bDriftNatGen|^n} \left(|\bDriftNatGen| - \frac{n - 1}{n + 1}\right) & \text{if}~|\bDriftNatGen| \ge 1~\text{and}~n \ne -1 
	\end{cases}}
\end{equation}

The mean of a random variable for a broadcast sample is weighted by the number of broadcasts that are in the sample. This leads to the biasing of the mean {toward} high drift rate broadcasts:
 \begin{equation}
\label{eqn:BoxChordMean}
{\jjMeanGen{f(\bDurationGen)}} = \int_{-\infty}^{\infty} {\jjMeanGen{f(\bDurationGen) | \bDriftRate}} {\jjPDFGen{\bDriftRate}} d\bDriftRate = \frac{1}{\oBandwidthGen + \oDurationGen {\jjMean{|\bDriftRate|}}} \int_{-\infty}^{\infty} (\oBandwidthGen + \oDurationGen |\bDriftRate|) \Mean{f(\bDurationGen) | \bDriftRate} {\jjPDF{\bDriftRate}} d\bDriftRate .
\end{equation}
for the random variable defined by applying the function $f$ to $\bDurationGen$. This immediately gives us {the ${\GenLabel}$-relative mean}
\begin{equation}
{\jjMean{\bDurationGen}} = \frac{\oDurationGen}{1 + \oDurationGen {\jjMean{|\bDriftRate|}} / \oBandwidthGen}
\end{equation}

For the uniform drift rate distribution, I find:
\begin{multline}
{\jjMean{\bDurationGen^2}} = \oDurationGen {\jjMean{\bDurationGen}} \cdot~\\
\begin{cases}
			                     \displaystyle 1 - (1/6)({\bjjDriftNatBARGen} + ({\bjjDriftNatMidGen})^2/{\bjjDriftNatBARGen}) & \begin{aligned}
													 \text{if}~& |{\bjjDriftNatMidGen}| \le {\bjjDriftNatBARGen}~\text{and}\\
													           & {\bjjDriftNatBARGen} + |{\bjjDriftNatMidGen}| \le 1\end{aligned}\\
													\displaystyle \frac{1}{4{\bjjDriftNatBARGen}}\left[ 1 + 2({\bjjDriftNatBARGen} - |{\bjjDriftNatMidGen}|)  - (1/3)({\bjjDriftNatBARGen} - |{\bjjDriftNatMidGen}|)^2 + 2 \ln ({\bjjDriftNatBARGen} + |{\bjjDriftNatMidGen}|) + \frac{(2/3)}{{\bjjDriftNatBARGen} + |{\bjjDriftNatMidGen}|} \right] & \begin{aligned}
													\text{if}~& |{\bjjDriftNatMidGen}| \le {\bjjDriftNatBARGen}~\text{and}\\
                                    & {|1 - \bjjDriftNatBARGen| \le |\bjjDriftNatMidGen|}\end{aligned}\\
													 \displaystyle \frac{1}{2{\bjjDriftNatBARGen}} \left[1 + \ln ({\bjjDriftNatBARGen}^2 - |{\bjjDriftNatMidGen}|^2) + \frac{(1/3)}{{\bjjDriftNatBARGen} + |{\bjjDriftNatMidGen}|} + \frac{(1/3)}{{\bjjDriftNatBARGen} - |{\bjjDriftNatMidGen}|}\right] & \begin{aligned}
													\text{if}~& |{\bjjDriftNatMidGen}| \le {\bjjDriftNatBARGen}~\text{and}\\
													          & 1 \le {\bjjDriftNatBARGen} - |{\bjjDriftNatMidGen}|\end{aligned}\\				
													 \displaystyle 1 - (1/3)|{\bjjDriftNatMidGen}| & \begin{aligned}
													\text{if}~& |{\bjjDriftNatMidGen}| \ge {\bjjDriftNatBARGen}~\text{and}\\
													          & {\bjjDriftNatBARGen} + |{\bjjDriftNatMidGen}| \le 1\end{aligned}\\
													 \displaystyle \frac{1}{4{\bjjDriftNatBARGen}} \left[1 - 2(|{\bjjDriftNatMidGen}| - {\bjjDriftNatBARGen}) + (1/3)(|{\bjjDriftNatMidGen}| - {\bjjDriftNatBARGen})^2 + 2 \ln ({\bjjDriftNatBARGen} + |{\bjjDriftNatMidGen}|) + \frac{(2/3)}{{\bjjDriftNatBARGen} + |{\bjjDriftNatMidGen}|}\right] & \begin{aligned}
													\text{if}~& |{\bjjDriftNatMidGen}| \ge {\bjjDriftNatBARGen}~\text{and}\\
                                    & {|1 - |\bjjDriftNatMidGen|| \le |\bjjDriftNatBARGen|}\end{aligned}\\
													 \displaystyle \frac{1}{2 {\bjjDriftNatBARGen}} \left[\ln \frac{|{\bjjDriftNatMidGen}| + {\bjjDriftNatBARGen}}{|{\bjjDriftNatMidGen}| - {\bjjDriftNatBARGen}} + \frac{(1/3)}{|{\bjjDriftNatMidGen}| + {\bjjDriftNatBARGen}} - \frac{(1/3)}{|{\bjjDriftNatMidGen}| - {\bjjDriftNatBARGen}}\right] & \begin{aligned}
													\text{if}~& |{\bjjDriftNatMidGen}| \ge {\bjjDriftNatBARGen}~\text{and}\\
													          & 1 \le |{\bjjDriftNatMidGen}| - {\bjjDriftNatBARGen} \end{aligned}
		\end{cases} .
\end{multline} 

\subsection{Emission and fluence}
\label{sec:BoxChordModelFluence}

The effective isotropic energy is found using the (source-frame) effective luminosity and the (source-frame) chord crossing time:
\begin{equation}
\bEisoGen = \bFPolGen \bLiso \bDurationGen .
\end{equation}
The polarization, luminosity, and chord duration are all presumed independent, so the mean energy fluence is
\begin{equation}
{\jjMean{\bEisoGen}} = \frac{|\oPolSetGen|}{2} {\jjMeanGen{\bLiso}} \frac{\oDurationGen}{1 + \oDurationGen {\jjMean{|\bDriftRate|}} / \oBandwidthGen} .
\end{equation}
If the uniform drift rate distribution is adopted and the mean drift rate is near zero ($|{\bjjDriftRateMid}| \ll {\bjjDriftRateBAR}$),
\begin{equation}
{\jjMean{\bEisoGen^2}} = {\jjMean{\bFPolGen^2}} {\jjMeanGen{\bLiso^2}} {\jjMean{\bDurationGen^2}} \approx {\jjMean{\bFPolGen^2}} {\jjMeanGen{\bLiso^2}} \oDurationGen^2 \cdot~\begin{cases}
                       \displaystyle \frac{1 - (1/6) {\bjjDriftNatBARGen}}{1 + (1/2) {\bjjDriftNatBARGen}} & \text{if}~{\bjjDriftNatBARGen \le 1}\\
											 \displaystyle \frac{1}{2{\bjjDriftNatBARGen}} \frac{1 + \ln {\bjjDriftNatBARGen} + (2/3) {\bjjDriftNatBARGen}^{-1}}{1 + (1/2) {\bjjDriftNatBARGen}} & \text{if}~{\bjjDriftNatBARGen \ge 1}
\end{cases} .
\end{equation}
From this windowed energy, it is easy to find ${\jjMean{\lFluenceEGen}}$ (equation~\ref{eqn:EmissionToFluence}):
\begin{equation}
{\jjMean{\lFluenceEGen}} = \frac{|\oPolSetGen|}{2} {\jjMeanGen{\bLiso}} {\jjMeanGen{{\lTransmittanceEGen} \lDilutionE}} \frac{\oDurationGen}{1 + \oDurationGen {\jjMean{|\bDriftRate|}} / \oBandwidthGen} .
\end{equation}

In the chord model, if ${\bjjDriftRateBAR} \gg \oBandwidthGen / \oDurationGen$, the typical broadcast in a sample is a high drift rate signal that slashes through the sample ``box'' from low to high frequency or {vice versa}. Fixing broadcast effective isotropic luminosity, these typical broadcasts are much fainter than the much slower {drifting} broadcasts with $\bDriftRate \approx 0$. Now, the mean total fluence is found by integrating the sampled drift rate distribution,
\begin{equation}
\frac{d\Mean{{\mjjFluenceEGen}}}{d\bDriftRate} = \frac{d\Mean{{\bjjNGen}}}{d\bDriftRate} {\jjMeanGen{\lFluenceEGen | \bDriftRate}} = \frac{|\oPolSetGen|}{2} {\jjMean{\bLiso}} {\jjMeanGen{{\lTransmittanceEGen} \lDilutionE}} {\bjjAbundnu} {\Mean{\hjjNGen}} \oDurationGen \oBandwidthGen {\jjPDF{\bDriftRate}}.
\end{equation}
We find $\Mean{{\mjjFluenceEGen} (|\bDriftRate| \le \bDriftRateCore)} = \Mean{{\mjjFluenceEGen}} {{\jjCDF{|\bDriftRate|}} (\bDriftRateCore)}$ {(note the use of the unbiased CDF)}. The fastest drifting half of the \emph{population} contributes half the expected fluence. Yet this half is overrepresented in a sample. In the uniform drift rate model, if $|\bDriftRate| \gg \oBandwidthGen / \oDurationGen$, the bottom quartile of \emph{sampled} broadcasts in $|\bDriftRate|$ contribute half the fluence.

As a result, the observed ${\mjjFluenceEGen}$ typically underestimates $\Mean{{\mjjFluenceEGen}}$ because small samples tend to miss the slowest {drifting} broadcasts. For a uniform drift rate distribution, a few broadcasts suffice to counteract this effect. Biasing becomes more of an issue if the distribution has a long tail.  For an exponential distribution ${\jjPDF{\bDriftRate}} = (1/{\bjjDriftRateBAR}) \exp(-\bDriftRate/{\bjjDriftRateBAR})$, the fraction of sampled broadcasts that contribute half the fluence falls to $(1 - \ln 2)/2 \approx 15\%$ as ${{\bjjDriftRateBAR}} \to \infty$. Thus, we might need to sample $\sim 5 \endash 10$ broadcasts until ${\mjjFluenceEGen}$ starts converging to $\Mean{{\mjjFluenceEGen}}$. Heavy-tailed distributions like power laws would require still more broadcasts to be sampled before {the aggregate emission converges to expectations from} $\Mean{{\mjjFluenceEGen}}$.

\bibliographystyle{aasjournal}
\bibliography{ETIPopulations_1_Formalism_arXiv}

\begin{thebibliography}{}
\expandafter\ifx\csname natexlab\endcsname\relax\def\natexlab#1{#1}\fi
\providecommand{\url}[1]{\href{#1}{#1}}
\providecommand{\dodoi}[1]{doi:~\href{http://doi.org/#1}{\nolinkurl{#1}}}
\providecommand{\doeprint}[1]{\href{http://ascl.net/#1}{\nolinkurl{http://ascl.net/#1}}}
\providecommand{\doarXiv}[1]{\href{https://arxiv.org/abs/#1}{\nolinkurl{https://arxiv.org/abs/#1}}}

\bibitem[{{Abbot} \& {Switzer}(2011)}]{Abbot11}
{Abbot}, D.~S., \& {Switzer}, E.~R. 2011, \apjl, 735, L27,
  \dodoi{10.1088/2041-8205/735/2/L27}

\bibitem[{{Adams} \& {Laughlin}(1997)}]{Adams97}
{Adams}, F.~C., \& {Laughlin}, G. 1997, Reviews of Modern Physics, 69, 337,
  \dodoi{10.1103/RevModPhys.69.337}

\bibitem[{{Adelson}(1966)}]{Adelson66}
{Adelson}, R.~M. 1966, Journal of the Operational Research Society, 17, 73

\bibitem[{{Annis}(1999)}]{Annis99}
{Annis}, J. 1999, Journal of the British Interplanetary Society, 52, 19.
\newblock \doarXiv{astro-ph/9901322}

\bibitem[{{Armstrong} \& {Sandberg}(2013)}]{Armstrong13}
{Armstrong}, S., \& {Sandberg}, A. 2013, Acta Astronautica, 89, 1,
  \dodoi{10.1016/j.actaastro.2013.04.002}

\bibitem[{{Arnold}(2005)}]{Arnold05}
{Arnold}, L. F.~A. 2005, \apj, 627, 534, \dodoi{10.1086/430437}

\bibitem[{Baddeley(2007)}]{Baddeley07}
Baddeley, A. 2007, in Stochastic Geometry, ed. W.~{Weil} (Berlin: Springer),
  1--75, \dodoi{10.1007/978-3-540-38175-4_1}

\bibitem[{{Badescu}(2011)}]{Badescu11}
{Badescu}, V. 2011, \icarus, 216, 485, \dodoi{10.1016/j.icarus.2011.09.013}

\bibitem[{{Badescu} \& {Cathcart}(2006)}]{Badescu06}
{Badescu}, V., \& {Cathcart}, R.~B. 2006, Acta Astronautica, 58, 119,
  \dodoi{10.1016/j.actaastro.2005.09.005}

\bibitem[{{Bailyn}(1995)}]{Bailyn95}
{Bailyn}, C.~D. 1995, \araa, 33, 133,
  \dodoi{10.1146/annurev.aa.33.090195.001025}

\bibitem[{{Balbi} \& {{\'C}irkovi{\'c}}(2021)}]{Balbi21}
{Balbi}, A., \& {{\'C}irkovi{\'c}}, M.~M. 2021, \aj, 161, 222,
  \dodoi{10.3847/1538-3881/abec48}

\bibitem[{{Balbi} \& {Tombesi}(2017)}]{Balbi17}
{Balbi}, A., \& {Tombesi}, F. 2017, Scientific Reports, 7, 16626,
  \dodoi{10.1038/s41598-017-16110-0}

\bibitem[{{Ball}(1973)}]{Ball73}
{Ball}, J.~A. 1973, \icarus, 19, 347, \dodoi{10.1016/0019-1035(73)90111-5}

\bibitem[{{Barbour} \& {Chryssaphinou}(2001)}]{Barbour01}
{Barbour}, A.~D., \& {Chryssaphinou}, O. 2001, Annals of Applied Probability,
  964, \dodoi{10.1214/aoap/1015345355}

\bibitem[{{Bas}(2019)}]{Bas19}
{Bas}, E. 2019, Basics of Probability and Stochastic Processes (Berlin:
  Springer), \dodoi{10.1007/978-3-030-32323-3}

\bibitem[{{Benford} \& {Benford}(2016)}]{Benford16}
{Benford}, J.~N., \& {Benford}, D.~J. 2016, \apj, 825, 101,
  \dodoi{10.3847/0004-637X/825/2/101}

\bibitem[{{Blair} \& {Zadnik}(1993)}]{Blair93}
{Blair}, D.~G., \& {Zadnik}, M.~G. 1993, \aap, 278, 669

\bibitem[{{Borra}(2012)}]{Borra12}
{Borra}, E.~F. 2012, \aj, 144, 181, \dodoi{10.1088/0004-6256/144/6/181}

\bibitem[{{Bracewell}(1975)}]{Bracewell75}
{Bracewell}, R.~N. 1975, {The Galactic Club: Intelligent Life in Outer Space}
  (San Francisco: W. H. Freeman and Company)

\bibitem[{{Brillinger}(1969)}]{Brillinger69}
{Brillinger}, D.~R. 1969, Annals of the Institute of Statistical Mathematics,
  21, 215, \dodoi{10.1007/BF02532246}

\bibitem[{{Brin}(1983)}]{Brin83}
{Brin}, G.~D. 1983, \qjras, 24, 283

\bibitem[{{Carrigan}(2012)}]{Carrigan12}
{Carrigan}, R.~A. 2012, Acta Astronautica, 78, 121,
  \dodoi{10.1016/j.actaastro.2011.12.002}

\bibitem[{{Carroll-Nellenback} {et~al.}(2019){Carroll-Nellenback}, {Frank},
  {Wright}, \& {Scharf}}]{CarrollNellenback19}
{Carroll-Nellenback}, J., {Frank}, A., {Wright}, J., \& {Scharf}, C. 2019, \aj,
  158, 117, \dodoi{10.3847/1538-3881/ab31a3}

\bibitem[{{Carter}(1983)}]{Carter83}
{Carter}, B. 1983, Philosophical Transactions of the Royal Society of London
  Series A, 310, 347, \dodoi{10.1098/rsta.1983.0096}

\bibitem[{{Castillo} {et~al.}(2005){Castillo}, {Hadi}, {Balakrishnan}, \&
  {Sarabia}}]{Castillo05}
{Castillo}, E., {Hadi}, A.~S., {Balakrishnan}, N., \& {Sarabia}, J.~M. 2005,
  {Extreme Value and Related Models with Applications in Engineering and
  Science} (Hoboken, NJ: Wiley-Interscience)

\bibitem[{{Chabrier}(2003)}]{Chabrier03}
{Chabrier}, G. 2003, \apjl, 586, L133, \dodoi{10.1086/374879}

\bibitem[{{Chen} \& {Garrett}(2021)}]{Chen21}
{Chen}, H., \& {Garrett}, M.~A. 2021, \mnras, 507, 3761,
  \dodoi{10.1093/mnras/stab2207}

\bibitem[{{Chennamangalam} {et~al.}(2015){Chennamangalam}, {Siemion},
  {Lorimer}, \& {Werthimer}}]{Chennamangalam15}
{Chennamangalam}, J., {Siemion}, A. P.~V., {Lorimer}, D.~R., \& {Werthimer}, D.
  2015, \na, 34, 245, \dodoi{10.1016/j.newast.2014.07.011}

\bibitem[{{Chiu} {et~al.}(2013){Chiu}, {Stoyan}, {Kendall}, \&
  {Mecke}}]{Chiu13}
{Chiu}, S.~N., {Stoyan}, D., {Kendall}, W.~S., \& {Mecke}, J. 2013, {Stochastic
  Geometry and its Applications: Third Edition} (New York: Wiley),
  \dodoi{10.1002/9781118658222}

\bibitem[{{{\'C}irkovi{\'c}}(2018{\natexlab{a}})}]{Cirkovic18-Catch}
{{\'C}irkovi{\'c}}, M.~M. 2018{\natexlab{a}}, Acta Astronautica, 152, 289,
  \dodoi{10.1016/j.actaastro.2018.07.051}

\bibitem[{{{\'C}irkovi{\'c}}(2018{\natexlab{b}})}]{Cirkovic18-Book}
---. 2018{\natexlab{b}}, {The Great Silence: Science and Philosophy of Fermi's
  Paradox} (New York: Oxford University Press)

\bibitem[{{{\'C}irkovi{\'c}} \& {Bradbury}(2006)}]{Cirkovic06-OuterGalaxy}
{{\'C}irkovi{\'c}}, M.~M., \& {Bradbury}, R.~J. 2006, \na, 11, 628,
  \dodoi{10.1016/j.newast.2006.04.003}

\bibitem[{{{\'C}irkovi{\'c}} \& {Vukoti{\'c}}(2008)}]{Cirkovic08}
{{\'C}irkovi{\'c}}, M.~M., \& {Vukoti{\'c}}, B. 2008, Origins of Life and
  Evolution of the Biosphere, 38, 535, \dodoi{10.1007/s11084-008-9149-y}

\bibitem[{{{\'C}irkovi{\'c}} \& {Vukoti{\'c}}(2020)}]{Cirkovic20}
---. 2020, arXiv e-prints, arXiv:2007.12645, \dodoi{10.48550/arXiv.2007.12645}

\bibitem[{{Coles} {et~al.}(2001){Coles}, {Bawa}, {Trenner}, \&
  {Dorazio}}]{Coles01}
{Coles}, S., {Bawa}, J., {Trenner}, L., \& {Dorazio}, P. 2001, {An Introduction
  to Statistical Modeling of Extreme Values}, Vol. 208 (Berlin: Springer),
  \dodoi{10.1007/978-1-4471-3675-0}

\bibitem[{{Conroy} {et~al.}(2014){Conroy}, {Graves}, \& {van
  Dokkum}}]{Conroy14}
{Conroy}, C., {Graves}, G.~J., \& {van Dokkum}, P.~G. 2014, \apj, 780, 33,
  \dodoi{10.1088/0004-637X/780/1/33}

\bibitem[{{Corbet}(1997)}]{Corbet97}
{Corbet}, R.~H.~D. 1997, Journal of the British Interplanetary Society, 50, 253

\bibitem[{{Corbet}(2003)}]{Corbet03}
{Corbet}, R. H.~D. 2003, Astrobiology, 3, 305,
  \dodoi{10.1089/153110703769016398}

\bibitem[{{Cordes} {et~al.}(1997){Cordes}, {Lazio}, \& {Sagan}}]{Cordes97}
{Cordes}, J.~M., {Lazio}, J.~W., \& {Sagan}, C. 1997, \apj, 487, 782,
  \dodoi{10.1086/304620}

\bibitem[{{Crick} \& {Orgel}(1973)}]{Crick73}
{Crick}, F.~H.~C., \& {Orgel}, L.~E. 1973, \icarus, 19, 341,
  \dodoi{10.1016/0019-1035(73)90110-3}

\bibitem[{{Czech} {et~al.}(2021){Czech}, {Isaacson}, {Pearce}, {Cox}, {Sheikh},
  {Brzycki}, {Buchner}, {Croft}, {DeBoer}, {DeMarines}, {Drew}, {Gajjar},
  {Lacki}, {Lebofsky}, {MacMahon}, {Ng}, {de Pater}, {Price}, {Siemion}, {Van
  Rooyen}, \& {Pete Worden}}]{Czech21}
{Czech}, D., {Isaacson}, H., {Pearce}, L., {et~al.} 2021, \pasp, 133, 064502,
  \dodoi{10.1088/1538-3873/abf329}

\bibitem[{{Daley} \& {Vere-Jones}(2003)}]{Daley03}
{Daley}, D.~J., \& {Vere-Jones}, D. 2003, {An Introduction to the Theory of
  Point Processes. Volume I: Elementary Theory and Methods} (New York:
  Springer), \dodoi{10.1007/b97277}

\bibitem[{{Daley} \& {Vere-Jones}(2008)}]{Daley08}
---. 2008, {An Introduction to the Theory of Point Processes. Volume II:
  General Theory and Structure} (New York: Springer),
  \dodoi{10.1007/978-0-387-49835-5}

\bibitem[{{Davies} \& {Wagner}(2013)}]{Davies13}
{Davies}, P.~C.~W., \& {Wagner}, R.~V. 2013, Acta Astronautica, 89, 261,
  \dodoi{10.1016/j.actaastro.2011.10.022}

\bibitem[{{Dayal} {et~al.}(2015){Dayal}, {Cockell}, {Rice}, \&
  {Mazumdar}}]{Dayal15}
{Dayal}, P., {Cockell}, C., {Rice}, K., \& {Mazumdar}, A. 2015, \apjl, 810, L2,
  \dodoi{10.1088/2041-8205/810/1/L2}

\bibitem[{{Di Stefano} \& {Ray}(2016)}]{DiStefano16}
{Di Stefano}, R., \& {Ray}, A. 2016, \apj, 827, 54,
  \dodoi{10.3847/0004-637X/827/1/54}

\bibitem[{{Djorgovski} {et~al.}(2013){Djorgovski}, {Mahabal}, {Drake},
  {Graham}, \& {Donalek}}]{Djorgovski13}
{Djorgovski}, S.~G., {Mahabal}, A., {Drake}, A., {Graham}, M., \& {Donalek}, C.
  2013, in Planets, Stars and Stellar Systems. Volume 2: Astronomical
  Techniques, Software and Data, ed. T.~D. {Oswalt} \& H.~E. {Bond} (Springer),
  223, \dodoi{10.1007/978-94-007-5618-2_5}

\bibitem[{{Drake} {et~al.}(1973){Drake}, {Kardashev}, {Troitsky}, {Gindilis},
  {Petrovich}, {Pariisky}, {Moroz}, {Oliver}, \& {Townes}}]{Drake73-L}
{Drake}, {Kardashev}, N.~S., {Troitsky}, {et~al.} 1973, in Communication With
  Extraterrestrial Intelligence, ed. C.~{Sagan} (Cambridge, MA: MIT Press), 230

\bibitem[{{Drake}(1961)}]{Drake61}
{Drake}, F.~D. 1961, Physics Today, 14, 40, \dodoi{10.1063/1.3057500}

\bibitem[{{Drake}(1965)}]{Drake65}
---. 1965, in Current Aspects of Exobiology, ed. G.~{Mamikunian} \& M.~H.
  {Briggs} (Oxford: Pergamon Press), 323--345,
  \dodoi{10.1016/b978-1-4832-0047-7.50015-0}

\bibitem[{{Drake} \& {Sagan}(1973)}]{Drake73-Freq}
{Drake}, F.~D., \& {Sagan}, C. 1973, \nat, 245, 257, \dodoi{10.1038/245257a0}

\bibitem[{{Dreher}(2004)}]{Dreher04}
{Dreher}, J.~W. 2004, in Bioastronomy 2002: Life Among the Stars, ed.
  R.~{Norris} \& F.~{Stootman}, Vol. 213 (San Francisco: Astronomical Society
  of the Pacific), 467, \dodoi{10.1017/s0074180900193726}

\bibitem[{{Dyson}(1960)}]{Dyson60}
{Dyson}, F.~J. 1960, Science, 131, 1667, \dodoi{10.1126/science.131.3414.1667}

\bibitem[{{Dyson}(1963)}]{Dyson63}
{Dyson}, F.~J. 1963, in Interstellar Communication, ed. A.~G.~W. {Cameron} (New
  York: W. A. Benjamin), 115--120

\bibitem[{{Embrechts} {et~al.}(2013){Embrechts}, {Kl{\"u}ppelberg}, \&
  {Mikosch}}]{Embrechts13}
{Embrechts}, P., {Kl{\"u}ppelberg}, C., \& {Mikosch}, T. 2013, {Modelling
  Extremal Events: for Insurance and Finance}, Vol.~33 (Berlin: Springer),
  \dodoi{10.1007/978-3-642-33483-2}

\bibitem[{{Enriquez} {et~al.}(2017){Enriquez}, {Siemion}, {Foster}, {Gajjar},
  {Hellbourg}, {Hickish}, {Isaacson}, {Price}, {Croft}, {DeBoer}, {Lebofsky},
  {MacMahon}, \& {Werthimer}}]{Enriquez17}
{Enriquez}, J.~E., {Siemion}, A., {Foster}, G., {et~al.} 2017, \apj, 849, 104,
  \dodoi{10.3847/1538-4357/aa8d1b}

\bibitem[{{Fogg}(1987)}]{Fogg87}
{Fogg}, M.~J. 1987, \icarus, 69, 370, \dodoi{10.1016/0019-1035(87)90112-6}

\bibitem[{{Forgan}(2017)}]{Forgan17}
{Forgan}, D.~H. 2017, International Journal of Astrobiology, 16, 349,
  \dodoi{10.1017/S1473550416000392}

\bibitem[{{Forgan}(2019)}]{Forgan19}
---. 2019, {Solving Fermi's Paradox} (Cambridge: Cambridge University Press),
  \dodoi{10.1017/9781316681510}

\bibitem[{{Freitas} \& {Valdes}(1985)}]{Freitas85}
{Freitas}, R.~A., J., \& {Valdes}, F. 1985, Acta Astronautica, 12, 1027,
  \dodoi{10.1016/0094-5765(85)90031-1}

\bibitem[{{Gajjar} {et~al.}(2021){Gajjar}, {Perez}, {Siemion}, {Foster},
  {Brzycki}, {Chatterjee}, {Chen}, {Cordes}, {Croft}, {Czech}, {DeBoer},
  {DeMarines}, {Drew}, {Gowanlock}, {Isaacson}, {Lacki}, {Lebofsky},
  {MacMahon}, {Morrison}, {Ng}, {de Pater}, {Price}, {Sheikh}, {Suresh},
  {Webb}, \& {Pete Worden}}]{Gajjar21}
{Gajjar}, V., {Perez}, K.~I., {Siemion}, A. P.~V., {et~al.} 2021, \aj, 162, 33,
  \dodoi{10.3847/1538-3881/abfd36}

\bibitem[{{Gallazzi} {et~al.}(2005){Gallazzi}, {Charlot}, {Brinchmann},
  {White}, \& {Tremonti}}]{Gallazzi05}
{Gallazzi}, A., {Charlot}, S., {Brinchmann}, J., {White}, S. D.~M., \&
  {Tremonti}, C.~A. 2005, \mnras, 362, 41,
  \dodoi{10.1111/j.1365-2966.2005.09321.x}

\bibitem[{{Garrett}(2015)}]{Garrett15}
{Garrett}, M.~A. 2015, \aap, 581, L5, \dodoi{10.1051/0004-6361/201526687}

\bibitem[{{Garrett} \& {Siemion}(2023)}]{Garrett23}
{Garrett}, M.~A., \& {Siemion}, A.~P.~V. 2023, \mnras, 519, 4581,
  \dodoi{10.1093/mnras/stac2607}

\bibitem[{{Geringer-Sameth} {et~al.}(2015){Geringer-Sameth}, {Koushiappas}, \&
  {Walker}}]{GeringerSameth15}
{Geringer-Sameth}, A., {Koushiappas}, S.~M., \& {Walker}, M.~G. 2015, \prd, 91,
  083535, \dodoi{10.1103/PhysRevD.91.083535}

\bibitem[{{Glade} {et~al.}(2012){Glade}, {Ballet}, \& {Bastien}}]{Glade12}
{Glade}, N., {Ballet}, P., \& {Bastien}, O. 2012, International Journal of
  Astrobiology, 11, 103, \dodoi{10.1017/S1473550411000413}

\bibitem[{{Gonzalez} {et~al.}(2001){Gonzalez}, {Brownlee}, \&
  {Ward}}]{Gonzalez01}
{Gonzalez}, G., {Brownlee}, D., \& {Ward}, P. 2001, \icarus, 152, 185,
  \dodoi{10.1006/icar.2001.6617}

\bibitem[{{Gowanlock}(2016)}]{Gowanlock16}
{Gowanlock}, M.~G. 2016, \apj, 832, 38, \dodoi{10.3847/0004-637X/832/1/38}

\bibitem[{{Gray} \& {Ellingsen}(2002)}]{Gray02}
{Gray}, R.~H., \& {Ellingsen}, S. 2002, \apj, 578, 967, \dodoi{10.1086/342646}

\bibitem[{{Gray} \& {Mooley}(2017)}]{Gray17}
{Gray}, R.~H., \& {Mooley}, K. 2017, \aj, 153, 110,
  \dodoi{10.3847/1538-3881/153/3/110}

\bibitem[{{Griffith} {et~al.}(2015){Griffith}, {Wright}, {Maldonado}, {Povich},
  {Sigur{\dj}sson}, \& {Mullan}}]{Griffith15}
{Griffith}, R.~L., {Wright}, J.~T., {Maldonado}, J., {et~al.} 2015, \apjs, 217,
  25, \dodoi{10.1088/0067-0049/217/2/25}

\bibitem[{{Gulkis}(1985)}]{Gulkis85}
{Gulkis}, S. 1985, in IAU Symposium, Vol. 112, The Search for Extraterrestrial
  Life: Recent Developments, ed. M.~D. {Papagiannis} (Dordrecht: D. Reidel
  Publishing Co.), 411--417, \dodoi{10.1007/978-94-009-5462-5_52}

\bibitem[{{Gumbel}(1958)}]{Gumbel58}
{Gumbel}, E.~J. 1958, Statistics of Extremes (New York: Columbia University
  Press), \dodoi{10.7312/gumb92958}

\bibitem[{{Haenggi}(2013)}]{Haenggi13}
{Haenggi}, M. 2013, Stochastic Geometry for Wireless Networks (Cambridge:
  Cambridge University Press), \dodoi{10.1017/cbo9781139043816}

\bibitem[{{Haigh}(2013)}]{Haigh13}
{Haigh}, J. 2013, Probability Models (London: Springer London),
  \dodoi{10.1007/978-1-4471-5343-6}

\bibitem[{{Hair}(2011)}]{Hair11}
{Hair}, T.~W. 2011, International Journal of Astrobiology, 10, 131,
  \dodoi{10.1017/S1473550411000024}

\bibitem[{{Hanski}(1998)}]{Hanski98}
{Hanski}, I. 1998, \nat, 396, 41, \dodoi{10.1038/23876}

\bibitem[{{Hanson} {et~al.}(2021){Hanson}, {Martin}, {McCarter}, \&
  {Paulson}}]{Hanson21}
{Hanson}, R., {Martin}, D., {McCarter}, C., \& {Paulson}, J. 2021, \apj, 922,
  182, \dodoi{10.3847/1538-4357/ac2369}

\bibitem[{{Haqq-Misra} \& {Baum}(2009)}]{HaqqMisra09}
{Haqq-Misra}, J.~D., \& {Baum}, S.~D. 2009, Journal of the British
  Interplanetary Society, 62, 47.
\newblock \doarXiv{0906.0568}

\bibitem[{{Harp} {et~al.}(2018){Harp}, {Ackermann}, {Astorga}, {Arbunich},
  {Barrios}, {Hightower}, {Meitzner}, {Barott}, {Nolan}, {Messerschmitt},
  {Vakoch}, {Shostak}, \& {Tarter}}]{Harp18}
{Harp}, G.~R., {Ackermann}, R.~F., {Astorga}, A., {et~al.} 2018, \apj, 869, 66,
  \dodoi{10.3847/1538-4357/aaeb98}

\bibitem[{{Harris}(1986)}]{Harris86}
{Harris}, M.~J. 1986, \apss, 123, 297, \dodoi{10.1007/BF00653949}

\bibitem[{{Hart}(1975)}]{Hart75}
{Hart}, M.~H. 1975, \qjras, 16, 128

\bibitem[{{Hart}(1979)}]{Hart79}
---. 1979, \icarus, 37, 351, \dodoi{10.1016/0019-1035(79)90141-6}

\bibitem[{{Harwit}(1981)}]{Harwit81}
{Harwit}, M. 1981, {Cosmic discovery: The search, scope, and heritage of
  astronomy} (New York: Basic Books, Inc.)

\bibitem[{{Hippke}(2018)}]{Hippke18-Messenger}
{Hippke}, M. 2018, Acta Astronautica, 151, 53,
  \dodoi{10.1016/j.actaastro.2018.05.038}

\bibitem[{{Hippke} \& {Forgan}(2017)}]{Hippke17-XRay}
{Hippke}, M., \& {Forgan}, D.~H. 2017, arXiv e-prints, arXiv:1711.05761,
  \dodoi{10.48550/arXiv.1711.05761}

\bibitem[{{Hogg}(1999)}]{Hogg99}
{Hogg}, D.~W. 1999, arXiv e-prints, astro.
\newblock \doarXiv{astro-ph/9905116}

\bibitem[{{Horowitz} \& {Sagan}(1993)}]{Horowitz93}
{Horowitz}, P., \& {Sagan}, C. 1993, \apj, 415, 218, \dodoi{10.1086/173157}

\bibitem[{{Howard} {et~al.}(2004){Howard}, {Horowitz}, {Wilkinson}, {Coldwell},
  {Groth}, {Jarosik}, {Latham}, {Stefanik}, {Willman}, {Wolff}, \&
  {Zajac}}]{Howard04}
{Howard}, A.~W., {Horowitz}, P., {Wilkinson}, D.~T., {et~al.} 2004, \apj, 613,
  1270, \dodoi{10.1086/423300}

\bibitem[{{Huang}(1959)}]{Huang59}
{Huang}, S.-S. 1959, \pasp, 71, 421, \dodoi{10.1086/127417}

\bibitem[{{Imara} \& {Di Stefano}(2018)}]{Imara18}
{Imara}, N., \& {Di Stefano}, R. 2018, \apj, 859, 40,
  \dodoi{10.3847/1538-4357/aab903}

\bibitem[{{Inoue} \& {Yokoo}(2011)}]{Inoue11}
{Inoue}, M., \& {Yokoo}, H. 2011, Journal of the British Interplanetary
  Society, 64, 59

\bibitem[{{Isaacson} {et~al.}(2017){Isaacson}, {Siemion}, {Marcy}, {Lebofsky},
  {Price}, {MacMahon}, {Croft}, {DeBoer}, {Hickish}, {Werthimer}, {Sheikh},
  {Hellbourg}, \& {Enriquez}}]{Isaacson17}
{Isaacson}, H., {Siemion}, A. P.~V., {Marcy}, G.~W., {et~al.} 2017, \pasp, 129,
  054501, \dodoi{10.1088/1538-3873/aa5800}

\bibitem[{{Johnson} {et~al.}(2010){Johnson}, {Aller}, {Howard}, \&
  {Crepp}}]{Johnson10}
{Johnson}, J.~A., {Aller}, K.~M., {Howard}, A.~W., \& {Crepp}, J.~R. 2010,
  \pasp, 122, 905, \dodoi{10.1086/655775}

\bibitem[{{Jones}(1981)}]{Jones81}
{Jones}, E.~M. 1981, \icarus, 46, 328, \dodoi{10.1016/0019-1035(81)90136-6}

\bibitem[{{Jusup} {et~al.}(2022){Jusup}, {Holme}, {Kanazawa}, {Takayasu},
  {Romi{\'c}}, {Wang}, {Ge{\v{c}}ek}, {Lipi{\'c}}, {Podobnik}, {Wang}, {Luo},
  {Klanj{\v{s}}{\v{c}}ek}, {Fan}, {Boccaletti}, \& {Perc}}]{Jusup22}
{Jusup}, M., {Holme}, P., {Kanazawa}, K., {et~al.} 2022, \physrep, 948, 1,
  \dodoi{10.1016/j.physrep.2021.10.005}

\bibitem[{{Kardashev}(1985)}]{Kardashev85}
{Kardashev}, N.~S. 1985, in IAU Symposium, Vol. 112, The Search for
  Extraterrestrial Life: Recent Developments, ed. M.~D. {Papagiannis}
  (Dordrecht: D. Reidel Publishing Co.), 497--504,
  \dodoi{10.1007/978-94-009-5462-5_65}

\bibitem[{{Karlis} \& {Xekalaki}(2005)}]{Karlis05}
{Karlis}, D., \& {Xekalaki}, E. 2005, International Statistical Review/Revue
  Internationale de Statistique, 73, 35,
  \dodoi{10.1111/j.1751-5823.2005.tb00250.x}

\bibitem[{{Kasting} {et~al.}(1993){Kasting}, {Whitmire}, \&
  {Reynolds}}]{Kasting93}
{Kasting}, J.~F., {Whitmire}, D.~P., \& {Reynolds}, R.~T. 1993, \icarus, 101,
  108, \dodoi{10.1006/icar.1993.1010}

\bibitem[{{Kemp}(1967)}]{Kemp67}
{Kemp}, C.~D. 1967, Journal of the Statistical and Social Inquiry Society of
  Ireland, 21, 151

\bibitem[{{Kingman}(1993)}]{Kingman93}
{Kingman}, J. F.~C. 1993, {Poisson Processes} (Oxford: Clarendon Press),
  \dodoi{10.1093/oso/9780198536932.001.0001}

\bibitem[{{Kipping}(2021)}]{Kipping21}
{Kipping}, D. 2021, Research Notes of the American Astronomical Society, 5, 44,
  \dodoi{10.3847/2515-5172/abeb7b}

\bibitem[{{Kipping} {et~al.}(2020){Kipping}, {Frank}, \& {Scharf}}]{Kipping20}
{Kipping}, D., {Frank}, A., \& {Scharf}, C. 2020, International Journal of
  Astrobiology, 19, 430, \dodoi{10.1017/S1473550420000208}

\bibitem[{{Kipping} \& {Gray}(2022)}]{Kipping22}
{Kipping}, D., \& {Gray}, R. 2022, \mnras, 515, 1122,
  \dodoi{10.1093/mnras/stac1807}

\bibitem[{{Kirsten} {et~al.}(2022){Kirsten}, {Marcote}, {Nimmo}, {Hessels},
  {Bhardwaj}, {Tendulkar}, {Keimpema}, {Yang}, {Snelders}, {Scholz},
  {Pearlman}, {Law}, {Peters}, {Giroletti}, {Paragi}, {Bassa}, {Hewitt},
  {Bach}, {Bezrukovs}, {Burgay}, {Buttaccio}, {Conway}, {Corongiu}, {Feiler},
  {Forss{\'e}n}, {Gawro{\'n}ski}, {Karuppusamy}, {Kharinov}, {Lindqvist},
  {Maccaferri}, {Melnikov}, {Ould-Boukattine}, {Possenti}, {Surcis}, {Wang},
  {Yuan}, {Aggarwal}, {Anna-Thomas}, {Bower}, {Blaauw}, {Burke-Spolaor},
  {Cassanelli}, {Clarke}, {Fonseca}, {Gaensler}, {Gopinath}, {Kaspi}, {Kassim},
  {Lazio}, {Leung}, {Li}, {Lin}, {Masui}, {Mckinven}, {Michilli}, {Mikhailov},
  {Ng}, {Orbidans}, {Pen}, {Petroff}, {Rahman}, {Ransom}, {Shin}, {Smith},
  {Stairs}, \& {Vlemmings}}]{Kirsten22}
{Kirsten}, F., {Marcote}, B., {Nimmo}, K., {et~al.} 2022, \nat, 602, 585,
  \dodoi{10.1038/s41586-021-04354-w}

\bibitem[{{Klenke}(2020)}]{Klenke20}
{Klenke}, A. 2020, Probability Theory: A Comprehensive Course (Cham,
  Switzerland: Springer Cham), \dodoi{10.1007/978-3-030-56402-5}

\bibitem[{{Kuiper} \& {Morris}(1977)}]{Kuiper77}
{Kuiper}, T.~B.~H., \& {Morris}, M. 1977, Science, 196, 616,
  \dodoi{10.1126/science.196.4290.616}

\bibitem[{{Lacki}(2016)}]{Lacki16-K3}
{Lacki}, B.~C. 2016, arXiv e-prints, arXiv:1604.07844.
\newblock \doarXiv{1604.07844}

\bibitem[{{Lacki}(2019)}]{Lacki19-Sunscreen}
---. 2019, \pasp, 131, 024102, \dodoi{10.1088/1538-3873/aaf3df}

\bibitem[{{Lacki}(2020)}]{Lacki20-LensFlare}
---. 2020, \apj, 905, 18, \dodoi{10.3847/1538-4357/abc1e3}

\bibitem[{{Lacki}(2021)}]{Lacki21-Traversability}
---. 2021, International Journal of Astrobiology, 20, 359,
  \dodoi{10.1017/S1473550421000252}

\bibitem[{{Landis}(1998)}]{Landis98}
{Landis}, G.~A. 1998, Journal of the British Interplanetary Society, 51, 163

\bibitem[{{Last} \& {Penrose}(2017)}]{Last17}
{Last}, G., \& {Penrose}, M. 2017, {Lectures on the Poisson Process}
  (Cambridge: Cambridge University Press), \dodoi{10.1017/9781316104477}

\bibitem[{{Laughlin} {et~al.}(1997){Laughlin}, {Bodenheimer}, \&
  {Adams}}]{Laughlin97}
{Laughlin}, G., {Bodenheimer}, P., \& {Adams}, F.~C. 1997, \apj, 482, 420,
  \dodoi{10.1086/304125}

\bibitem[{{Lawrence} {et~al.}(2017){Lawrence}, {Vander Wiel}, {Law}, {Burke
  Spolaor}, \& {Bower}}]{Lawrence17}
{Lawrence}, E., {Vander Wiel}, S., {Law}, C., {Burke Spolaor}, S., \& {Bower},
  G.~C. 2017, \aj, 154, 117, \dodoi{10.3847/1538-3881/aa844e}

\bibitem[{{Learned} {et~al.}(2012){Learned}, {Kudritzki}, {Pakvasa}, \&
  {Zee}}]{Learned08}
{Learned}, J.~G., {Kudritzki}, R.~P., {Pakvasa}, S., \& {Zee}, A. 2012,
  Contemporary Physics, 53, 113, \dodoi{10.1080/00107514.2011.640142}

\bibitem[{{Learned} {et~al.}(1994){Learned}, {Pakvasa}, {Simmons}, \&
  {Tata}}]{Learned94}
{Learned}, J.~G., {Pakvasa}, S., {Simmons}, W.~A., \& {Tata}, X. 1994, \qjras,
  35, 321

\bibitem[{Leibold {et~al.}(2004)Leibold, Holyoak, Mouquet, Amarasekare, Chase,
  Hoopes, Holt, Shurin, Law, Tilman, {et~al.}}]{Leibold04}
Leibold, M.~A., Holyoak, M., Mouquet, N., {et~al.} 2004, Ecology Letters, 7,
  601, \dodoi{10.1111/j.1461-0248.2004.00608.x}

\bibitem[{{Lin} {et~al.}(2014){Lin}, {Gonzalez Abad}, \& {Loeb}}]{Lin14}
{Lin}, H.~W., {Gonzalez Abad}, G., \& {Loeb}, A. 2014, \apjl, 792, L7,
  \dodoi{10.1088/2041-8205/792/1/L7}

\bibitem[{{Lineweaver} {et~al.}(2004){Lineweaver}, {Fenner}, \&
  {Gibson}}]{Lineweaver04}
{Lineweaver}, C.~H., {Fenner}, Y., \& {Gibson}, B.~K. 2004, Science, 303, 59,
  \dodoi{10.1126/science.1092322}

\bibitem[{{Lingam} {et~al.}(2019){Lingam}, {Ginsburg}, \& {Bialy}}]{Lingam19}
{Lingam}, M., {Ginsburg}, I., \& {Bialy}, S. 2019, \apj, 877, 62,
  \dodoi{10.3847/1538-4357/ab1b2f}

\bibitem[{{Lingam} \& {Loeb}(2017)}]{Lingam17}
{Lingam}, M., \& {Loeb}, A. 2017, \apjl, 837, L23,
  \dodoi{10.3847/2041-8213/aa633e}

\bibitem[{{Lingam} \& {Loeb}(2020)}]{Lingam20}
---. 2020, \apj, 894, 36, \dodoi{10.3847/1538-4357/ab7dc7}

\bibitem[{{Lingam} \& {Loeb}(2021)}]{Lingam21}
---. 2021, {Life in the Cosmos: From Biosignatures to Technosignatures}
  (Cambridge, MA: Harvard University Press), \dodoi{10.4159/9780674259959}

\bibitem[{{Livio}(1999)}]{Livio99}
{Livio}, M. 1999, \apj, 511, 429, \dodoi{10.1086/306668}

\bibitem[{{Maire} {et~al.}(2019){Maire}, {Wright}, {Barrett}, {Dexter},
  {Dorval}, {Duenas}, {Drake}, {Hultgren}, {Isaacson}, {Marcy}, {Meyer},
  {Ramos}, {Shirman}, {Siemion}, {Stone}, {Tallis}, {Tellis}, {Treffers}, \&
  {Werthimer}}]{Maire19}
{Maire}, J., {Wright}, S.~A., {Barrett}, C.~T., {et~al.} 2019, \aj, 158, 203,
  \dodoi{10.3847/1538-3881/ab44d3}

\bibitem[{{Makovetskii}(1977)}]{Makovetskii77}
{Makovetskii}, P.~V. 1977, \sovast, 21, 251

\bibitem[{{Mart{\'\i}nez} \& {Saar}(2002)}]{Martinez02}
{Mart{\'\i}nez}, V.~J., \& {Saar}, E. 2002, {Statistics of the Galaxy
  Distribution} (Boca Raton, FL: Chapman \& Hall/CRC Press),
  \dodoi{10.1201/9781420036169}

\bibitem[{{Messerschmitt}(2015)}]{Messerschmitt15}
{Messerschmitt}, D.~G. 2015, Acta Astronautica, 107, 20,
  \dodoi{10.1016/j.actaastro.2014.11.007}

\bibitem[{{Miller} \& {Scalo}(1979)}]{Miller79}
{Miller}, G.~E., \& {Scalo}, J.~M. 1979, \apjs, 41, 513, \dodoi{10.1086/190629}

\bibitem[{{Moustakas} {et~al.}(2013){Moustakas}, {Coil}, {Aird}, {Blanton},
  {Cool}, {Eisenstein}, {Mendez}, {Wong}, {Zhu}, \& {Arnouts}}]{Moustakas13}
{Moustakas}, J., {Coil}, A.~L., {Aird}, J., {et~al.} 2013, \apj, 767, 50,
  \dodoi{10.1088/0004-637X/767/1/50}

\bibitem[{{Moyal}(1962{\natexlab{a}})}]{Moyal62-Mult}
{Moyal}, J.~E. 1962{\natexlab{a}}, Proceedings of the Royal Society of London.
  Series A. Mathematical and Physical Sciences, 266, 518,
  \dodoi{10.1098/rspa.1962.0075}

\bibitem[{{Moyal}(1962{\natexlab{b}})}]{Moyal62-Gen}
---. 1962{\natexlab{b}}, Acta Mathematica, 1, 1, \dodoi{10.1007/bf02545761}

\bibitem[{{Mulders} {et~al.}(2015){Mulders}, {Pascucci}, \& {Apai}}]{Mulders15}
{Mulders}, G.~D., {Pascucci}, I., \& {Apai}, D. 2015, \apj, 814, 130,
  \dodoi{10.1088/0004-637X/814/2/130}

\bibitem[{{Muno} {et~al.}(2005){Muno}, {Pfahl}, {Baganoff}, {Brandt}, {Ghez},
  {Lu}, \& {Morris}}]{Muno05}
{Muno}, M.~P., {Pfahl}, E., {Baganoff}, F.~K., {et~al.} 2005, \apjl, 622, L113,
  \dodoi{10.1086/429721}

\bibitem[{{Napier}(2004)}]{Napier04}
{Napier}, W.~M. 2004, \mnras, 348, 46, \dodoi{10.1111/j.1365-2966.2004.07287.x}

\bibitem[{{Neyman} \& {Scott}(1952)}]{Neyman52}
{Neyman}, J., \& {Scott}, E.~L. 1952, \apj, 116, 144, \dodoi{10.1086/145599}

\bibitem[{{Neyman} \& {Scott}(1958)}]{Neyman58}
---. 1958, Journal of the Royal Statistical Society: Series B (Methodological),
  20, 1

\bibitem[{{Nishino} \& {Seto}(2018)}]{Nishino18}
{Nishino}, Y., \& {Seto}, N. 2018, \apjl, 862, L21,
  \dodoi{10.3847/2041-8213/aad33d}

\bibitem[{{Oliver} \& {Billingham}(1971)}]{Oliver71}
{Oliver}, B.~M., \& {Billingham}, J. 1971, {Project Cyclops: A Design Study of
  a System for Detecting Extraterrestrial Intelligent Life}, Vol.
  NASA-CR-114445 (Mountain View, CA: NASA Ames Research Center)

\bibitem[{{Olson}(2015)}]{Olson15}
{Olson}, S.~J. 2015, Classical and Quantum Gravity, 32, 215025,
  \dodoi{10.1088/0264-9381/32/21/215025}

\bibitem[{{Olson}(2016)}]{Olson16}
---. 2016, \jcap, 2016, 021, \dodoi{10.1088/1475-7516/2016/04/021}

\bibitem[{{Osmanov}(2016)}]{Osmanov16}
{Osmanov}, Z. 2016, International Journal of Astrobiology, 15, 127,
  \dodoi{10.1017/S1473550415000257}

\bibitem[{{Pace} \& {Walker}(1975)}]{Pace75}
{Pace}, G.~W., \& {Walker}, J.~C.~G. 1975, \nat, 254, 400,
  \dodoi{10.1038/254400a0}

\bibitem[{{Pooley} {et~al.}(2003){Pooley}, {Lewin}, {Anderson}, {Baumgardt},
  {Filippenko}, {Gaensler}, {Homer}, {Hut}, {Kaspi}, {Makino}, {Margon},
  {McMillan}, {Portegies Zwart}, {van der Klis}, \& {Verbunt}}]{Pooley03}
{Pooley}, D., {Lewin}, W. H.~G., {Anderson}, S.~F., {et~al.} 2003, \apjl, 591,
  L131, \dodoi{10.1086/377074}

\bibitem[{{Prantzos}(2020)}]{Prantzos20}
{Prantzos}, N. 2020, \mnras, 493, 3464, \dodoi{10.1093/mnras/staa512}

\bibitem[{{Price}(2021)}]{Price16}
{Price}, D.~C. 2021, in The WSPC Handbook of Astronomical Instrumentation,
  Volume 1: Radio Astronomic al Instrumentation, ed. A.~{Wolszczan} (Singapore:
  World Scientific), 159--179, \dodoi{10.1142/9789811203770_0007}

\bibitem[{{Price} {et~al.}(2020){Price}, {Enriquez}, {Brzycki}, {Croft},
  {Czech}, {DeBoer}, {DeMarines}, {Foster}, {Gajjar}, {Gizani}, {Hellbourg},
  {Isaacson}, {Lacki}, {Lebofsky}, {MacMahon}, {Pater}, {Siemion}, {Werthimer},
  {Green}, {Kaczmarek}, {Maddalena}, {Mader}, {Drew}, \& {Worden}}]{Price20}
{Price}, D.~C., {Enriquez}, J.~E., {Brzycki}, B., {et~al.} 2020, \aj, 159, 86,
  \dodoi{10.3847/1538-3881/ab65f1}

\bibitem[{{Rose} \& {Wright}(2004)}]{Rose04}
{Rose}, C., \& {Wright}, G. 2004, \nat, 431, 47, \dodoi{10.1038/nature02884}

\bibitem[{{Ross}(1996)}]{Ross96}
{Ross}, S.~M. 1996, {Stochastic Processes} (New York: John Wiley \& Sons)

\bibitem[{{Sagan}(1973)}]{Sagan73}
{Sagan}, C. 1973, \icarus, 19, 350, \dodoi{10.1016/0019-1035(73)90112-7}

\bibitem[{{Sagan}(1985)}]{Sagan85}
---. 1985, {Contact: A Novel} (New York: Simon \& Schuster, Inc.)

\bibitem[{{Scheffer}(1994)}]{Scheffer94}
{Scheffer}, L.~K. 1994, \qjras, 35, 157

\bibitem[{{Scheffer}(2014)}]{Scheffer14}
---. 2014, International Journal of Astrobiology, 13, 62,
  \dodoi{10.1017/S147355041300030X}

\bibitem[{{Scheuer}(1957)}]{Scheuer57}
{Scheuer}, P.~A.~G. 1957, Proceedings of the Cambridge Philosophical Society,
  53, 764, \dodoi{10.1017/S0305004100032825}

\bibitem[{{Schmidt} \& {Frank}(2019)}]{Schmidt19}
{Schmidt}, G.~A., \& {Frank}, A. 2019, International Journal of Astrobiology,
  18, 142, \dodoi{10.1017/S1473550418000095}

\bibitem[{{Schwartz} \& {Townes}(1961)}]{Schwartz61}
{Schwartz}, R.~N., \& {Townes}, C.~H. 1961, \nat, 190, 205,
  \dodoi{10.1038/190205a0}

\bibitem[{{Semiz} \& {O{\u{g}}ur}(2015)}]{Semiz15}
{Semiz}, {\.I}., \& {O{\u{g}}ur}, S. 2015, arXiv e-prints, arXiv:1503.04376,
  \dodoi{10.48550/arXiv.1503.04376}

\bibitem[{{Sheikh} {et~al.}(2019){Sheikh}, {Wright}, {Siemion}, \&
  {Enriquez}}]{Sheikh19}
{Sheikh}, S.~Z., {Wright}, J.~T., {Siemion}, A., \& {Enriquez}, J.~E. 2019,
  \apj, 884, 14, \dodoi{10.3847/1538-4357/ab3fa8}

\bibitem[{{Sheikh} {et~al.}(2021){Sheikh}, {Smith}, {Price}, {DeBoer}, {Lacki},
  {Czech}, {Croft}, {Gajjar}, {Isaacson}, {Lebofsky}, {MacMahon}, {Ng},
  {Perez}, {Siemion}, {Webb}, {Zic}, {Drew}, \& {Worden}}]{Sheikh21}
{Sheikh}, S.~Z., {Smith}, S., {Price}, D.~C., {et~al.} 2021, Nature Astronomy,
  5, 1153, \dodoi{10.1038/s41550-021-01508-8}

\bibitem[{{Shields} {et~al.}(2016){Shields}, {Ballard}, \&
  {Johnson}}]{Shields16}
{Shields}, A.~L., {Ballard}, S., \& {Johnson}, J.~A. 2016, \physrep, 663, 1,
  \dodoi{10.1016/j.physrep.2016.10.003}

\bibitem[{{Shostak} {et~al.}(1996){Shostak}, {Ekers}, \& {Vaile}}]{Shostak96}
{Shostak}, S., {Ekers}, R., \& {Vaile}, R. 1996, \aj, 112, 164,
  \dodoi{10.1086/117996}

\bibitem[{{Shvartsman} {et~al.}(1993){Shvartsman}, {Beskin}, {Mitronova},
  {Neizvestny}, {Plakhotnichenko}, \& {Pustil'nik}}]{Shvartsman93}
{Shvartsman}, V., {Beskin}, G., {Mitronova}, S., {et~al.} 1993, in Astronomical
  Society of the Pacific Conference Series, Vol.~47, Third Decennial US-USSR
  Conference on SETI, ed. G.~S. {Shostak} (San Francisco: Astronomical Society
  of the Pacific), 381

\bibitem[{{Siemion} {et~al.}(2010){Siemion}, {Von Korff}, {McMahon}, {Korpela},
  {Werthimer}, {Anderson}, {Bower}, {Cobb}, {Foster}, {Lebofsky}, {van
  Leeuwen}, \& {Wagner}}]{Siemion10}
{Siemion}, A., {Von Korff}, J., {McMahon}, P., {et~al.} 2010, Acta
  Astronautica, 67, 1342, \dodoi{10.1016/j.actaastro.2010.01.016}

\bibitem[{{Smart}(2009)}]{Smart09}
{Smart}, J.~M. 2009, in Cosmos \& Culture: Cultural Evolution in a Cosmic
  Context, ed. S.~J. {Dick} \& M.~L. {Lupisella}, Vol. 4802 (National
  Aeronautics and Space Administration), 201

\bibitem[{{Stevenson}(1999)}]{Stevenson99}
{Stevenson}, D.~J. 1999, \nat, 400, 32, \dodoi{10.1038/21811}

\bibitem[{{Suazo} {et~al.}(2022){Suazo}, {Zackrisson}, {Wright}, {Korn}, \&
  {Huston}}]{Suazo22}
{Suazo}, M., {Zackrisson}, E., {Wright}, J.~T., {Korn}, A.~J., \& {Huston}, M.
  2022, \mnras, 512, 2988, \dodoi{10.1093/mnras/stac280}

\bibitem[{{Subotowicz}(1979)}]{Subotowicz79}
{Subotowicz}, M. 1979, Acta Astronautica, 6, 213,
  \dodoi{10.1016/0094-5765(79)90157-7}

\bibitem[{{Tarter}(2001)}]{Tarter01}
{Tarter}, J. 2001, \araa, 39, 511, \dodoi{10.1146/annurev.astro.39.1.511}

\bibitem[{{Tarter}(1984)}]{Tarter84}
{Tarter}, J.~C. 1984, Acta Astronautica, 11, 387,
  \dodoi{10.1016/0094-5765(84)90079-1}

\bibitem[{{Tarter}(2007)}]{Tarter07}
---. 2007, Highlights of Astronomy, 14, 14, \dodoi{10.1017/S1743921307009829}

\bibitem[{{Tellis} \& {Marcy}(2017)}]{Tellis17}
{Tellis}, N.~K., \& {Marcy}, G.~W. 2017, \aj, 153, 251,
  \dodoi{10.3847/1538-3881/aa6d12}

\bibitem[{{Tipler}(1980)}]{Tipler80}
{Tipler}, F.~J. 1980, \qjras, 21, 267

\bibitem[{{Tonry} \& {Schneider}(1988)}]{Tonry88}
{Tonry}, J., \& {Schneider}, D.~P. 1988, \aj, 96, 807, \dodoi{10.1086/114847}

\bibitem[{{Tremblay} \& {Tingay}(2020)}]{Tremblay20}
{Tremblay}, C.~D., \& {Tingay}, S.~J. 2020, \pasa, 37, e035,
  \dodoi{10.1017/pasa.2020.27}

\bibitem[{{Vidal}(2011)}]{Vidal11}
{Vidal}, C. 2011, arXiv e-prints, arXiv:1104.4362.
\newblock \doarXiv{1104.4362}

\bibitem[{{Ward} \& {Brownlee}(2000)}]{Ward00}
{Ward}, P., \& {Brownlee}, D. 2000, {Rare Earth: Why Complex Life is Uncommon
  in the Universe} (New York: Copernicus Books), \dodoi{10.1007/b97646}

\bibitem[{{Wasserman}(2004)}]{Wasserman04}
{Wasserman}, L. 2004, All of Statistics: A Concise Course in Statistical
  Inference (New York: Springer New York), \dodoi{10.1007/978-0-387-21736-9}

\bibitem[{{Webb}(2015)}]{Webb15}
{Webb}, S. 2015, {If the Universe Is Teeming with Aliens... Where is
  Everybody?} (Cham, Switzerland: Springer Cham),
  \dodoi{10.1007/978-3-319-13236-5}

\bibitem[{{Whitmire} \& {Wright}(1980)}]{Whitmire80}
{Whitmire}, D.~P., \& {Wright}, D.~P. 1980, \icarus, 42, 149,
  \dodoi{10.1016/0019-1035(80)90253-5}

\bibitem[{{Wiegand} \& {Moloney}(2013)}]{Wiegand13}
{Wiegand}, T., \& {Moloney}, K.~A. 2013, {Handbook of Spatial Point-Pattern
  Analysis in Ecology} (New York: Chapman and Hall/CRC Press),
  \dodoi{10.1201/b16195}

\bibitem[{{Wiley}(2011)}]{Wiley11}
{Wiley}, K.~B. 2011, arXiv e-prints, arXiv:1111.6131,
  \dodoi{10.48550/arXiv.1111.6131}

\bibitem[{{Wlodarczyk-Sroka} {et~al.}(2020){Wlodarczyk-Sroka}, {Garrett}, \&
  {Siemion}}]{WlodarczykSroka20}
{Wlodarczyk-Sroka}, B.~S., {Garrett}, M.~A., \& {Siemion}, A.~P.~V. 2020,
  \mnras, 498, 5720, \dodoi{10.1093/mnras/staa2672}

\bibitem[{{Worden} {et~al.}(2017){Worden}, {Drew}, {Siemion}, {Werthimer},
  {DeBoer}, {Croft}, {MacMahon}, {Lebofsky}, {Isaacson}, {Hickish}, {Price},
  {Gajjar}, \& {Wright}}]{Worden17}
{Worden}, S.~P., {Drew}, J., {Siemion}, A., {et~al.} 2017, Acta Astronautica,
  139, 98, \dodoi{10.1016/j.actaastro.2017.06.008}

\bibitem[{{Wright}(2018)}]{Wright18-PITS}
{Wright}, J.~T. 2018, International Journal of Astrobiology, 17, 96,
  \dodoi{10.1017/S1473550417000143}

\bibitem[{{Wright} {et~al.}(2016){Wright}, {Cartier}, {Zhao}, {Jontof-Hutter},
  \& {Ford}}]{Wright16}
{Wright}, J.~T., {Cartier}, K. M.~S., {Zhao}, M., {Jontof-Hutter}, D., \&
  {Ford}, E.~B. 2016, \apj, 816, 17, \dodoi{10.3847/0004-637X/816/1/17}

\bibitem[{{Wright} {et~al.}(2022){Wright}, {Haqq-Misra}, {Frank}, {Kopparapu},
  {Lingam}, \& {Sheikh}}]{Wright22}
{Wright}, J.~T., {Haqq-Misra}, J., {Frank}, A., {et~al.} 2022, \apjl, 927, L30,
  \dodoi{10.3847/2041-8213/ac5824}

\bibitem[{{Wright} {et~al.}(2018){Wright}, {Kanodia}, \& {Lubar}}]{Wright18}
{Wright}, J.~T., {Kanodia}, S., \& {Lubar}, E. 2018, \aj, 156, 260,
  \dodoi{10.3847/1538-3881/aae099}

\bibitem[{{Wright} {et~al.}(2014){Wright}, {Mullan}, {Sigurdsson}, \&
  {Povich}}]{Wright14-Paradox}
{Wright}, J.~T., {Mullan}, B., {Sigurdsson}, S., \& {Povich}, M.~S. 2014, \apj,
  792, 26, \dodoi{10.1088/0004-637X/792/1/26}

\bibitem[{{Zackrisson} {et~al.}(2015){Zackrisson}, {Calissendorff}, {Asadi}, \&
  {Nyholm}}]{Zackrisson15}
{Zackrisson}, E., {Calissendorff}, P., {Asadi}, S., \& {Nyholm}, A. 2015, \apj,
  810, 23, \dodoi{10.1088/0004-637X/810/1/23}

\bibitem[{{Zackrisson} {et~al.}(2018){Zackrisson}, {Korn}, {Wehrhahn}, \&
  {Reiter}}]{Zackrisson18}
{Zackrisson}, E., {Korn}, A.~J., {Wehrhahn}, A., \& {Reiter}, J. 2018, \apj,
  862, 21, \dodoi{10.3847/1538-4357/aac386}

\bibitem[{{Zwicky}(1957)}]{Zwicky57}
{Zwicky}, F. 1957, {Morphological astronomy} (Berlin: Springer-Verlag),
  \dodoi{10.1007/978-3-642-87544-1}

\end{thebibliography}

\end{document}